\documentclass[aps,prd,showpacs,floats,onecolumn,nofootinbib]{revtex4}
\pdfoutput=1  
\usepackage{graphicx,color}
\usepackage{appendix}
\usepackage{subfigure}
\usepackage{latexsym,amsmath,amssymb,graphicx,booktabs}
\usepackage{hyperref}

\definecolor{MyBlue}{rgb}{0.15,0.15,0.70}

\hypersetup{
colorlinks=true,
citecolor=MyBlue,
linkcolor=MyBlue,
urlcolor=MyBlue
}

\usepackage{amssymb}
\usepackage{amsmath}
\usepackage{amsfonts}
\usepackage{upgreek}
\usepackage{latexsym}
 \usepackage{placeins}

\newcommand{\bk}{{\mathbf k}}
\newcommand{\bfe}{{\mathbf e}}

\newcommand{\HH}{{\cal H}}

\newcommand{\LL}{{\cal L}}

\newcommand{\VV}{{\cal V}}

\newcommand{\de}{\delta}

\newcommand{\ep}{\epsilon}

\newcommand{\La}{\Lambda}

\newcommand{\Om}{\Omega}
\newcommand{\om}{\omega}

\newcommand{\ra}{\rightarrow}

\newcommand{\be}{\begin{equation}}
\newcommand{\ee}{\end{equation}}
\newcommand{\gsim}{\stackrel{>}{\sim}}

\newcommand{\bea}{\begin{eqnarray}}
\newcommand{\eea}{\end{eqnarray}}
\newcommand{\bean}{\begin{eqnarray*}}
\newcommand{\eean}{\end{eqnarray*}}

\def\id{{\rm 1\kern -2.5pt I}}

\graphicspath{ {./Graphs-final/} }

\begin{document}

\vspace*{2cm}

\title{Gravitational waves in bigravity cosmology}
\author{Giulia Cusin, Ruth Durrer, Pietro Guarato and  Mariele Motta}
\affiliation{D\'epartement de Physique Th\'eorique and Center for Astroparticle Physics, Universit\'e de Gen\`eve, 24 quai Ansermet, CH--1211 Gen\`eve 4, Switzerland}
\email{giulia.cusin@unige.ch, ruth.durrer@unige.ch, pietro.guarato@unige.ch, mariele.motta@unige.ch}

\date{\today}

\begin{abstract}
In this paper we study gravitational wave perturbations in a cosmological setting of bigravity which can reproduce the $\La$CDM background and large scale structure. We show that in general gravitational wave perturbations are unstable and only for very fine tuned initial conditions such a cosmology is viable. We quantify this fine tuning. We argue that similar fine tuning is also required in the scalar sector in order to prevent the tensor instability  to be induced by second order scalar perturbations. Finally, we show that due to this power law instability, models of bigravity can lead to a large tensor to scalar  ratio even for low scale inflation.
\end{abstract}


\maketitle

\section{Introduction}
\label{s:intro}
The problem of dark energy is one of the fundamental problems not only of cosmology but of theoretical physics. How can vacuum energy, or equivalently a cosmological constant, be as small as the measured density of dark energy? (Fine tuning problem.) And why should it be of the same order of magnitude as the present mean density of matter in the Universe? (Coincidence problem.) These questions have led cosmologists to search for alternative explanations of the accelerated expansion of the Universe.

The possibility of giving the graviton a mass has attracted considerable attention. Such a mass leads to  `degravitation' which can solve the cosmological constant problem~\cite{Dvali:2007kt,deRham:2007rw,deRham:2009rm}. If the graviton is massive, the range of gravity is finite and a cosmological constant does not gravitate. Furthermore, if one fine tunes the graviton mass to $m_g \sim H_0$, where $H_0 \simeq h\times 2.1\times 10^{-42}$ GeV is the value of the Hubble constant and $h=H_0/(100$km/sec/Mpc$)$, gravity weakens around this scale, which might lead to accelerated expansion.

Adding a mass term to gravity is a non-trivial problem. It removes diffeomorphism invariance and hence the metric has six degrees of freedom (four being absorbed by the Bianchi identities). Five of these are the massive graviton while the sixth is usually a ghost, the so-called Boulware-Deser ghost~\cite{Boulware:1973my}. To remove this ghost one has to make sure to obtain an additional constraint.  This was shown to be possible with a very specific form of the potential for the gravitational field, the dRGT (de Rham, Gabadadze, Tolley) potential ~\cite{deRham:2010ik,deRham:2010kj,Hassan:2011hr}, which has been the basis for a large amount of work on this topic (see, e.g.,~\cite{Hassan:2011vm,Hassan:2011tf,Koyama:2011yg,Guarato:2013gba} and refs. therein). 
Applications to cosmology, however, have shown that it is not possible to find homogeneous and isotropic solutions of massive gravity which resemble our Universe. In massive gravity, a fixed reference metric has to be chosen with respect to which the mass term is defined. The possible solutions of course strongly depend on this reference metric, but even when choosing the reference metric to be Friedmann, the resulting solutions either do not show the well known cosmological behavior going from a radiation dominated to a matter dominated Universe followed by a late dark energy dominated phase, or they are unstable~\cite{Gumrukcuoglu:2011zh,Langlois:2012hk,Fasiello:2012rw}, see~\cite{deRham:2014zqa} for a review and \cite{Comelli:2013tja,deRham:2014gla} for the study of the cosmology in the contest of the so-called \emph{generalized} massive gravity models.  

Apart from these problems, it is somewhat disappointing to introduce the reference metric as an  `absolute element', i.e., a non-dynamical field in the theory. From this point of view, bimetric (or more general multi-metric) theories, where also the reference metric is dynamical are better motivated~\cite{Hassan:2011zd, Hassan:2011ea, Hassan:2012wr}. Interesting discussions about theoretical aspects of bimetric massive gravity can be found in \cite{Akrami:2014lja, Hassan:2014vja, deRham:2014fha, Cusin:2014zoa, Noller:2014sta, Akrami:2013ffa}. 

It has been shown that bigravity theories can by stable and well behaved~\cite{Fasiello:2013woa}, and that cosmological solutions of bimetric theories can actually fit the expansion history of the accelerating Universe~\cite{Volkov:2011an,Comelli:2011zm,Konnig:2014dna, Tamanini:2013xia}. Observational tests of various  models of bigravity are discussed in \cite{Solomon:2014dua,vonStrauss:2011mq,Berg:2012kn,2013JHEP...03..099A,Konnig:2013gxa}. The study of the cosmology of  models of bigravity where matter is coupled to a combination of the two metrics is addressed in a series of recent papers~\cite{Gumrukcuoglu:2015nua, Comelli:2015pua, Enander:2014xga}. The analysis of cosmological perturbations have been studied in different settings and different models of bigravity in \cite{Comelli:2012db, Comelli:2014bqa, DeFelice}. Recently, scalar perturbations of these models have been investigated and it has been shown that there exists a sub-class of models of bigravity that admit solutions with  well behaved scalar perturbations of the physical metric, while other sub-classes of models are unstable~\cite{Amendola_pert}.

In this paper we want to study the behavior of tensor perturbations, i.e., gravitational waves in bimetric gravity theories which fit the expansion history of the Universe and which lead to physically acceptable scalar perturbations, i.e., scalar perturbations which just exhibit the usual Jeans instability of Newtonian gravity which leads to cosmological structure formation but no significant additional instability. A more generic study of instabilities in bimetric theories can be found in Ref.~\cite{Kuhnel:2012gh}.

Examining cosmological tensor perturbations, we show that gravitational waves are unstable in this theory. We study this instability in detail. We find that it changes the spectrum and strongly boosts the amplitude of gravitational waves. In order to prevent conflict with observations we have to fine tune the initial conditions for tensor perturbations.  

While we were working on this, a preliminary study of gravitational waves in this model has appeared~\cite{Lagos:2014lca}. In Ref.~\cite{Lagos:2014lca} the authors investigate all perturbation modes, scalar, vector and tensor for two cosmological solutions, namely the expanding branch and the bouncing branch, called \textit{infinite-branch bigravity} (IBB) in Ref.~\cite{Amendola_pert}.  This latter has cosmologically acceptable scalar perturbations and is therefore of particular interest.

Our analysis goes beyond the results presented in~\cite{Lagos:2014lca}. We numerically solve the gravitational wave equations and study the resulting gravitational wave spectrum for different initial conditions for the perturbations. We finally argue that even if we would initially set gravitational wave perturbations to zero, non-linearities which induce small tensor perturbations are sufficient to trigger the instability and  the model can only be saved if we also fine tune the initial conditions of  scalar perturbations.

The rest of the paper is organised as follows. In the next section we write the Lagrangian and the equations of motion of bimetric gravity in general and for cosmological (i.e. homogeneous and isotropic) spacetimes. We then specialise to the physically viable model which gives an acceptable expansion history. In Section~\ref{s:scal} we comment on the scalar sector of perturbations and in  Section~\ref{s:ten} we study tensor perturbations and compute the gravitational wave spectrum for different initial conditions. In  Section~\ref{s:con} we discuss our results and conclude.

{\bf Notation:} We set $c=\hbar=k_{\rm Boltzmann}=1$. $M_g=1/\sqrt{8\pi G}\equiv M_p\simeq 2.4\times 10^{18}$GeV is the reduced Planck mass.
 
\section{Cosmological solutions of massive bigravity}
\subsection{The Lagrangian}
We start from a massive bigravity theory defined by the action
\be\label{startingaction}
S =   - \int d^4x\sqrt{-g}\left[\frac{M_g^2}{2}(R(g) -2 m^2V(g,f)) +\LL_m(g,\Phi)\right] - \int d^4x\sqrt{-f}\frac{M_f^2}{2}R(f)  \,.
\ee
Here $f$ and $g$ are the two metrics while $M_f$ and $M_g=1/\sqrt{8\pi G}\equiv M_p$ are the respective Planck masses with dimensionless ratio $m_*=M_f/M_g$. We assume the matter fields $\Phi$ to be coupled to $g$ only.  We use the notation of~\cite{vonStrauss:2011mq}. The potential is a function of the tensor field $X = \sqrt{g^{-1}f}$ given by
\be
V(g,f) = \sum_{n=0}^{4}\beta_{n}e_{n}(X)\,,
\ee
where the polynomials $e_{i}(X)$ are
\bea
e_{0} &=& \mathbb{I},\, \quad e_{1}=[X],\, \\ e_{2} &=&\frac{1}{2}([X]^{2}-[X^{2}]),\\ 
e_{3} &=& \frac{1}{6}([X]^{3}-3[X][X^{2}]+2[X^{3}],\\
e_{4} &=& \frac{1}{24}([X]^{4}-6[X]^{2}[X^{2}]+8[X][X^{3}] +3[X^{2}]^{2}-6[X^{4}]) = \det X \,.
\eea
The square bracket $\left[\cdots\right]$ denotes the trace.
The equations of motion of this theory are
\bea
 \frac{1}{M_g^2}\, T_{\mu\nu} &=& R_{\mu\nu}-\frac{1}{2}g_{\mu\nu}\,R+\frac{m^2}{2}\sum_{n=0}^3(-1)^n\, \beta_n \left[ g_{\mu\lambda}Y_{(n)\nu}^{(g)\lambda}+g_{\nu\lambda}\, Y_{(n)\mu}^{(g)\lambda}\right]\,,   \label{eq:g} \\
 0 &=& R^f_{\mu\nu}-\frac{1}{2}f_{\mu\nu}R^f+\frac{m^2}{2m_*^2}\sum_{n=0}^3(-1)^n\, \beta_{4-n}\,\left[ f_{\mu\lambda}\, Y_{(n)\nu}^{(f)\lambda} +f_{\nu\lambda}\, Y_{(n)\mu}^{(f)\lambda}\right]\,, 
 \label{eq:f} 
\eea
where the superscript $f$ indicates the  curvature of the metric $f_{\mu\nu}$.
The definition of the matrices $Y_{(n)\mu}^{(g)\nu} = Y_{(n)\mu}^\nu\left(\sqrt{g^{-1}f}\right)$ and $Y_{(n)\mu}^{(f)\nu} = Y_{(n)\mu}^\nu\left(\sqrt{f^{-1}g}\right)$
is as follows
\bean
Y_{(0)}\left(X\right) &=&\mathbb{I}\,,\hspace{0.5 cm} Y_{(1)}\left(X\right)=X -\mathbb{I}\left[X\right]\,,\\
Y_{(2)}\left(X\right) &=& X^2 -X \left[X\right]+\frac{1}{2}\mathbb{I}\left(\left[X\right]^2-\left[X^2\right]\right)\,,\\
Y_{(3)}\left(X\right) &=&X^3 -X^2\left[X\right]+\frac{1}{2}X\left(\left[X\right]^2-\left[X^2\right]\right) 
-\frac{1}{6}\mathbb{I}\left(\left[X\right]^3-3\left[X\right]\left[X^2\right]+2\left[X^3\right]\right)\,.
\eean

As a consequence of the Bianchi identities and of the covariant conservation of $T_{\mu\nu}$, we obtain the following Bianchi constraints for each  of  the two metrics
\be
\nabla^g_{\mu}\sum_{n=0}^3(-)^n\, \beta_n\, \left[Y_{(n)}^{(g)\nu\mu}+ Y_{(n)}^{(g)\mu\nu}\right]=0\,,
\ee
\be
{\nabla}^f_{\mu}\sum_{n=0}^3(-)^n\, \beta_{4-n}\,\left[ Y_{(n)}^{(f)\nu\mu}+Y_{(n)}^{(f)\mu\nu}\right]=0\,,
\ee
where we raise and lower indices of $Y^{(g)\cdots}_{(n)\cdots}$ and 
$Y^{(f)\cdots}_{(n)\cdots}$ with the metrics $g$ and $f$ respectively and the relevant metric is indicated in the covariant derivatives $\nabla^g$ and $\nabla^f$. Both these constraints follow from the invariance of the interaction term under the diagonal subgroup of the general coordinate transformations of both metrics.  Both constraints are equivalent and we will use only the first one. 

\subsection{Cosmological equations of motion}
We now consider solutions which are spatially homogeneous and isotropic. For simplicity we neglect spatial curvature. If curvature is included, it is easy to see that it has to be the same for both  metrics. The metrics can be written in the form
\be
g_{\mu\nu}dx^{\mu}dx^{\nu}=a^2(\tau)\left(-d\tau^2+\de_{ij}dx^idx^j\right)\,,
\ee
\be
f_{\mu\nu}dx^{\mu}dx^{\nu}=b^2(\tau)\left(-c^2(\tau) d\tau^2+\de_{ij}dx^idx^j\right)\,,
\ee
where $\tau$ is conformal time for the $g$-metric and $c(\tau)$ is a lapse function which parametrizes the difference between the conformal time $\tau_f$ for the $f$-metric and $\tau$, $d\tau_f= c(\tau)d\tau$.

It is convenient to define the conformal Hubble parameter ($\HH$) and the standard one ($H$) for the two metrics
\be\label{e:defH}
H=\frac{\mathcal{H}}{a}=\frac{a'}{a^2}\,,\hspace{0.5 cm} H_f=\frac{\mathcal{H}_f}{b}=\frac{b'}{b^2\,c}\,,
\ee
where with $'$ we denote the derivative with respect to the conformal time $\tau$. We also introduce the ratio between the two  scale factors
\be\label{e:defr}
r=\frac{b}{a}\,.
\ee

For symmetry reasons, the energy-momentum tensor has the form  of a perfect fluid with equation of state $p=w \rho$. Explicitly
\bea
T_{\mu\nu}&=&\left(p+\rho\right)\, u_{\mu} u_{\nu}+p \,g_{\mu\nu}\,,\\
\rho'&=&-3(\rho+p)\,\mathcal{H}\,,\\
p&=&w\rho\,.
\eea

Introducing a  `gravity fluid'  which represents the mass term in the Einstein equations, we define 
\bea
\rho_g &=&\frac{m^2}{8\pi G}\left(\beta_3\, r^3+3\beta_2\,r^2+3\beta_1\,r+\beta_0\right)\,, \\
p_g &=& -\frac{m^2}{8\pi G}\Big(\beta_3 c \,r^3+\beta_2(2c+1)r^2
  +\beta_1(c+2)r+\beta_0\Big)\,.
\eea
Note that  the gravity fluid becomes like a cosmological constant when $c=1$.

The Bianchi constraint in the cosmological ansatz can then be written as
\be
\rho_g'=-3\mathcal{H}\,\left(\rho_g+p_g\right)\,,
\ee
This constraint is equivalent to
\be
m^2\left(\beta_3r^2+2\beta_2r+\beta_1\right)\,(c\,b\,a'-a\,b')=0\,, \quad \mbox{ or } \quad m^2\left(\beta_3r^2+2\beta_2r+\beta_1\right)(\HH-\HH_f)=0\, .
\ee

The full set of equations of motion is given by the time-time and the space-space components of the modified Einstein equations. For the metric $g$ they are
\be\label{F11}
3H^2=8\pi G\,\left(\rho+\rho_g\right)\,,
\ee
\be\label{F12}
3H^2+\frac{2H'}{a}=-8\pi G\,\left(p+p_g\right)\,.
\ee
The modified Friedmann equations for the $f$ metric become 
\bea
3H_f^2 =& M_f^{-2}\rho_f  
=& \frac{1}{m_*^2}r^{-4}\left(\rho_g-m^2M_g^2\beta_0+m^2M_g^2\beta_4r^4\right) 
= \frac{m^2}{m_*^2}\left(\frac{\beta_1}{r^3}+\frac{3\beta_2}{r^2}+\frac{3\beta_3}{r}+\beta_4\right)\,,\label{F21}\\
\label{F22}
 3H_f^2+\frac{2\,H_f'}{a\,cr}-\frac{2\,c'\,H_f}{a\, c^2\,r} =& -M_f^{-2}p_f
=&\frac{m^2}{m_*^2}\!\left(\frac{\beta_1}{c\,r^3}+\frac{2\beta_2}{c\,r^2}+\frac{\beta_3}{c\,r}+\frac{\beta_2}{r^2}+\frac{2\beta_3}{r}+\beta_4\!\right)\,.
\eea

We consider the first Friedmann equation for both metrics, the Bianchi constraint, and in the matter sector, the  `energy conservation'  equation and the equation of state as independent equations which determine the 5 functions $a(\tau)$, $b(\tau)$, $c(\tau)$, $\rho(\tau)$ and $p(\tau)$, 
\be\label{eq1}
H^2=\frac{8\pi G}{3}\,\left(\rho+\rho_g\right)\,,
\ee
\be\label{eq2}
H_f^2=\frac{m^2}{3\,m_*^2}\left(\frac{\beta_1}{r^3}+\frac{3\beta_2}{r^2}+\frac{3\beta_3}{r}+\beta_4\right)\,,
\ee
\be\label{e:Bian}
\left(\beta_3\,r^2+2\beta_2\,r+\beta_1\right)(\HH-\HH_f)=0\,,
\ee
\be
\rho'=-3(\rho+p)\,\mathcal{H}\,,\hspace{0.5 cm} p=w\rho\,.
\ee
Eq.~(\ref{e:Bian}) allows for two branches of solutions. Either $r=\bar r$ is constant given by the solution of the quadratic equation
\be
\beta_3\bar r^2+2\beta_2\bar r+\beta_1=0\,,
\ee
or the second factor of eq.~(\ref{e:Bian}) vanishes implying
\be
\mathcal{H}_f=\mathcal{H} \,, \quad \mbox{ or } \quad  rH_f=H \,.
\ee
According to \cite{vonStrauss:2011mq, DeFelice} the first branch is equivalent to general relativity with an effective cosmological constant. Therefore we expect the usual $\Lambda$CDM phenomenology for this branch. We now concentrate on the second branch which is more interesting, hence we request $rH_f=H$.
Inserting this in (\ref{eq2}) yields
\be
H^2 =\frac{m^2}{3\,m_*^2}\,\left(\frac{\beta_1}{r}+3\beta_2+3\beta_3\,r+\beta_4\,r^2\right)\,.
\ee
With this, the Friedmann equation for $g$ can be written as
\bea
-\frac{\beta_3}{3}\,r^3+r^2\left(\frac{\beta_4}{3}\frac{1}{m_*^2}-\beta_2\right)+r\left(\frac{\beta_3}{m_*^2}-\beta_1\right) 
+\frac{\beta_1}{3m_*^2r}+\frac{\beta_2}{m_*^2}-\frac{\beta_0}{3}=\frac{8\pi G}{3m^2}\,\rho\,. \hspace*{0.5cm} &&\label{e:rrho}
\eea
This is a polynomial equation which can be solved for $r$ in terms of the energy density $\rho$ and the constants $\beta_i$, $m^2$, $M_g^2$ and $m_*^2$.
The detailed evolution depends on the parameters but we can already observe that at late time, when $\rho\rightarrow 0$, $r \rightarrow$ const  so that also $H^2\rightarrow$ const. Therefore we have a late-time de Sitter phase, independent of the choice of parameters, as long as these admit a real and positive solution $\bar r$ of the fourth order equation given by (\ref{e:rrho}) with $\rho=0$.

 A detailed study of the background cosmology in this cosmological setting can be found in \cite{Comelli:2011zm,Konnig:2013gxa}. Several possible branches of the solutions  are possible, depending on the initial values for $r$. In \cite{Konnig:2013gxa}, a series of conditions defining a viable cosmological evolution are elaborated. Violations of these conditions do not necessarily imply contradiction with observations if they occur outside the observable range, hence in principle they can be relaxed or lifted. However, when these conditions are satisfied, the cosmological evolution requires no special tuning and it is much safer. 
 
 In order not to mimic a cosmological constant we set $\beta_0=0$. We now study in detail the case $\beta_0=\beta_2=\beta_3=0$ (we call it the `$\beta_1$-$ \beta_4$ model') which has been identified as the only one which gives both, an acceptable background solution and viable scalar  perturbations~\cite{Konnig:2013gxa,Arkami:2013a1,Amendola_pert}.
 
 We consider a Universe containing matter and radiation with densities $\rho_m$ and $\rho_r$ and pressure $p_m=0$ and $p_r=\rho_r/3 = w_r\rho_r$ which we assume to be separately conserved such that $\rho_m =\rho_{m0}a^{-3}$ and 
  $\rho_r =\rho_{r0}a^{-4}$. Here we have normalized the scale factor to unity today, $a_0=1$. The Bianchi constraint can be rewritten as
\be\label{e:c}
\frac{r'}{r} = (c-1) \,\HH,
\ee 
Furthermore, under the rescaling $f_{\mu\nu} \ra m_*^{-2}f_{\mu\nu}$ and $\beta_n \ra m_*^n\beta_n$ the equations become independent of $m_*$ so that we can simply set $m_*=1$, see~\cite{Berg:2012kn} for a more detailed discussion.
With this, the background equations can be written as
\be\label{11}
3 \mathcal{H}^2=a^2 \left(3m^2 \beta_1 r+M_p^{-2}(\rho_ m+\rho_r)\right)\,,
\ee
\be
a^2m^2\left(\beta_1+\beta_4 r^3\right)-3 \mathcal{H}^2 r= 0\,, 
\ee
\be\label{33}
\mathcal{H}^2 + 2 \mathcal{H}' = a^2 \left(3m^2 \beta_ 1 r  -M_p^{-2} \rho_r/3 + m^2\beta_ 1\frac{ r'}{\HH}\right)\,,
\ee
We solve the first three equations for $\mathcal{H}$, $\rho_m$ and  $r'$,   
\be\label{H}
\mathcal{H}^2=a^2m^2\frac{\beta_1+\beta_4r^3}{3 r}\,,
\ee
\be\label{rm}
\rho_m=M_p^2m^2\left(\frac{\beta_1}{r} - 3 \beta_1 r + \beta_4 r^2\right) - \rho_r\,.
\ee
 \be\label{rprepre}
\frac{r'}{r}= \frac{-9 \beta_1  r^2+3\beta_1+3\beta_4r^3+ r M_p^{-2}m^{-2}\rho_r }{3 \beta_1 r^2+\beta_1-2 \beta_4 r^3}\HH\,,
\ee
We want to solve eq. (\ref{rprepre})  numerically for a given present value of $r$. Let us divide  eq. (\ref{H}) by  $\mathcal{H}_0=H_0$ so that
\be\label{Hnorm}
\frac{\mathcal{H}}{\mathcal{H}_0}=a\frac{\sqrt{\beta_1+\beta_4r^3}}{\sqrt{3r}}\left(\frac{m}{\HH_0}\right)\,,
\ee
We now evaluate eq. (\ref{Hnorm}) at $\tau_0$ and we solve the resulting equation expressing $r_0=r(\tau_0)$ as a function of the constants $\beta_i$. This equation has three real solutions. We choose the only one that, when used as `final' condition in eq. (\ref{rprepre}), gives an evolution for $r$ starting at very large values and decreasing  to a finite value at late times. In \cite{Konnig:2013gxa} and \cite{Amendola_pert}, it has been shown that this solution is the only one able to give rise to both a viable background cosmology and viable scalar perturbations in the $\beta_1$-$\beta_4$ model. 

\begin{figure}[ht!]
 \centering
   {\includegraphics[scale=0.4]{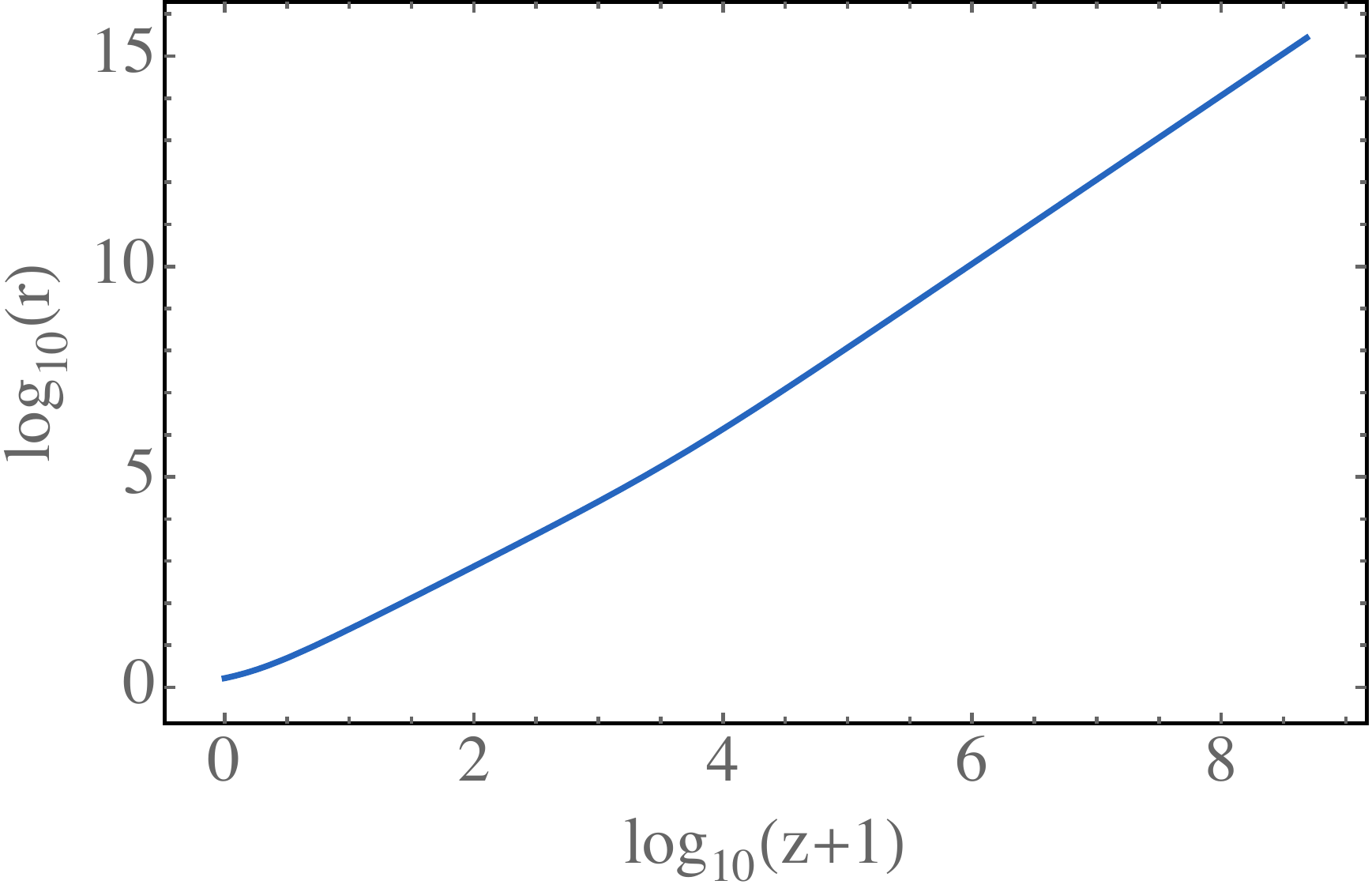}}
   \caption{\label{f:rplot}The evolution of the ratio of the two scale factors, $r=b/a$ is shown as function of the redshift $z+1= a^{-1}$.  We have chosen the parameters $m^2\beta_1=0.48\,\HH_0^2$, $m^2\beta_4 =0.94\,\HH_0^2$ and $\beta_0=\beta_2=\beta_3=0$.}
\end{figure}

We choose the best-fit values $\beta_1m^2=0.48H_0^2$ and $\beta_4m^2=0.94H_0^2$ obtained in \cite{Konnig:2013gxa} and \cite{Amendola_pert}  fitting measured growth data and type Ia supernovae (see also \cite{Arkami:2013a1} for an explanation of the procedure used to obtain these values from fits).  We can then solve eq. (\ref{rprepre}) numerically.  The evolution of $r$ is shown in Fig.~\ref{f:rplot}.
We observe that $r$ is very large at early times (the vertical axis is not $r$ but $\log_{10}(r)$), at the present time $r\simeq 1$ while the value $r= 1$ is a future attractor.

\begin{figure}[ht!]
 \centering
   {\includegraphics[scale=0.4]{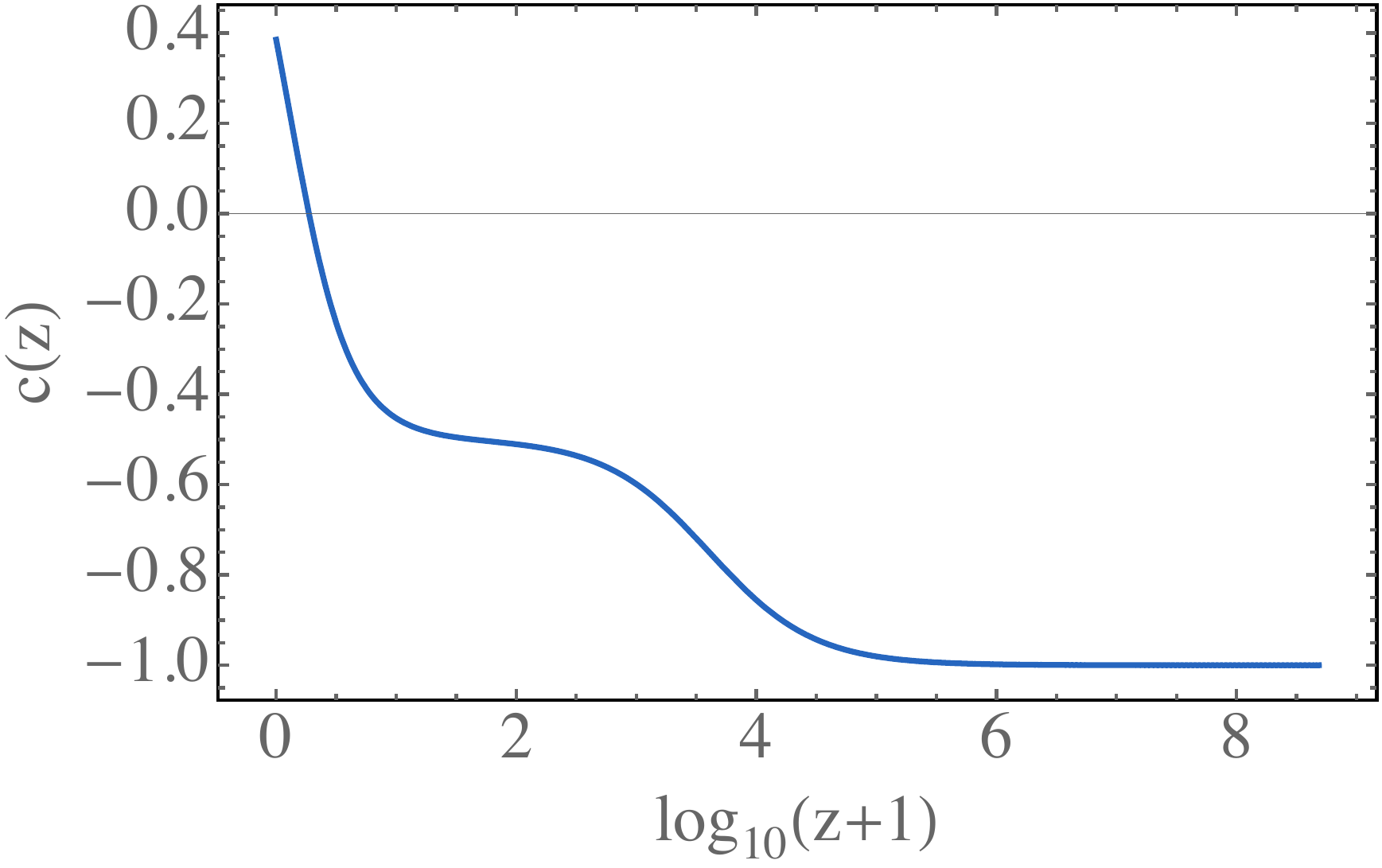}}
 \caption{\label{f:lapse}The evolution of the lapse of the $f$ metric for  $m^2\beta_1=0.48\,\HH_0^2$, $m^2\beta_4 =0.94\,\HH_0^2$ and $\beta_0=\beta_2=\beta_3=0$.}
 \end{figure}

The lapse $c$ of the $f$-metric is given by the Bianchi constraint (\ref{e:c}). Its time evolution is presented in Fig. \ref{f:lapse}. 
 
 The lapse function is $c\simeq -1 \simeq $ constant in the radiation dominated phase, so that $r'/r=-2\HH$, hence $r\propto a^{-2}$, see Eq.~(\ref{e:c}). It grows to a new plateau at the matter-radiation transition and stays $c\simeq -1/2$ during the matter dominated phase. Then again at the transition to the gravity-dominated phase in the future, $z<0$ the lapse grows to $c=1$ which is reached in the future de Sitter phase. 
 Interestingly, the lapse function changes sign  roughly at redshift $z_c\simeq 0.9$.
 This in principle signals a singularity in the $f$-metric where, for example, its determinant vanishes. However, since the Bianchi constraint requires that also $b'=0$ when $c=0$ and $a'\neq 0$, $\HH_f=b'/(bc) = \HH =a'/a$ remains finite and no physical observable diverges\footnote{This would be different if we would couple matter to the $f$-metric since e.g. its Ricci scalar $R_f$ which might then become observable diverges.}. 
 
 At this point we want to direct the attention of the reader to the fact that even though in the Langrangian the lapse function $c$ only appears as $\sqrt{c^2}$ which one naively might replace by $|c|$, this cannot lead to the background phenomenology 
    needed to mimic dark energy. At early times, when $r\gg 1$ and the Universe is radiation dominated, Eq.~(\ref{rprepre}) becomes
\be
\frac{r'}{r} = -\frac{3m^2\beta_4r^2+M_p^{-2}\rho_r}{2m^{2}\beta_4r^2}\HH = -2\HH \,.
\ee 
For the second equal sign we have used Eq. (\ref{H}) in the limit of large $r$.
 In order to satisfy both, Eq.~(\ref{e:c}) and  Eq.~(\ref{rprepre}), we therefore need $c=-1$ in the radiation dominated era and we cannot replace $c$ by $|c|$ in  Eq.~(\ref{e:c}). We can also not replace it by $-|c|$ since we need the factor $(c-1)\ra 0$ when the Universe becomes dark energy dominated. With our choice of the parameters $\beta_i$, a radiation dominated Universe at early time and a dark energy like solution at late time, requires that $c$ passes through zero, which $\pm|c|$ cannot.
 
\begin{figure}[ht]
 \centering
   \subfigure[ Complete evolution]
   {\includegraphics[scale=0.4]{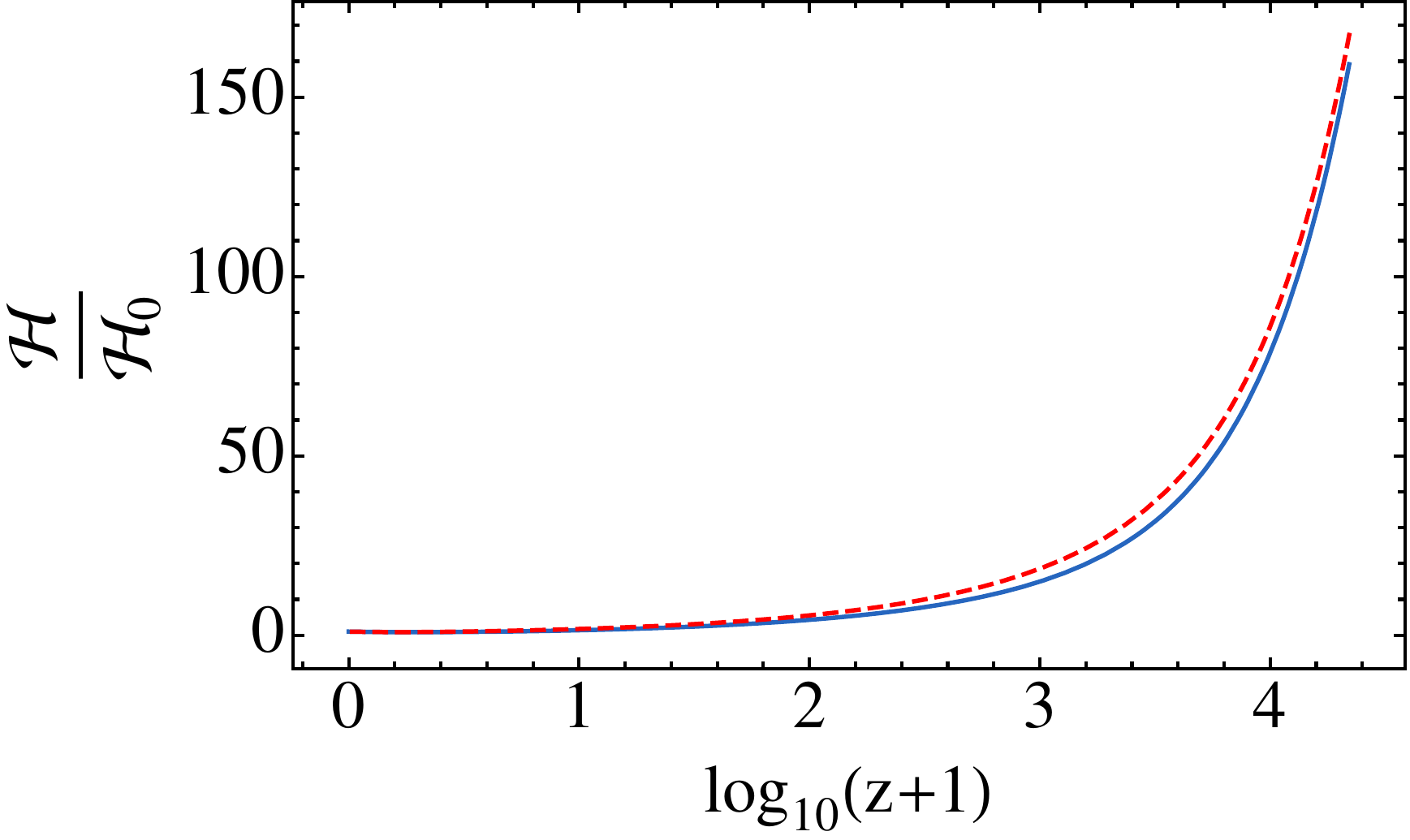}}
      \hspace{0.2 cm}
   \subfigure[ Low redshift behaviour]
      {\includegraphics[scale=0.4]{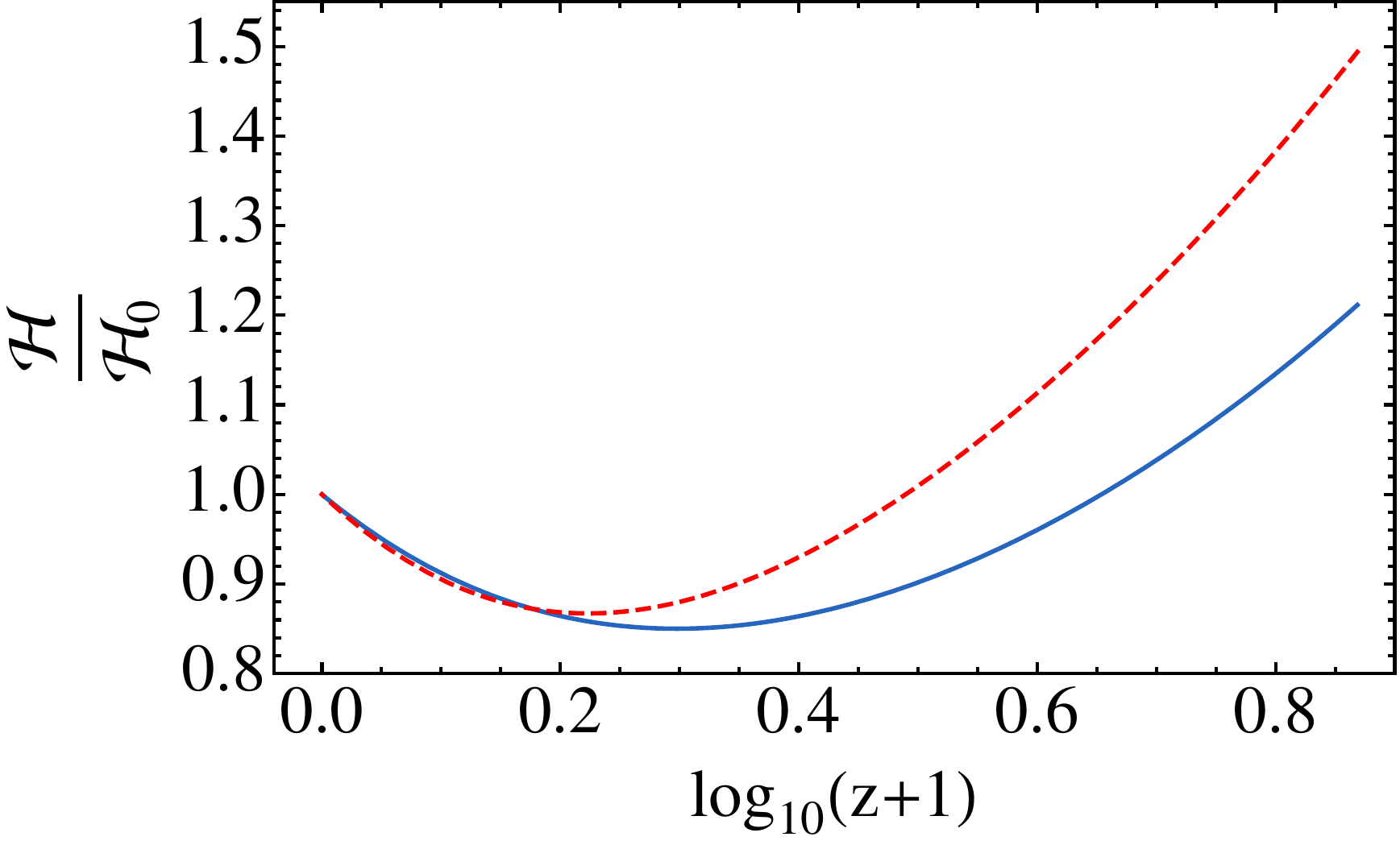}}
\caption{\label{Hplot} We show the evolution of the Hubble parameter $\mathcal{H}(z)$ for a $\La$CDM background with $\Omega_{\La}=0.7$  (red, dashed) and for a bimetric cosmology (blue, solid)  with $m^2\beta_1=0.48\,\HH_0^2$, $m^2\beta_4 =0.94\,\HH_0^2$ and $\beta_0=\beta_2=\beta_3=0$. }
 \end{figure}

 Substituting the numerical solution for $r$ in eq. (\ref{Hnorm}), we obtain the evolution of $\mathcal{H}/\mathcal{H}_0$. In Fig. \ref{Hplot} we have plotted  $\mathcal{H}/\mathcal{H}_0$  as a function of redshift in $\beta_1$-$\beta_4$ bigravity and in standard $\Lambda CDM$.\footnote{We have considered a scenario with radiation, matter and a cosmological constant with $\Omega_{\Lambda}=0.7$. }
  
\begin{figure}[ht]
 \centering
   {\includegraphics[scale=0.4]{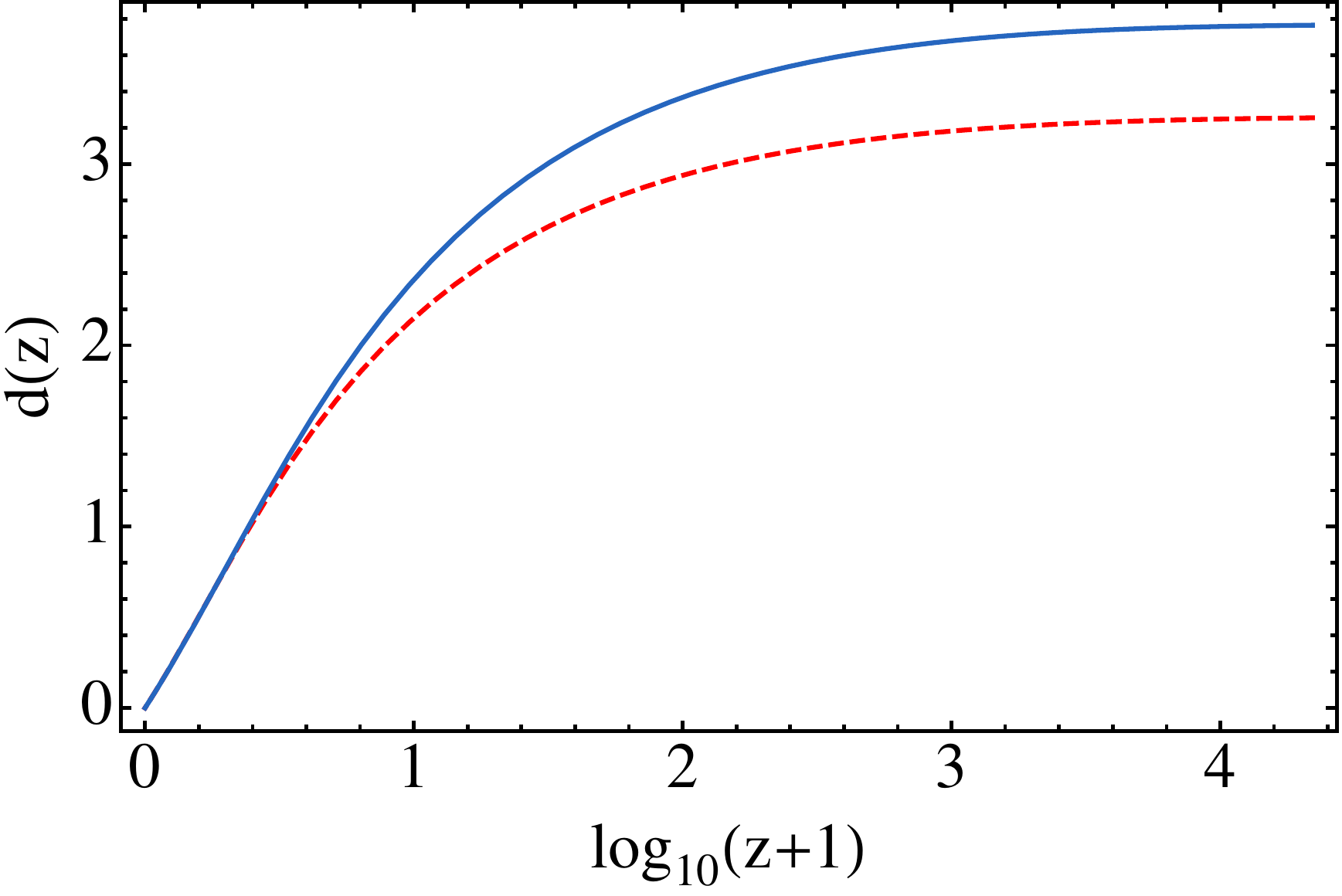}}
\caption{\label{f:dH} We show the evolution of the comoving distance $d(z)$ for a $\La$CDM background with $\Omega_{\La}=0.7$ (red, dashed) and for a bimetric cosmology (blue, solid) with $m^2\beta_1=0.48\,\HH_0^2$, $m^2\beta_4 =0.94\,\HH_0^2$ and $\beta_0=\beta_2=\beta_3=0$. (We do not optimise the cosmological parameters to obtain a good fit since this is not the point of the present work. The similarity of the behaviour obtained here is sufficient for our purpose.)}
 \end{figure}

 In Fig.~\ref{f:dH} we compare the background evolution of the comoving distance $d(z)=\int_0^z H(z')^{-1}dz'$ for the $\beta_1$-$\beta_4$ model with the best fit parameters $m^2\beta_1=0.48\HH_0^2$, $m^2\beta_4 =0.94\HH_0^2$ (which we shall also consider in the perturbation analysis) and for a $\La$CDM model with $\Omega_{\La}=0.7$. 

\section{Comments on scalar perturbations}
\label{s:scal}

Let us now consider perturbations of this cosmology around the homogeneous and isotropic background
\be
g_{\mu\nu}=\bar{g}_{\mu\nu}+a^2\, h_{g\mu\nu}\,,\ee
\be
f_{\mu\nu}=\bar{f}_{\mu\nu}+b^2\, h_{f\mu\nu}\,.
\ee
From now on, the use of an overbar indicates  background quantities. 
We parametrize the perturbations as follows
\be
(h_g)_{\mu\nu}=\left(
\begin{array}{cc}
-2 A_g&C^{(g)}_{j}-\partial_j B_g\\
C^{(g)}_{ i}-\partial_i B_g& h_{ij}^{(g)TT}+\partial_i\mathcal{V}^{(g)}_{j}+\partial_j\mathcal{V}^{(g)}_{i}+2\partial_i\partial_j S_g+2\delta_{ij} F_g\\
\end{array}
\right)\,,
\ee
 \be
(h_f)_{\mu\nu}=\left(
\begin{array}{cc}
-2 c^2 A_f&C^{(f)}_j-\partial_j B_f\\
C^{(f)}_{ i}-\partial_i B_f& h_{ij}^{(f)TT}+\partial_i\mathcal{V}^{(f)}_{j}+\partial_j\mathcal{V}^{(f)}_{i}+2\partial_i\partial_j S_f+2\delta_{ij} F_f\\
\end{array}
\right)\,,
\ee
with
\be
\partial_i C_\bullet^i=\partial_i\mathcal{V}_\bullet^i=\partial_i h_\bullet^{TT ij}=0\,,\hspace{1cm} \delta^{ij}h_{\bullet ij}^{TT}=0\,.
\ee
Spatial indices are raised and lowered using the flat spatial metric, $\de_{ij}$.
There are eight scalar perturbations, $A_\bullet,~B_\bullet,~S_\bullet$ and $F_\bullet$, eight vector perturbations, $C_{\bullet j}$ and $\VV_{\bullet j}$ and four tensor perturbations $h_{ ij}^{\bullet TT}$. Here $_\bullet$ denotes $_g$ or $_f$.
Two scalar and two vector modes can be removed by coordinate transformations, leaving six scalar, six vector and four tensor degrees of freedom.

In Ref.~\cite{Amendola_pert} scalar perturbations of the viable $\beta_1$-$\beta_4$ model have been analysed for perfect fluid matter (i.e. matter without anisotropic stress and with adiabatic perturbations) and it has been found that they can fit the growth rate of the observed perturbations during the matter and dark energy dominated eras\footnote{One of the  main conclusions of  Ref.~\cite{Amendola_pert} is that during \emph{matter domination} scalar perturbations do not exhibit \emph{exponential} instabilities. In that context, however, the stability of scalar perturbations at early times and the absence of power-low instabilities during matter is not analysed.}. In Ref.~\cite{Lagos:2014lca} a preliminary analysis of all, scalar, vector and tensor perturbations is presented and analytic solutions in limiting regimes are found, which all do not show exponential instabilities. 

In this section we discuss briefly scalar perturbations while the rest of this work is devoted to a detailed study of tensor perturbations. For the scalar sector, we derive analytic solutions for the propagating degrees of freedom valid in the radiation era and we compare them with the results of the numerical integration of the perturbation equations in radiation. The result of this analysis differs from the one of Ref.~\cite{Lagos:2014lca} and we find that an instability in the scalar sector of the $f$ metric shows up at early times and, if sufficiently large, it is transferred to the physical sector of the $g$ metric  through the coupling between the two sectors. 

The equations for the two propagating scalar degrees of freedom in the radiation dominated era can be approximated by \footnote{We adopt here the gauge choice of Ref.~\cite{Lagos:2014lca} to eliminate the redundant degrees of freedom in the scalar sector.}

\be\label{Sg}
S_{g}''+2\mathcal{H}S_{g}'+\frac{9\beta_{1}\mathcal{H}^{3}S_{f}'}{2\beta_{4}k^{2}r}-\frac{k^{2}S_{g}}{3}-\frac{1}{2}a^{2}m^2\beta_{1}r S_{f}=0\,,
\ee

\be\label{Sf}
S_{f}''+\frac{6\beta_{1}\mathcal{H}S_{f}'}{\beta_{4}r}-\frac{\beta_{1}k^{2}S_{g}'}{\beta_{4}\mathcal{H}r}+\frac{k^{2}S_{f}}{3}-\frac{2a^{2}m^2\beta_{1}k^{2}r S_{g}}{3\mathcal{H}^{2}}=0\,.
\ee

\begin{figure}[ht!]
 \centering
\subfigure[\label{Sg10-5}]
   {\includegraphics[scale=0.38]{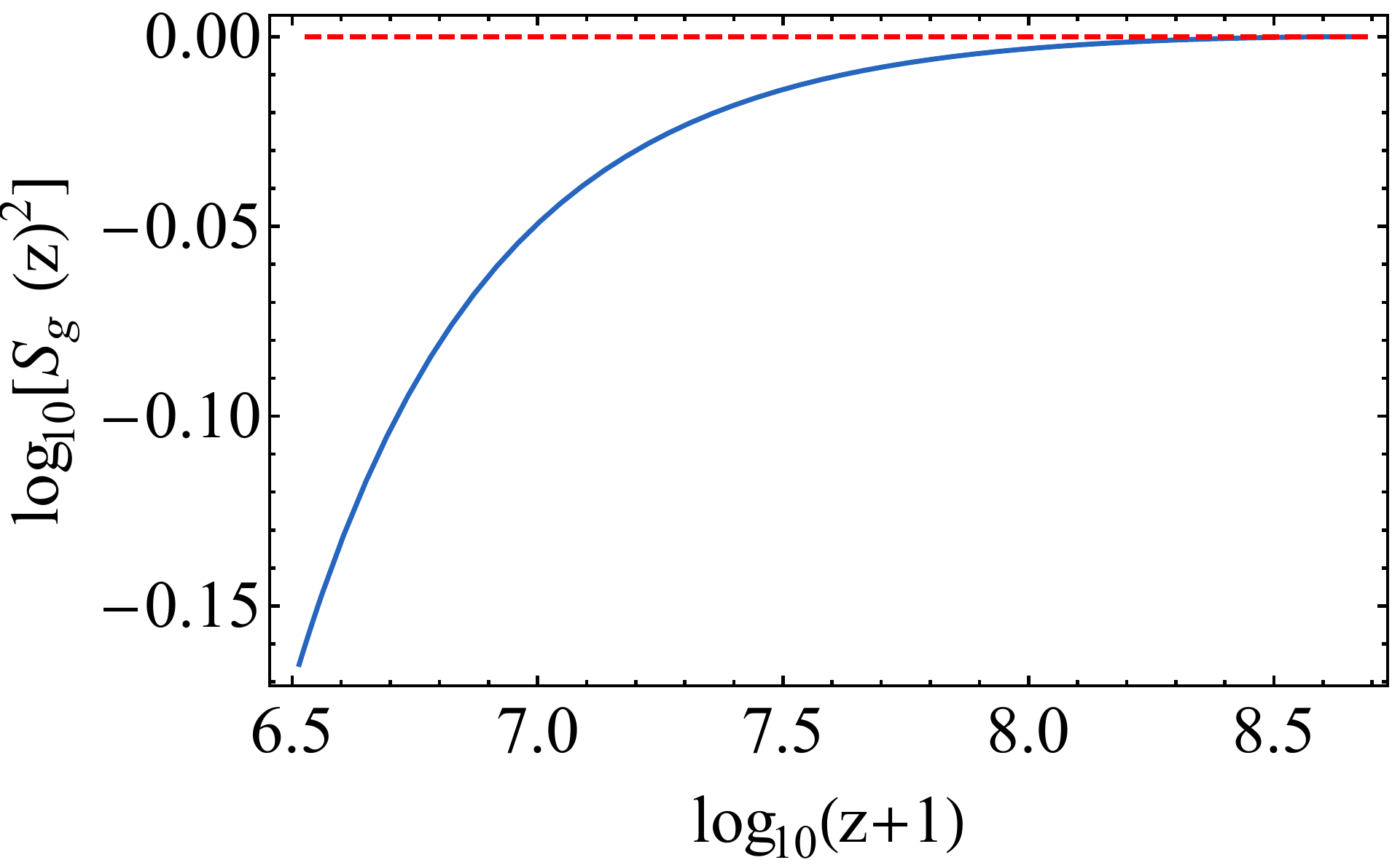}}\qquad
   \subfigure[\label{Sf10-5}]
      {\includegraphics[scale=0.38]{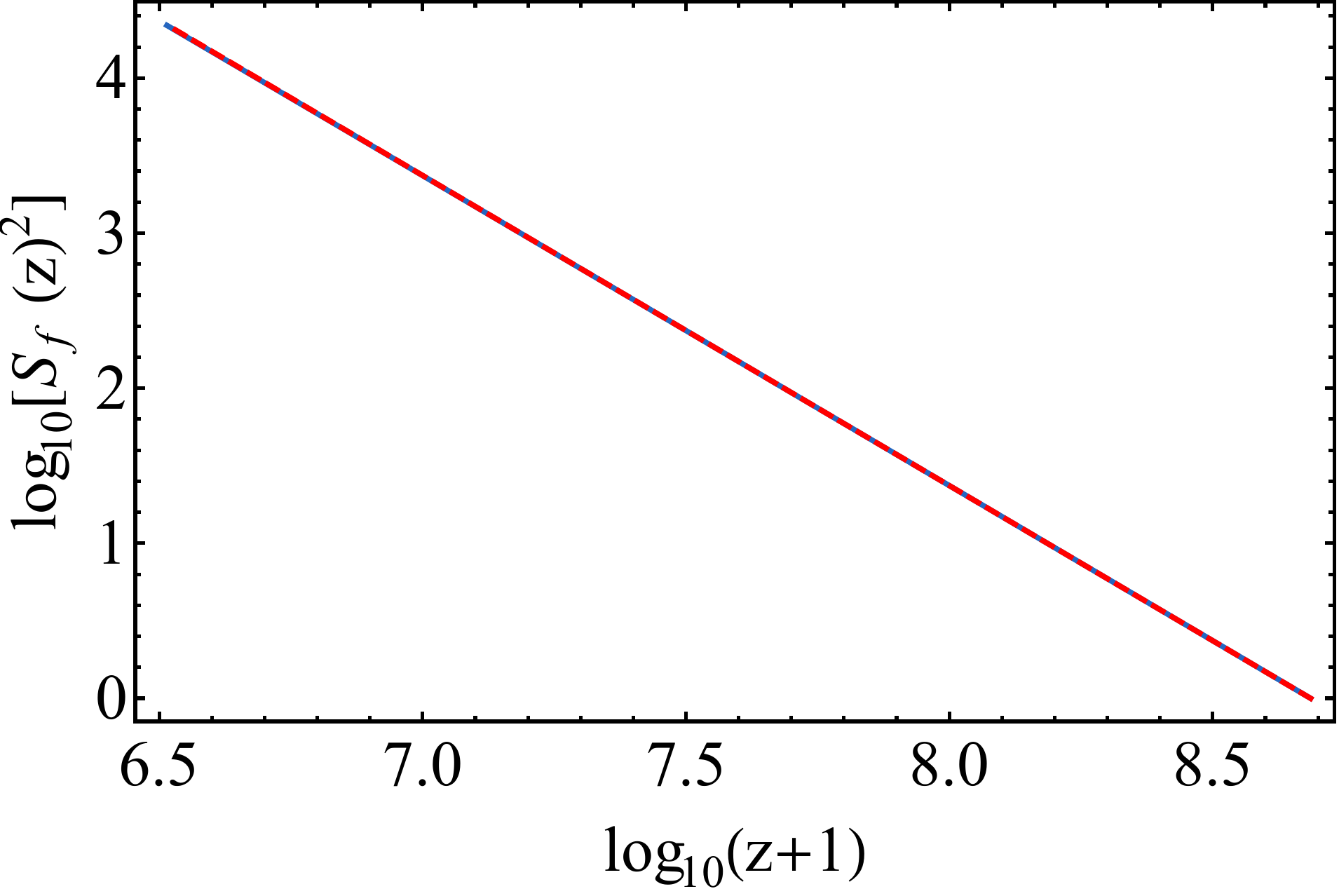}}
       \caption{\label{scalar1} Evolution of scalar perturbations for the metric $g$ and $f$, in the case $A=B=1$ for $k\simeq \HH_0$. The red-dashed line represent the analytical solution for the equation of scalar perturbations valid in the radiation dominated era.  For $S_f$ there is perfect agreement between the analytical and numerical solutions. }
 \end{figure}

For super-Hubble modes, $k\tau\ll1$ , we can neglect terms proportional to $k^{2}$. Furthermore, in the background branch under study, in the radiation era we have  $r\gg1$, $r'\simeq -2\mathcal{H}r$ and  $m^2\beta_1a^2r\simeq 0.48\HH_0^2\sqrt{\Om_r/0.94}\simeq 0.24\times 10^{-2}\HH_0^2=$ constant. Hence, in this regime, the last three terms in (\ref{Sg}) and (\ref{Sf}) can be dropped and the two equations decouple. 

The solutions of the resulting approximated equations can be written as \footnote{More precisely, the exact solution of the decoupled eq. (\ref{Sf}) has a constant mode and a growing one proportional to 
$\text{erf}\left(\sqrt{\frac{1.2\times 10^3 \beta^2 _1}{\beta _4}}\frac{m}{\HH_0(1+z)}\right)$. This function is growing roughly like $(1+z)^{-1}\propto \tau$ as long as the argument is smaller than $1$, hence during the entire radiation dominated epoch. It can therefore be approximated with the growing mode in (\ref{solSf}) to good precision.}
\be\label{solSf}
S_{g}=c_{1}+c_2\frac{\tau_{\rm in}}{\tau}\,,\hspace{1.5 cm}S_{f}=c_{3}+c_{4}\frac{\tau}{\tau_{\rm in}} \,.
\ee

 \begin{figure}[ht!]
 \centering
\subfigure[\label{B=10}]
   {\includegraphics[scale=0.38]{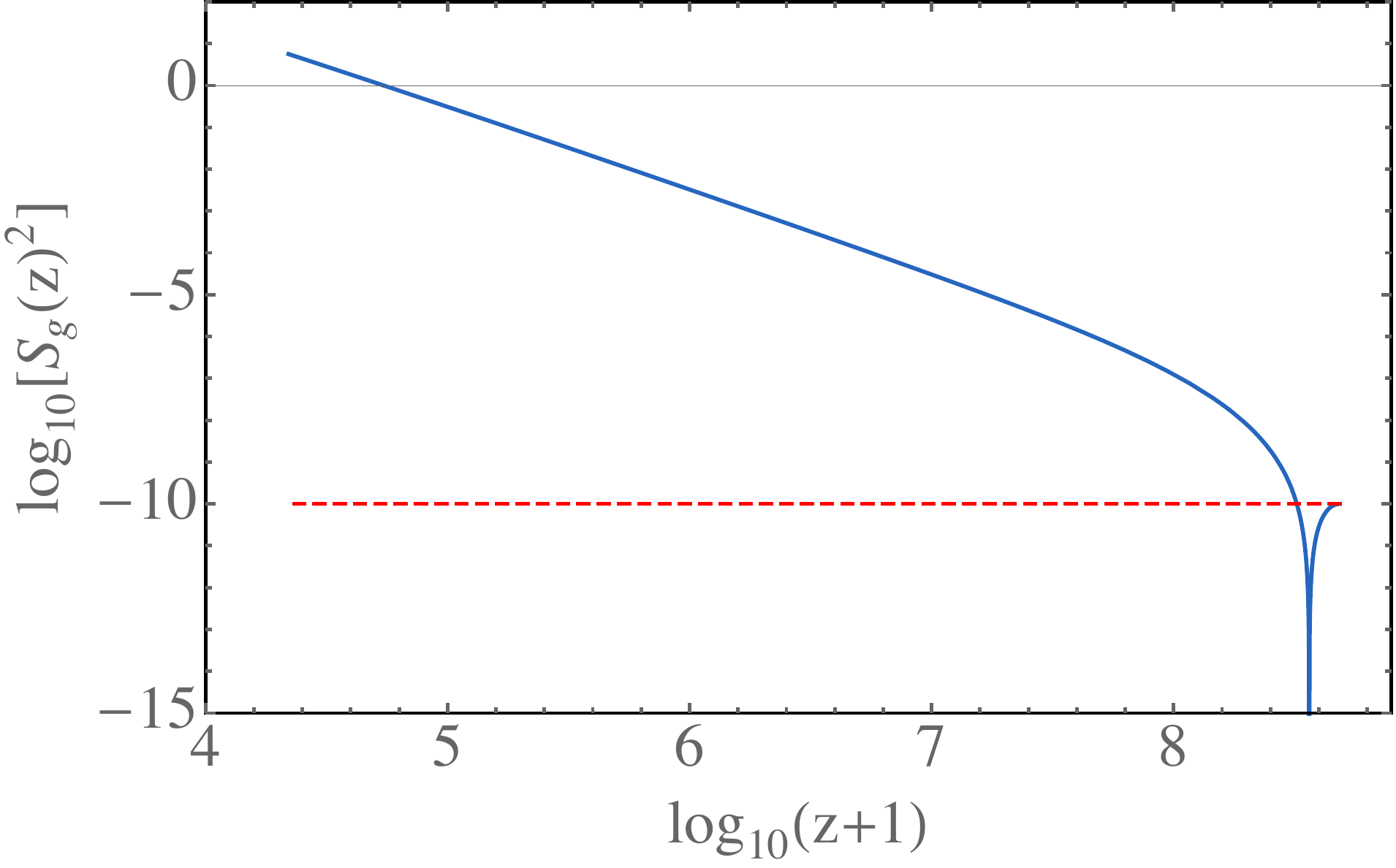}}
   \subfigure[\label{B=1}]
      {\includegraphics[scale=0.38]{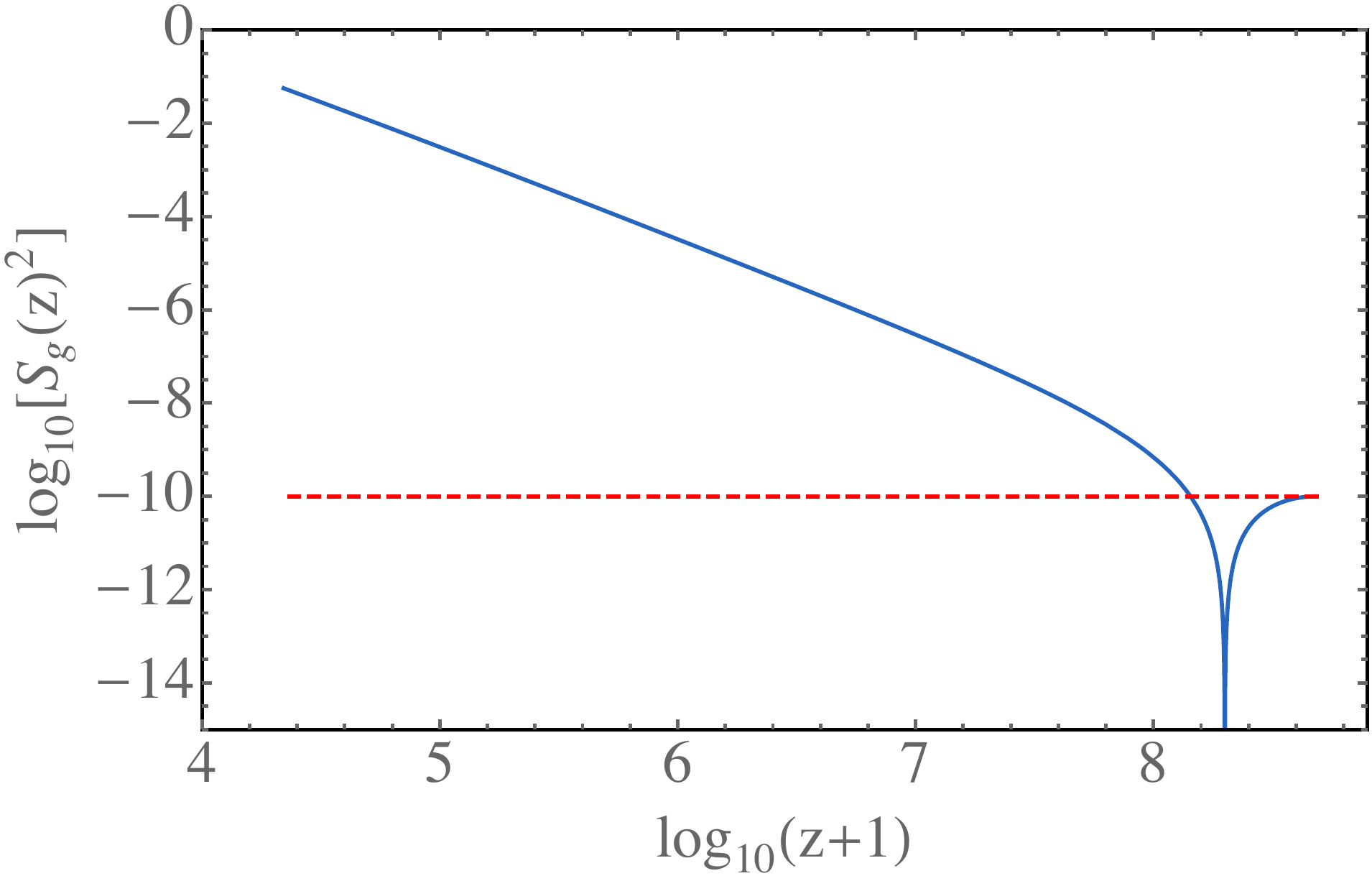}}
      \subfigure[\label{B=-1}]
   {\includegraphics[scale=0.38]{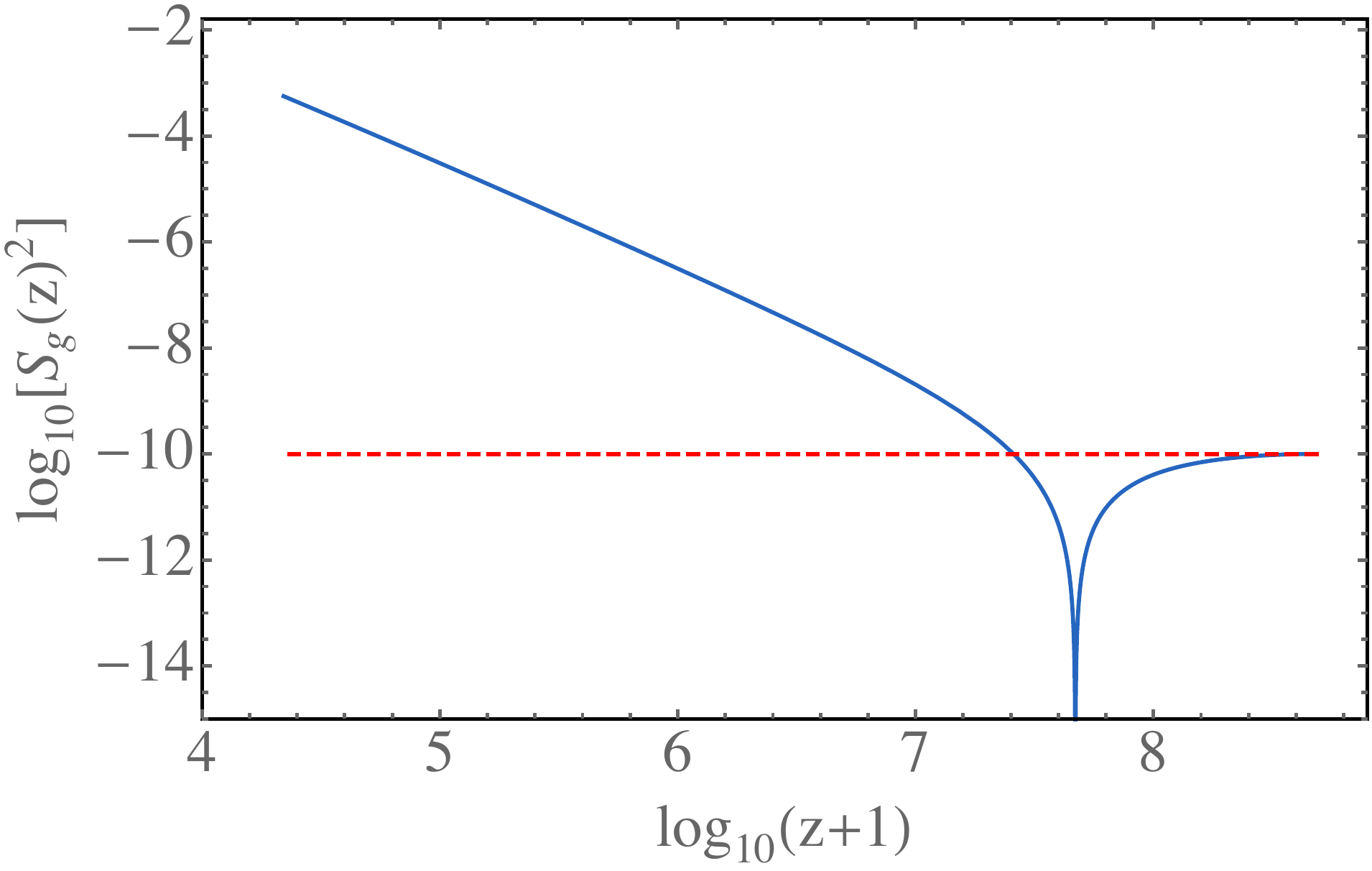}}
   \subfigure[\label{B=-3}]
      {\includegraphics[scale=0.38]{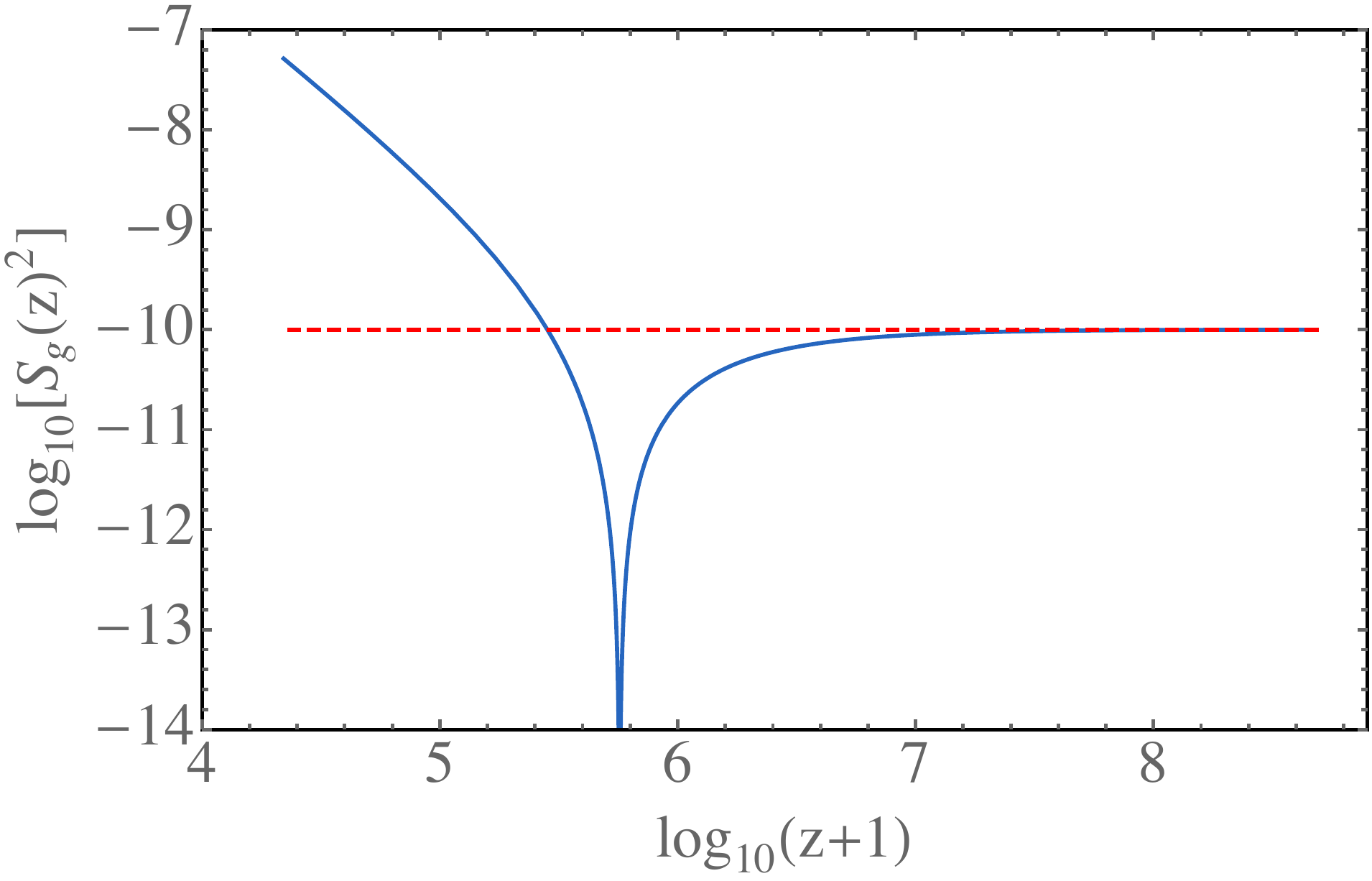}}
       \caption{\label{scalar2} Evolution of scalar perturbations for the metric $g$ for the case $A=10^{-5}$ and $k\simeq10\HH_0$, varying the initial condition for $S_f$. We have chosen the cases 
       $B=10,\, 1,\, 10^{-1},\,10^{-3}$, in Fig. \ref{B=10}, \ref{B=1}, \ref{B=-1} and \ref{B=-3},  respectively. The red-dashed line represent the analytical solution for the equation of scalar perturbations valid in the radiation dominated era.}
 \end{figure}

 \begin{figure}[ht!]
 \centering
\includegraphics[scale=0.38]{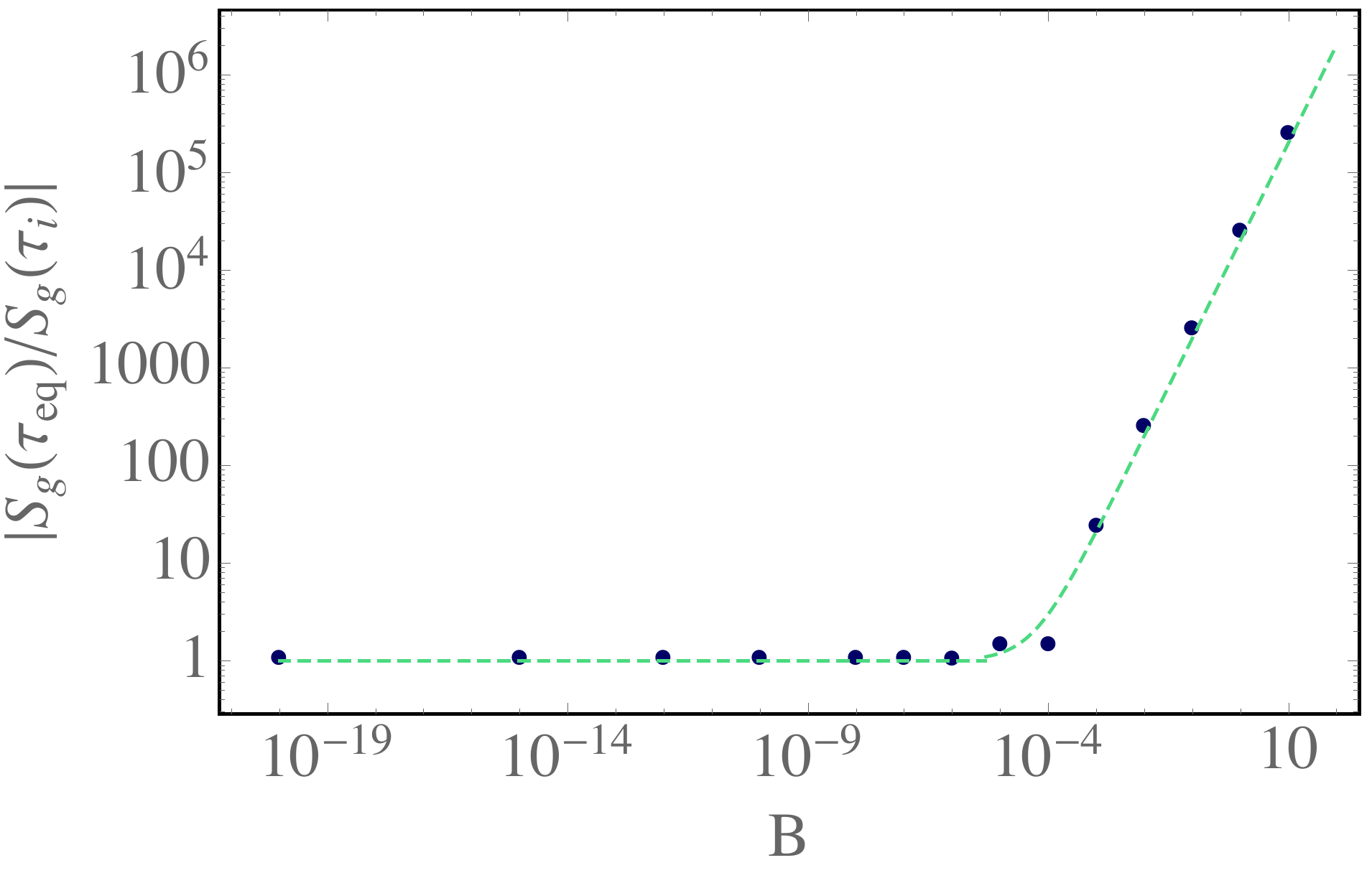}
       \caption{\label{scalar3} Amplification of scalar perturbation $S_g$ at equality as a function of $B$. The dotted line is the interpolating function $i(B)=0.3\, \frac{B}{A}\cdot \theta(B-A)+1$, where $\theta$ denotes the Heaviside function.  We have chosen 
       $A=10^{-5}$ and a mode with $k\simeq10\HH_0$.}
 \end{figure}

In the physical sector we recover the usual behavior of super-Hubble scalar perturbation in the radiation dominated era, while the perturbations of $f$ grow linearly.  Neglecting the constant mode for $S_f$ and the subdominant  decaying mode for $S_g$, we have
\be\label{initial}
S_{g}=A\,,\hspace{1.5 cm}S_{f}=B\frac{\tau}{\tau_{\rm in}}\,.
\ee

We solve Eqs. (\ref{Sg}) and (\ref{Sf}) numerically with initial conditions (\ref{initial}) and we compare the result with the analytical solution valid in the radiation era, see Fig. \ref{scalar1}. The analytical and numerical solutions for $S_f$ are in very good agreement. The solution for $S_g$, however is soon affected by the coupling term $\frac{9\beta_{1}\mathcal{H}^{3}}{2\beta_{4}k^{2}r}S_{f}'$ in (\ref{Sg}) which can be large for small values of $k$.

We then choose the initial condition for scalar perturbations in the physical sector compatible with the observational constraints from structure formation, $A=10^{-5}$, and we explore how the evolution changes varying $B$, i.e., the initial condition for $S_f$, see Fig.~\ref{scalar2}. If the ratio between the initial condition of $S_f$ and $S_g$ is  big, i.e. $B/A\gsim 1$, the solution for $S_g$ develops a growing mode in the radiation dominated epoch.  
In Fig. \ref{scalar3} we plot the amplification of $S_g$ at the end of the radiation era ($z\sim 10^4$) as a function of $B$. We see that the amplification is roughly proportional to the initial condition of $S_f$ for $B/A\gg1$. The amplification during the radiation era is absent for $B/A< 1$. 

Comparing the order of magnitude of the terms in eq.~(\ref{Sg}) we find 
in order for the instability to develop during the radiation dominated era we need
\be
\frac{B}{A}\gsim 100\frac{(1+z_{\rm eq})^2}{1+z_{\rm in}} \,.
\ee
For a realistic value of $1+z_{\rm eq}\simeq 3\times 10^3$ and our example plotted in Fig.~\ref{scalar2}, i.e., $1+z_{\rm in} =10^9$, this requires $B/A>1$.  For an early inflationary phase with reheat temperature $T_{\rm in}\simeq 10^{10}$GeV we obtain $1+z_{\rm in}\simeq 10^{23}$, hence in order to avoid this mild instability we need to require that
\be B < 10^{-14}A  \,. \ee
Hence for early inflation, only very fine tuned initial condition can avoid
to be affected by this instability in the scalar sector.

\section{Gravitational waves in massive bigravity cosmology}\label{s:ten}
 Tensor perturbations of a given $\bk$-mode can be written as 
 \be
 h^{TT}_{ij} =  h^+e^{(+2)}_{ij}  + h^{-}e^{(-2)}_{ij}
 \ee
 where $+$ and $-$ denote the two helicity-2 modes of the gravitational wave. For an orthonormal system $\widehat \bk,\bfe^{(1)},\bfe^{(2)}$ we have
 \be \bfe^{\pm} =\frac{1}{\sqrt{2}}\left(\bfe^{(1)}\pm i\bfe^{(2)}\right) \quad \mbox{ and }
 \quad e^{(+2)}_{ij} =\bfe^{+}_i\bfe^+_j \,, \quad  e^{(-2)}_{ij} =\bfe^{-}_i\bfe^-_j \,. 
 \ee
 
 For parity invariant perturbations 
 $$\langle h^+(\bk)(h^+(\bk'))^*\rangle =\langle h^-(\bk)(h^-(\bk'))^*\rangle = \delta(\bk-\bk')2\pi^2P_h(k)/k^3\,,$$
  and $\langle h^+h^-\rangle =0$. This is what we shall assume in the following and we shall consider just one mode, say $ h_f^+=h_fG $ and $h_g^+=h_gG$.  Here $G$ is a Gaussian random variable with vanishing mean and with variance 
  $\langle G(\bk)G(\bk')\rangle = \delta(\bk-\bk')2\pi^2/k^3$, so that $h_\bullet$ is the square root of the power spectrum. 
  All what follows is also valid for the modes $h^-_\bullet$ which are not correlated with  $h^+_\bullet$ in the parity symmetric situation which we consider.
 
 For the first order modified Einstein equation with a perfect fluid source term, i.e. no anisotropic stress, we obtain the following tensor perturbation equations
for our bimetric cosmology
 \be\label{e:hg}
h_g''+2\mathcal{H}\,h'_g+k^2 h_g+m^2a^2r\, \beta_1\left(h_g-h_f\right)=0\,,
\ee
\be\label{e:hf}
h_f''+\left[2\left(\mathcal{H}+\frac{r'}{r}\right)-\frac{c'}{c}\right]\,h_f'+c^2 k^2\,h_f-m^2\beta_1\frac{c\, a^2}{r}\, \left(h_g-h_f\right)=0\,.
\ee

At very early times, in the radiation dominated Universe where we want to define our initial conditions, $r$ is very large and
 $m^2\beta_1a^2r=0.48\HH_0^2\sqrt{3\rho_{r0}/(M_p^2m^2\beta_4)}=0.48\HH_0^2\sqrt{\Om_r/0.94}=0.24\times 10^{-2}\HH_0^2=$ constant. Furthermore, $c\simeq -1\simeq$ constant. This implies that in this limit the square bracket of eq.~(\ref{e:hf})
becomes $-2\HH$ and the coupling term is suppressed by a factor $1/r^2$ with respect to the coupling term in eq.~(\ref{e:hg}) and can be neglected. Choosing a super Hubble mode, $k \tau \ll 1$ and recalling that in the radiation era $\mathcal{H}=1/\tau$,  we can neglect the term proportional to $k^2$ in both the equations. To be consistent, in eq. ($\ref{e:hg}$), we then have to neglect also the coupling term, since $K^2=m^2\beta_1 a^2r \simeq (0.05\HH_0)^2<k^2$ for the best fit parameters with $m^2\beta_1 = 0.48 \HH_0^2$. On super Hubble scales in the radiation era we then obtain the solutions
\bea
h_g &=& c_1+c_2\left(\frac{\tau_{\text{in}}}{\tau}\right)\,, \label{e:hgsupsol}\\
h_f &=& c_3(k\tau)^2y_1(ck\tau) -3c_4\frac{(k\tau)^2}{(k\tau_{\rm in})^3}j_1(ck\tau) ~ \simeq ~  c_3 + c_4 \left(\frac{\tau}{\tau_{\text{in}}}\right)^3\,.\label{e:hfsupsol}
\eea
 The solution for $h_g$ differs from the one found in Ref.~\cite{Lagos:2014lca}: in this work when deriving the approximated equation valid in the radiation era for super Hubble modes, the term proportional to $K^2$ in eq. (\ref{e:hgsupsol}) is not neglected. As explained above, this approximation is not completely consistent.\footnote{This can be checked substituting the solutions found in Ref.~\cite{Lagos:2014lca} with coefficients $c_i$ expressed as functions of the initial conditions after inflation in the full equations for perturbations: the terms which do not cancel are negligible only in the specific case in which the initial condition for $h_f'$ after inflation is fine-tuned to be very small, $h_f'(\tau_{\text{in}})\ll \tau_{\text{in}}^3 K^4$.} 
 
 Interestingly, when neglecting the coupling term which for $h_f$ is never relevant, the first expression for the solution~(\ref{e:hfsupsol}) is valid both in the radiation and matter era on all scales as long as $c=$constant. Actually, in
 the matter dominated era the anti-damping term in eq.~(\ref{e:hf}) becomes $2\left(\mathcal{H}+\frac{r'}{r}\right)-\frac{c'}{c} \simeq -\HH$ and with $\HH=2/\tau$, the $h_f$ equation remains unchanged.  The functions $y_1$ and $j_1$ denote the spherical Bessel functions~\cite{Abramo} and $c=-1$ in the radiation era while $c=-1/2$ in the matter era. 
 Considering the growing mode proportional to $c_4$ we find that $h_f$ grows like $\tau^3$ on super Hubble scales and like $\tau$ on sub Hubble scales.
 
However, in general we can no longer  neglect the coupling term in the solution for $h_g$ since, depending on the initial condition $h_f$ may have grown too large to be neglected in its coupling to $h_g$. In contrary, since $h_g$ cannot grow more than $h_f$ and since the pre factor of the coupling term remains small, the coupling can be neglected in the $h_f$ equation and (\ref{e:hfsupsol}) remains a good approximation on super Hubble scales.

The solution for $h_g$ in the radiation dominated era agrees with the well know GR solution, but $h_f$ has a growing mode which indicates the presence of an instability. 
Neglecting the decaying modes we choose the initial conditions

\bea\label{incon}
h_g(\tau)=A\,,\qquad && h_f(\tau) = B  \left(\frac{\tau}{\tau_{\text{in}}}\right)^3\,.
\eea

\begin{figure}[ht!]
 \centering
\subfigure[\label{k10hglinA=B=1}]
   {\includegraphics[scale=0.38]{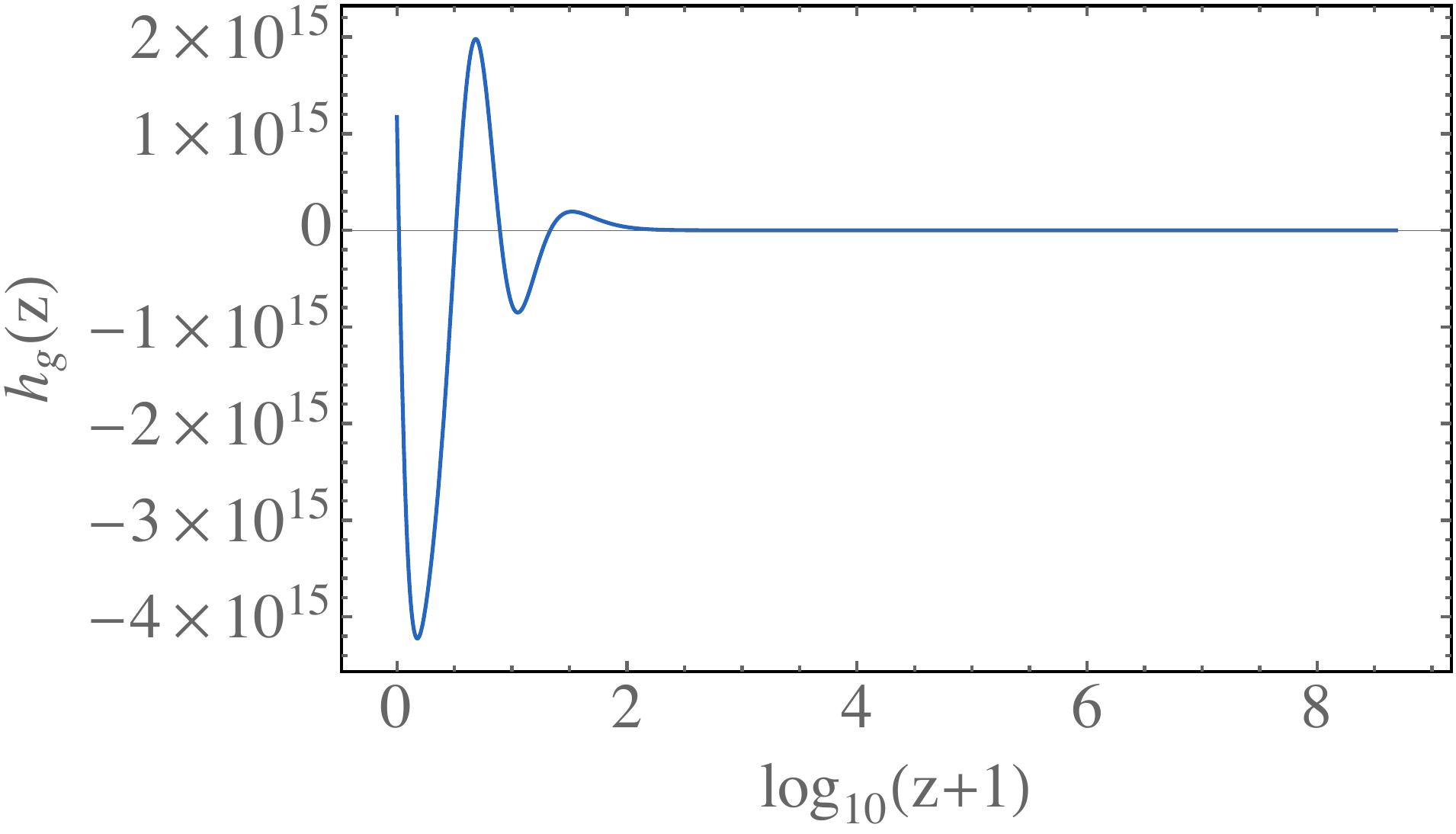}} \qquad
   \subfigure[\label{k10hflinA=B=1}]
      {\includegraphics[scale=0.38]{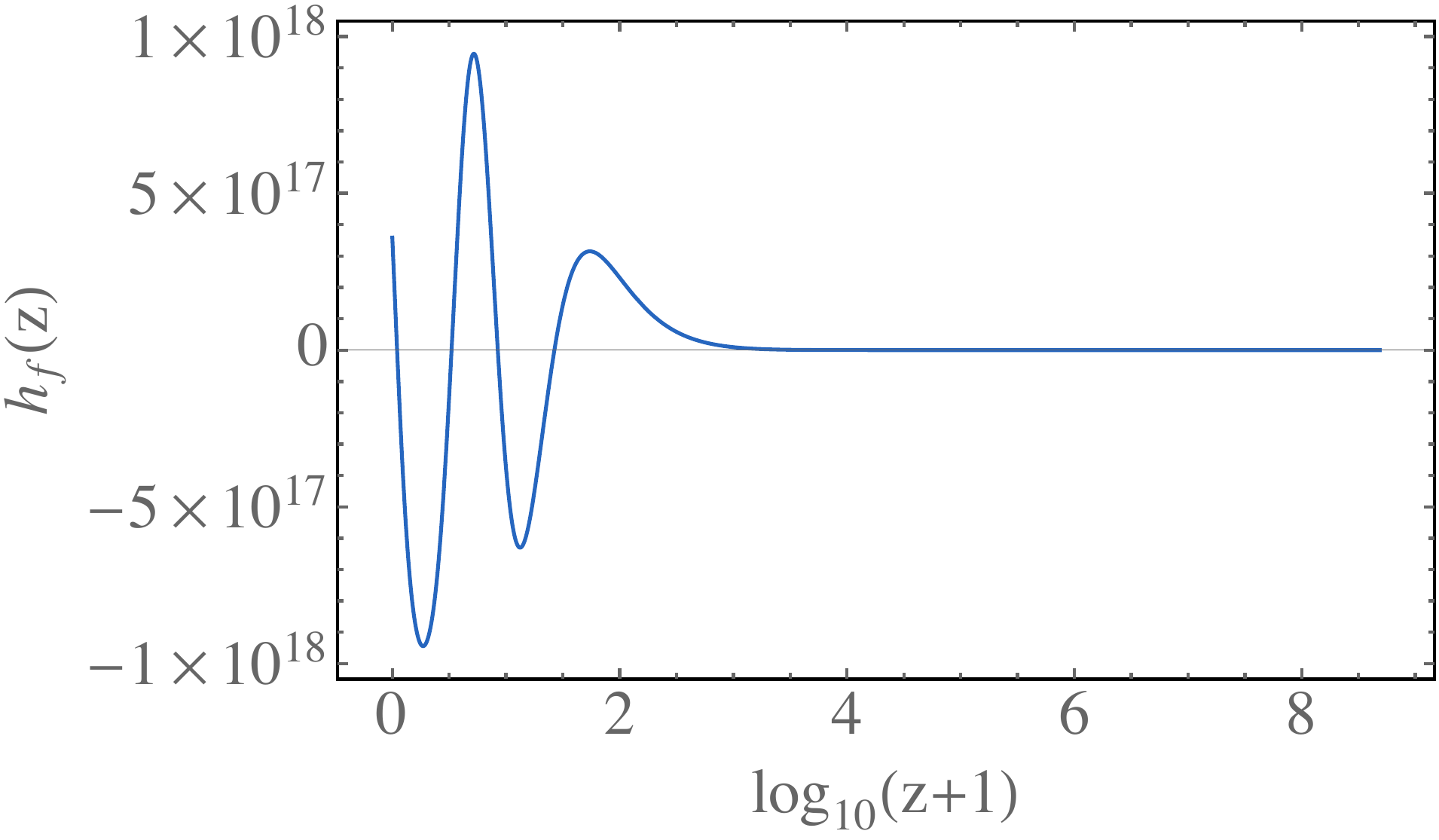}}
        \subfigure[\label{k100hglinA=B=1}]
   {\includegraphics[scale=0.38]{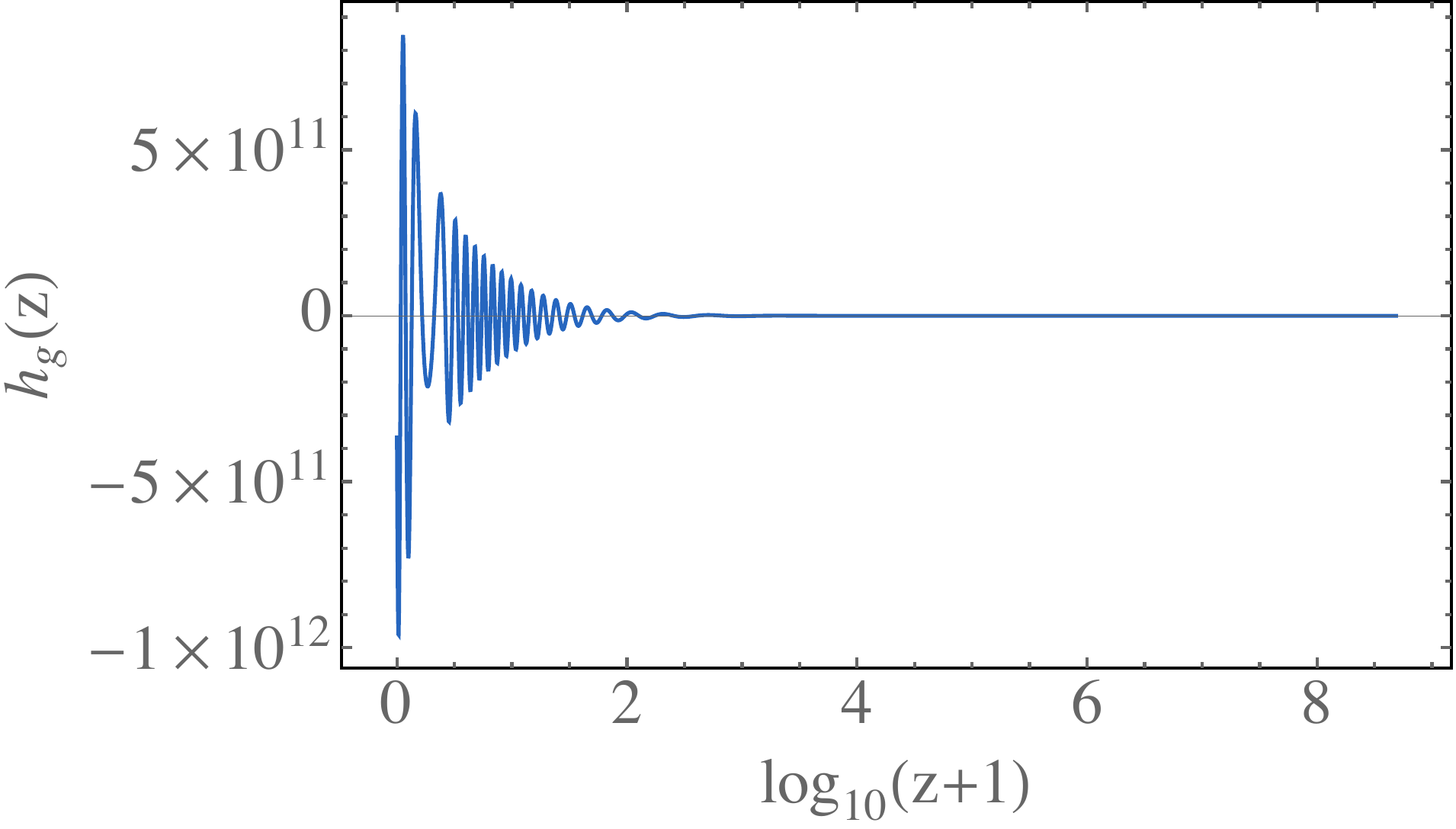}} \qquad
   \subfigure[\label{k100hflinA=B=1}]
      {\includegraphics[scale=0.38]{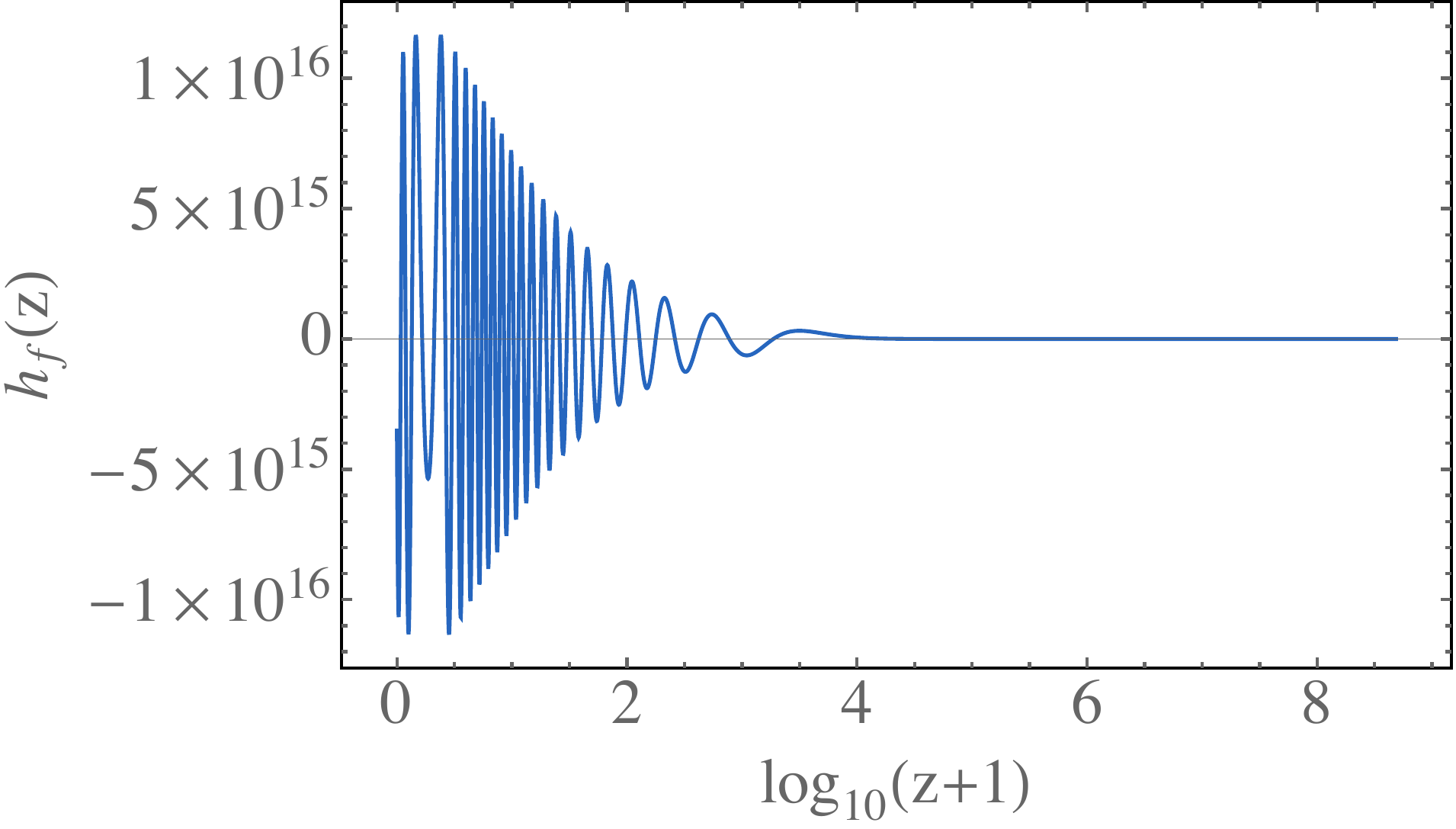}}
         \subfigure[\label{k200hglinA=B=1}]
   {\includegraphics[scale=0.38]{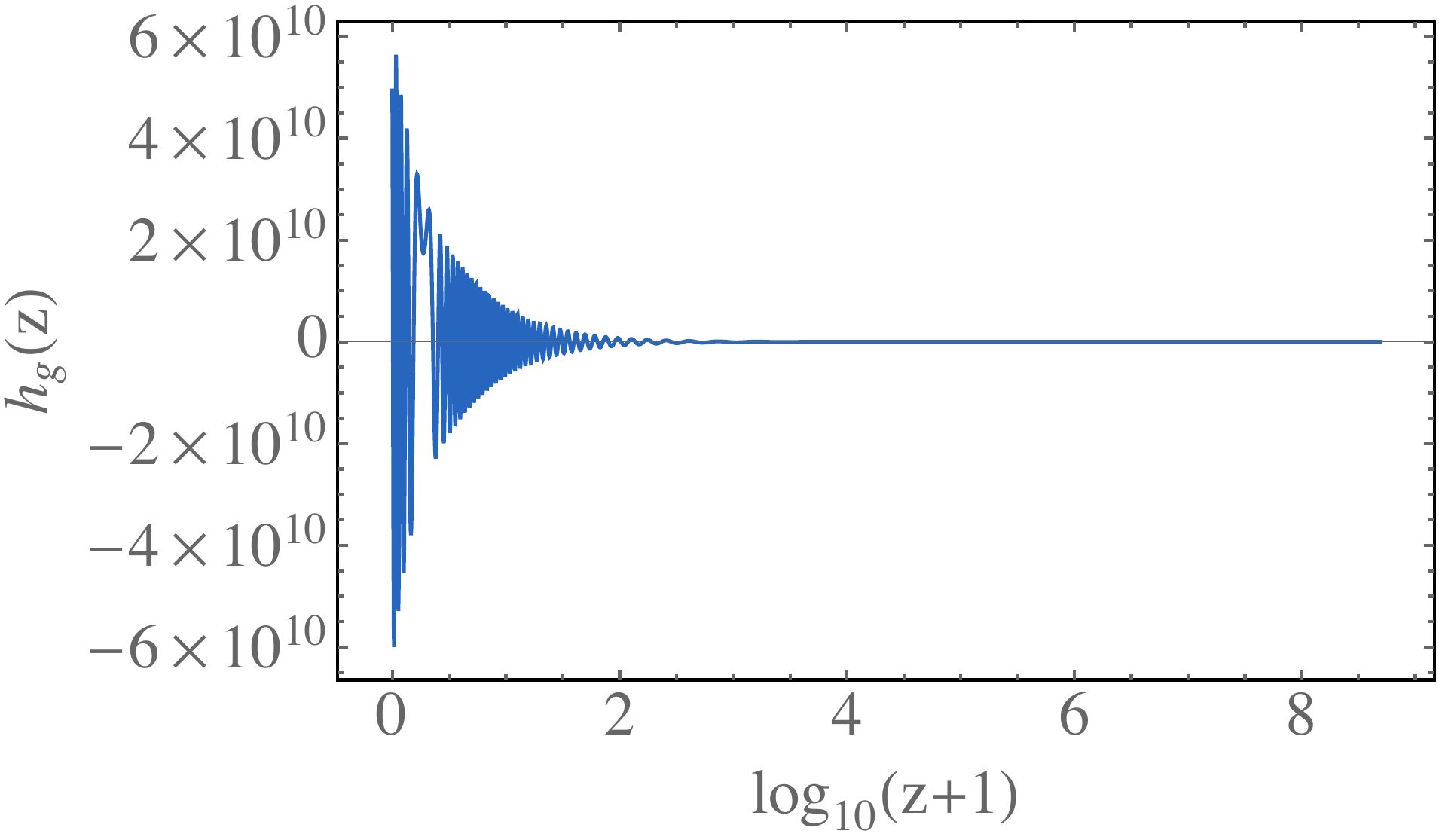}} \qquad
   \subfigure[\label{k200hflinA=B=1}]
      {\includegraphics[scale=0.38]{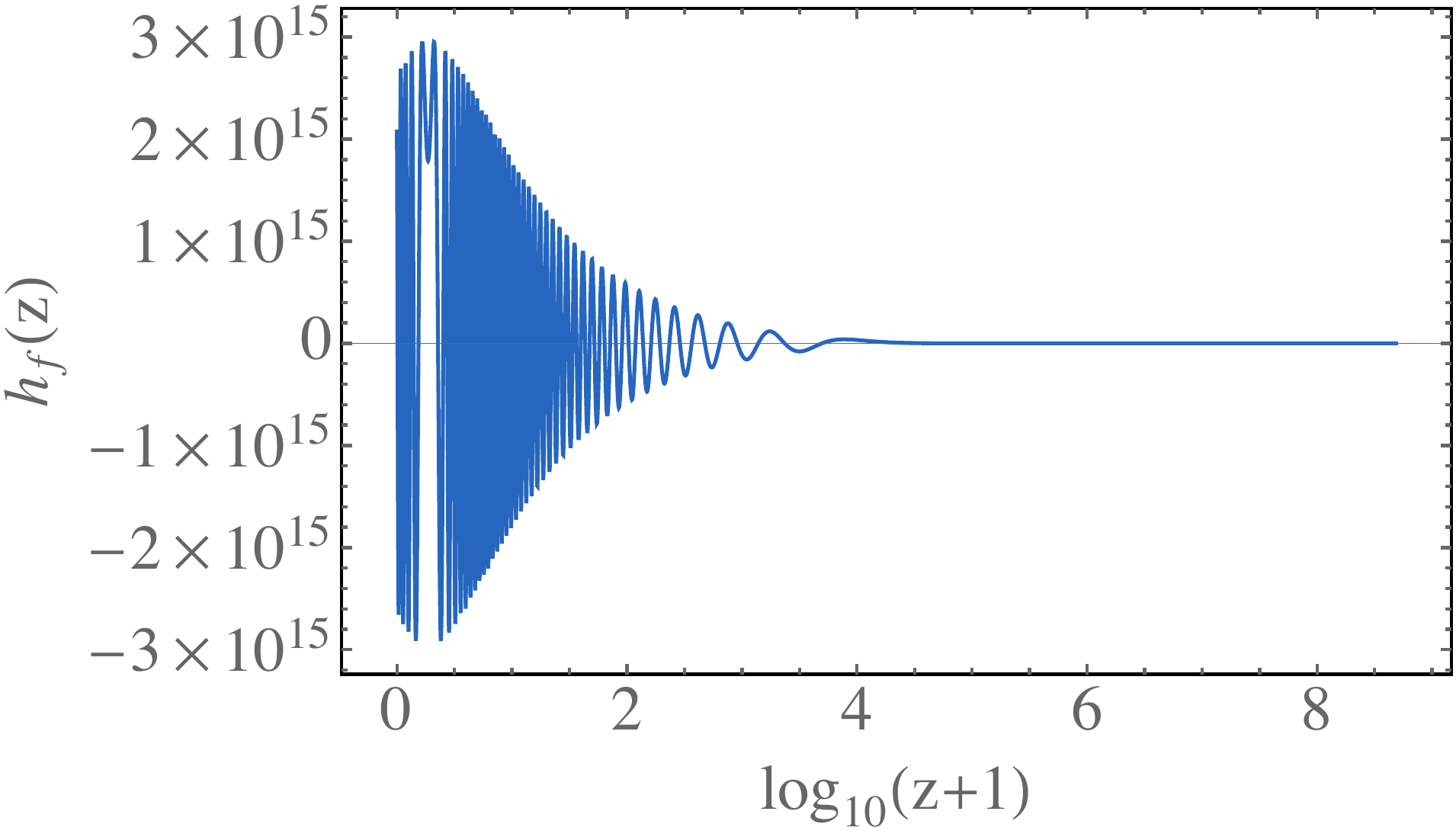}}
\caption{\label{f:gw1} Evolution of tensor perturbations for the metrics $g$ and $f$, in the case $A=B=1$ for the modes $k\simeq10\HH_0$, Figs.  \ref{k10hglinA=B=1},  \ref{k10hflinA=B=1},  $k\simeq100\HH_0$ Figs. \ref{k100hflinA=B=1},  \ref{k100hflinA=B=1} and  $k\simeq200\HH_0$ Figs. \ref{k200hflinA=B=1},  \ref{k200hflinA=B=1}. }
 \end{figure}

\begin{figure}[ht!]
    \centering
    \subfigure[\label{k10hganA=B=1}]
      {\includegraphics[scale=0.38]{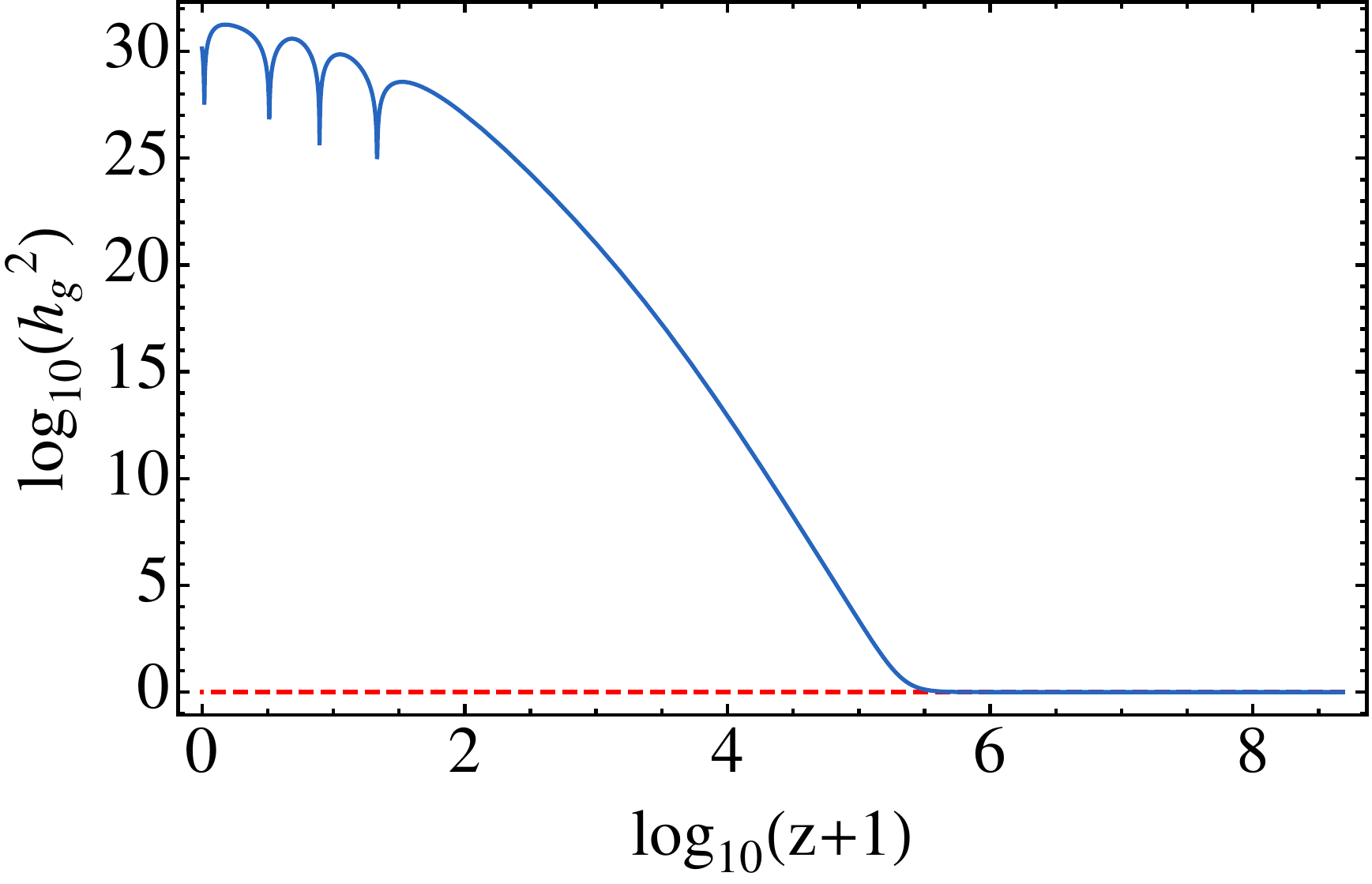}} \qquad
 \subfigure[\label{k10hfanA=B=1}]
      {\includegraphics[scale=0.38]{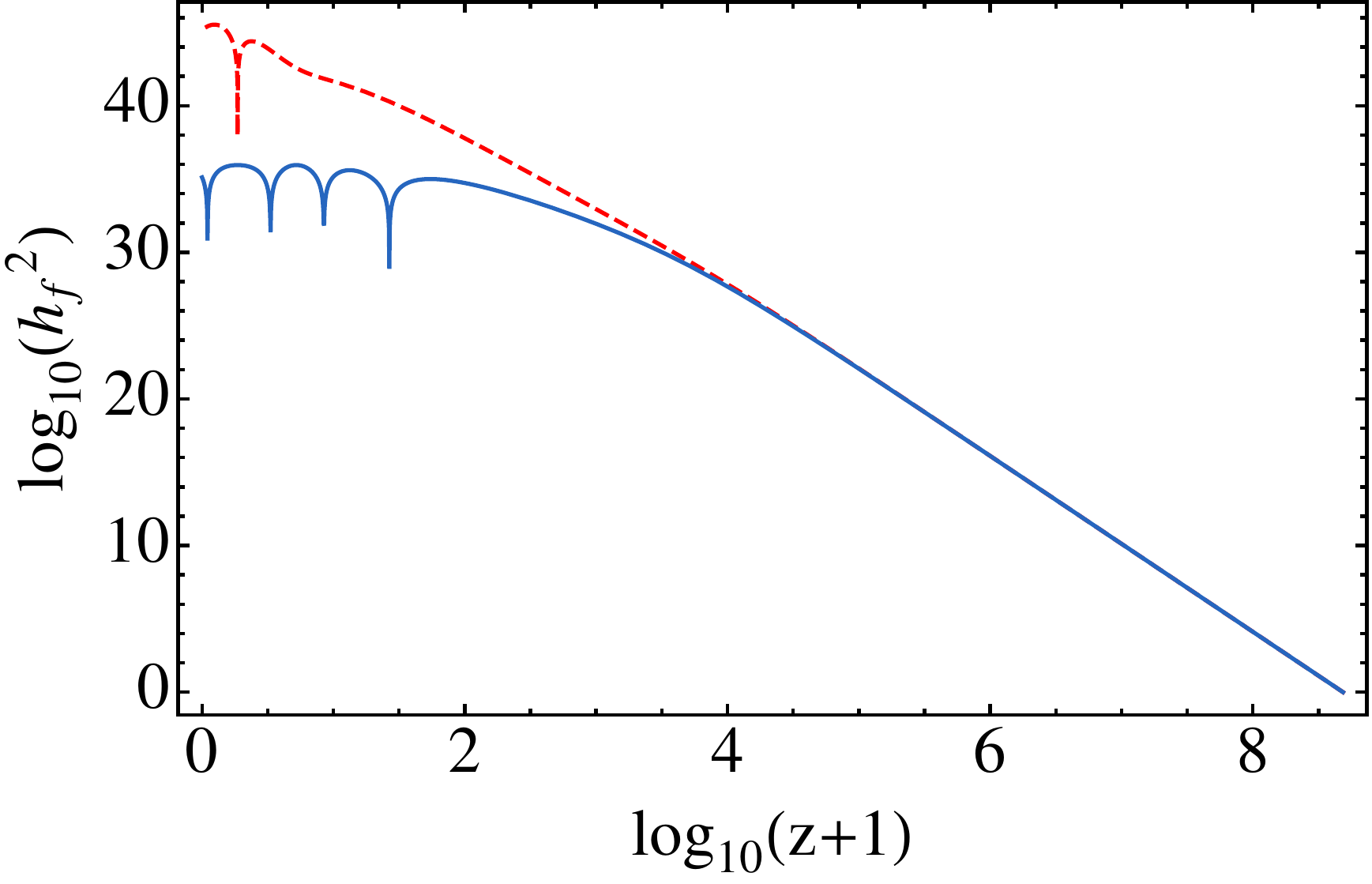}}
      \subfigure[\label{k100hganA=B=1}]
      {\includegraphics[scale=0.38]{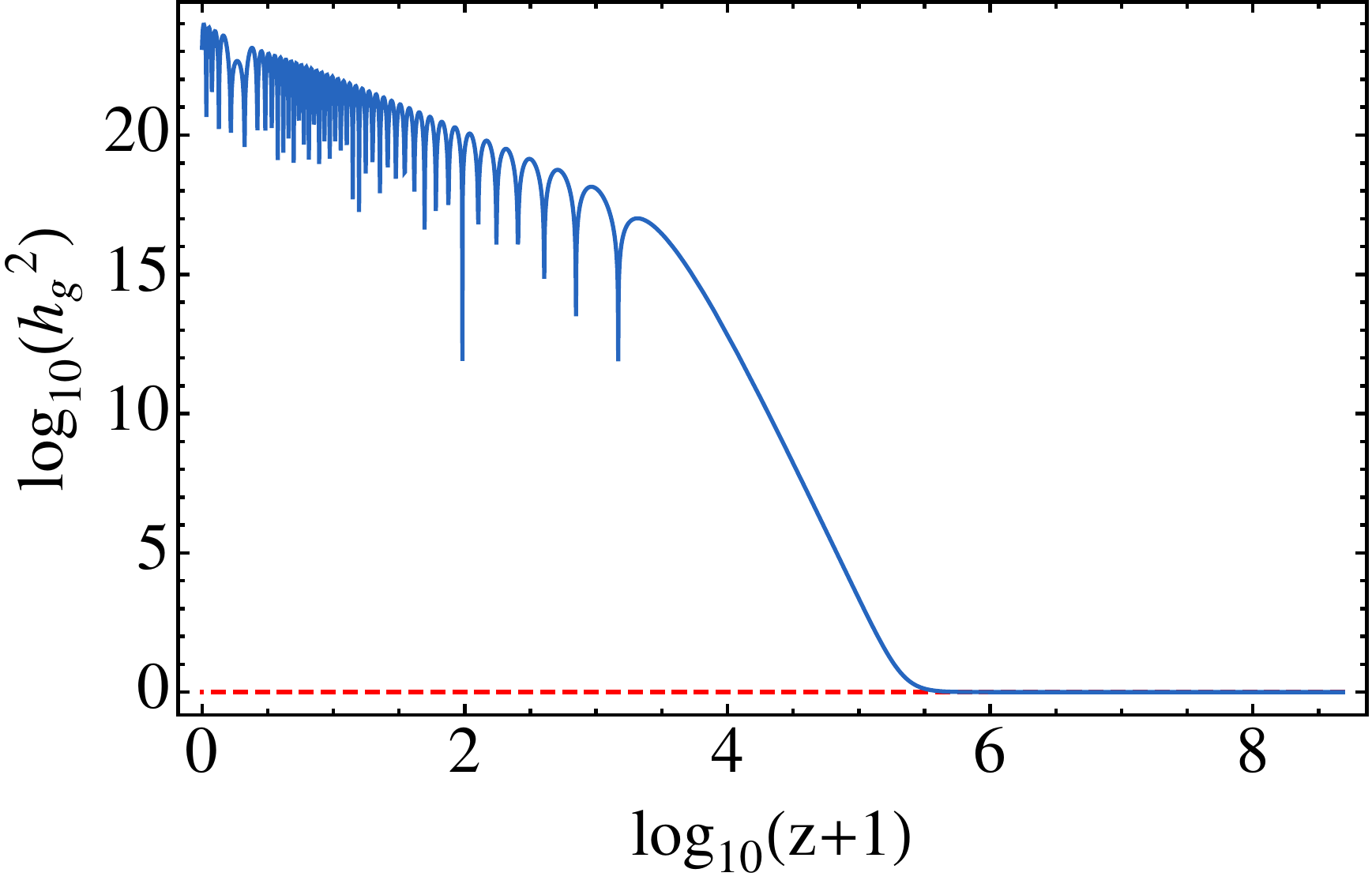}} \qquad
 \subfigure[\label{k100hfanA=B=1}]
      {\includegraphics[scale=0.38]{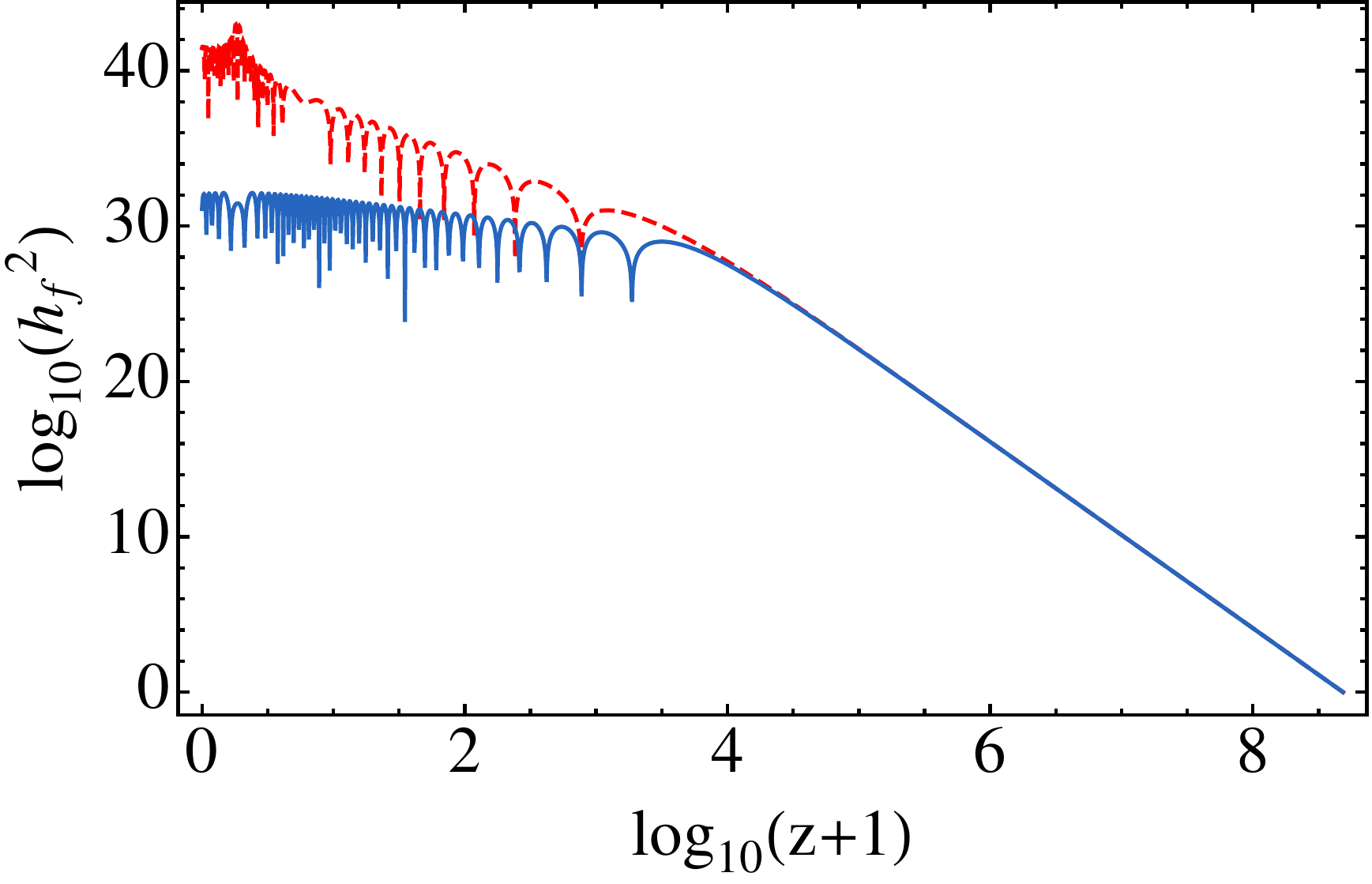}}
      \subfigure[\label{k200hganA=B=1}]
      {\includegraphics[scale=0.38]{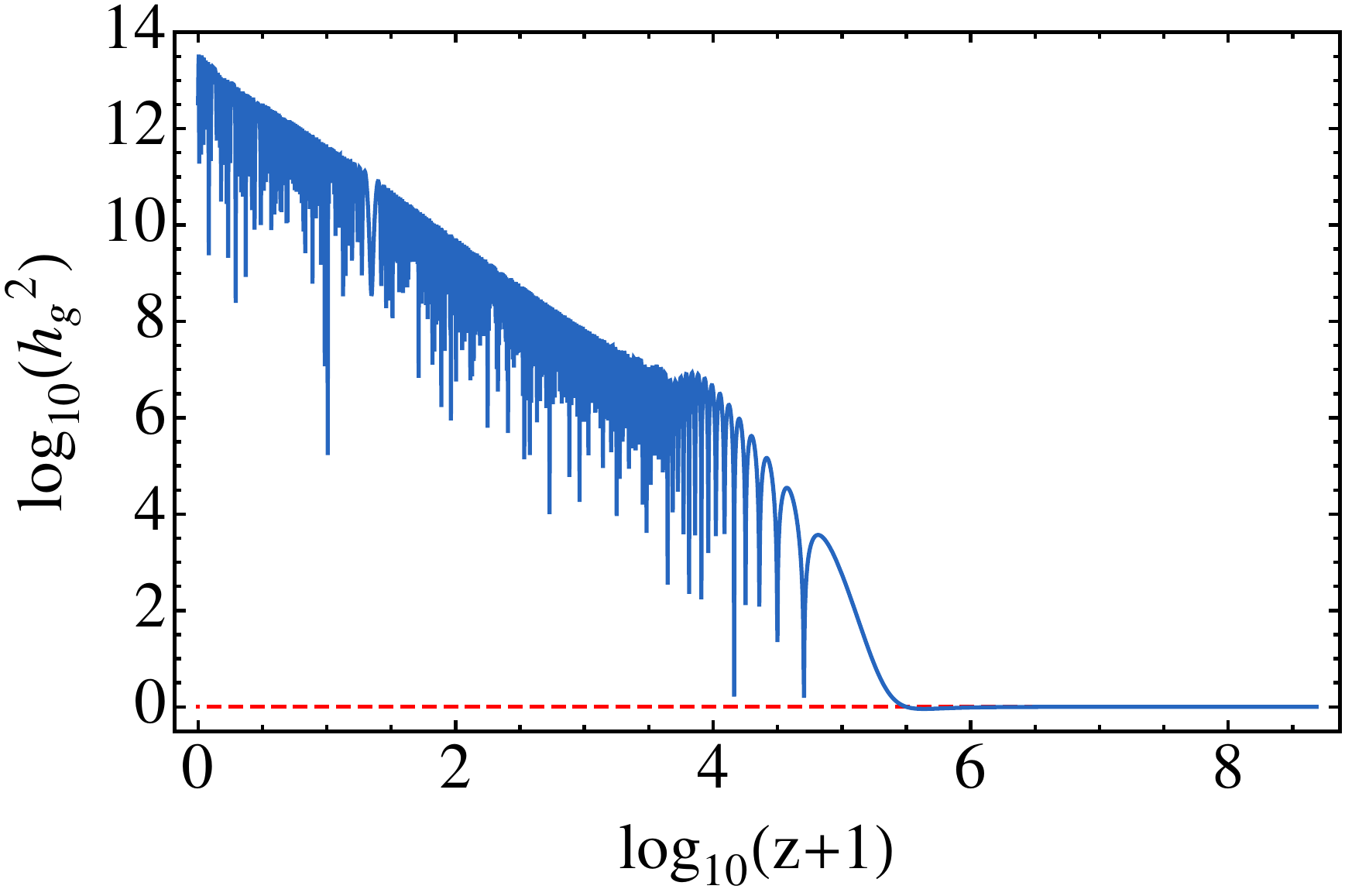}} \qquad
 \subfigure[\label{k200hfanA=B=1}]
      {\includegraphics[scale=0.38]{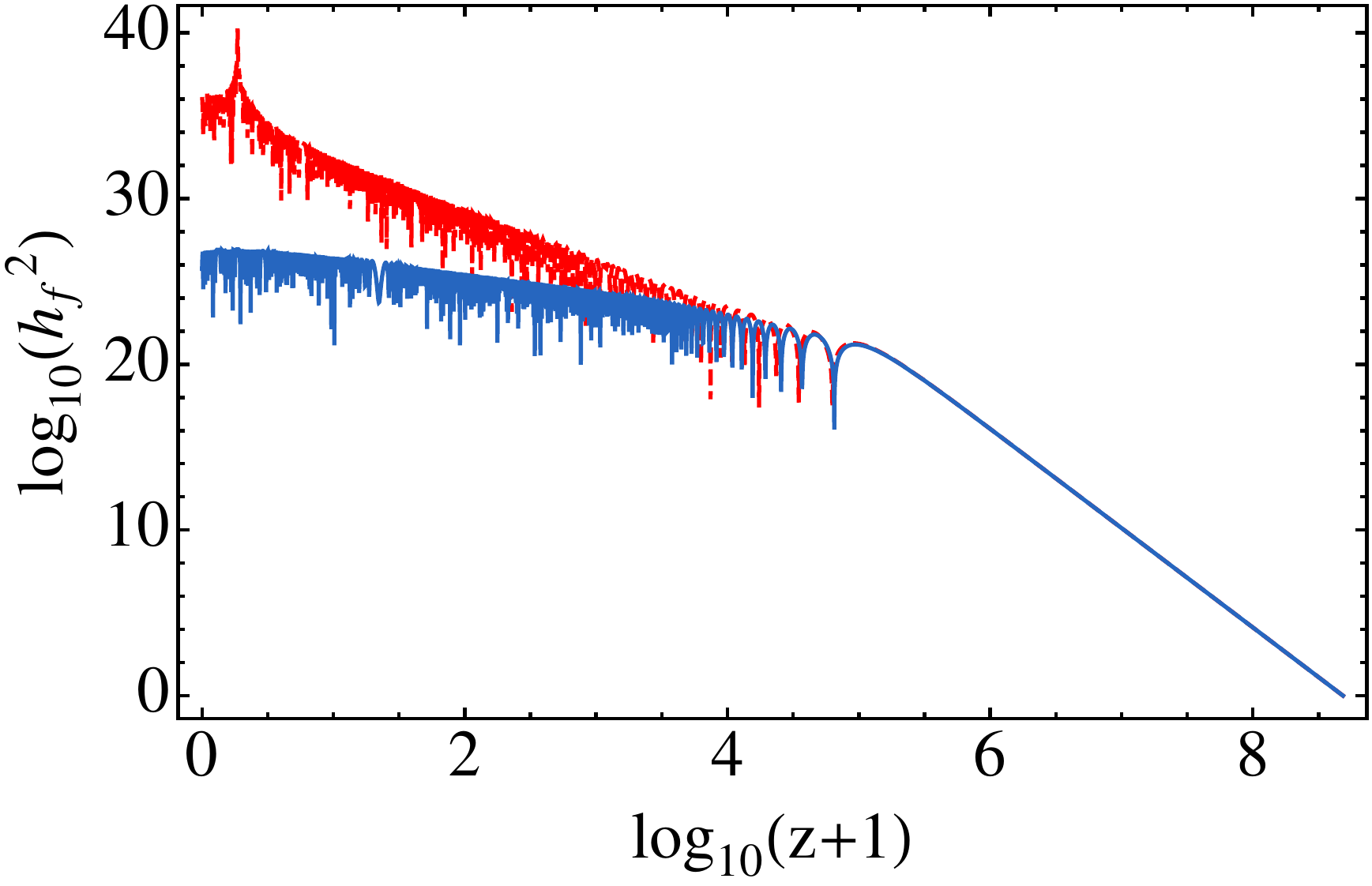}}
\caption{\label{f:gw2}Evolution of tensor perturbations for the metrics $g$ and $f$, in the case $A=B=1$ for the modes $k\simeq10\HH_0$, Figs.  \ref{k10hganA=B=1},  \ref{k10hfanA=B=1},  $k\simeq100\HH_0$ Figs. \ref{k100hganA=B=1},  \ref{k100hfanA=B=1} and  $k\simeq2000\HH_0$ Figs. \ref{k200hganA=B=1},  \ref{k200hfanA=B=1}. 
 The result of the numerical integration (blue, solid) is plotted together with the analytic approximation valid in the radiation era for $h_g$ and on super-Hubble scales for $h_f$ (red, dashed).}
 \end{figure}

The behaviour of the solution depends very sensitively on the initial condition, in the following we explore different possibilities. Naively, we might argue that initially, e.g., after inflation, both $h_f$ and $h_g$ are of the same order of magnitude, $A\simeq B$. The gravitational waves 
 $h_g(k,\tau)$ and $h_f(k,\tau)$ for these initial conditions found by solving numerically Eqs. (\ref{e:hg}) and  (\ref{e:hf}) for the wave numbers  $k\simeq 10\HH_0,~ 100\HH_0,~ 200\HH_0$ are shown in Fig.~\ref{f:gw1}. In a linear plot it looks as if $h_g$ and $h_f$ would be nearly constant during radiation, then $h_f$ starts oscillating with 
frequency $\om^2 = c^2k^2 + cK^2/r^2$ and with 
increasing amplitude at  redshift corresponding to the horizon crossing for the mode chosen. The instability is transferred to the $h_g$ mode trough the coupling. In Fig.~\ref{f:gw2}  we present a log-plot for the same modes  together with the analytic solutions (\ref{e:hgsupsol},\ref{e:hfsupsol}). The analytic solution for $h_f$ is a very good approximation on super-Hubble scales. There one sees that the $\tau^3$ growth on super Hubble scales turns into the milder growth $\propto \tau$ after Hubble entry.

\begin{figure}[ht!]
    \centering
      \includegraphics[scale=0.40]{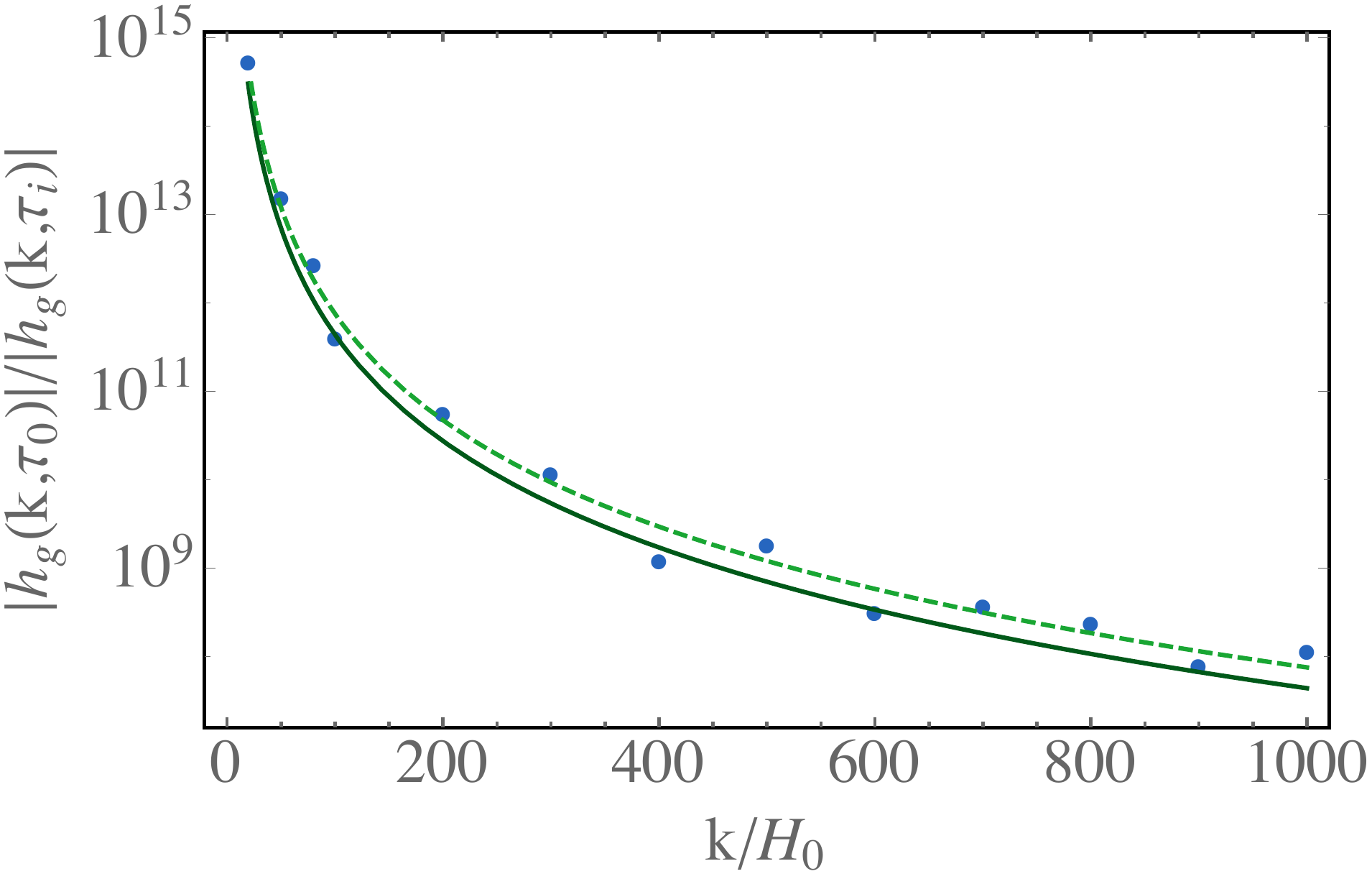}
\caption{\label{amplification}Amplification of tensor perturbations of the $g$ metric in the case $A=B$. The dots represent the value of the amplifications for different modes, the blue solid line is the interpolating function 
$f(k)=\HH(\tau_0)\HH(\tau_i)^3/k^4$, while the green dashed line is the best polynomial fit of the data points given by the software Mathematica.}. 
 \end{figure}

When $A\simeq B$, the gravitational wave  amplitude today is  amplified tremendously, for a mode with wavenumber $k$, roughly by a factor $f(k)=\HH(\tau_0) \HH(\tau_i)^3/k^4$, as shown in Fig.~\ref{amplification}. Therefore, in any case, if the initial amplitudes are not very small, gravitational wave perturbations will grow very large at late time.

\begin{figure}[ht!]
    \centering
 \subfigure[\label{A1-B10^-2-k50-hg}]
      {\includegraphics[scale=0.40]{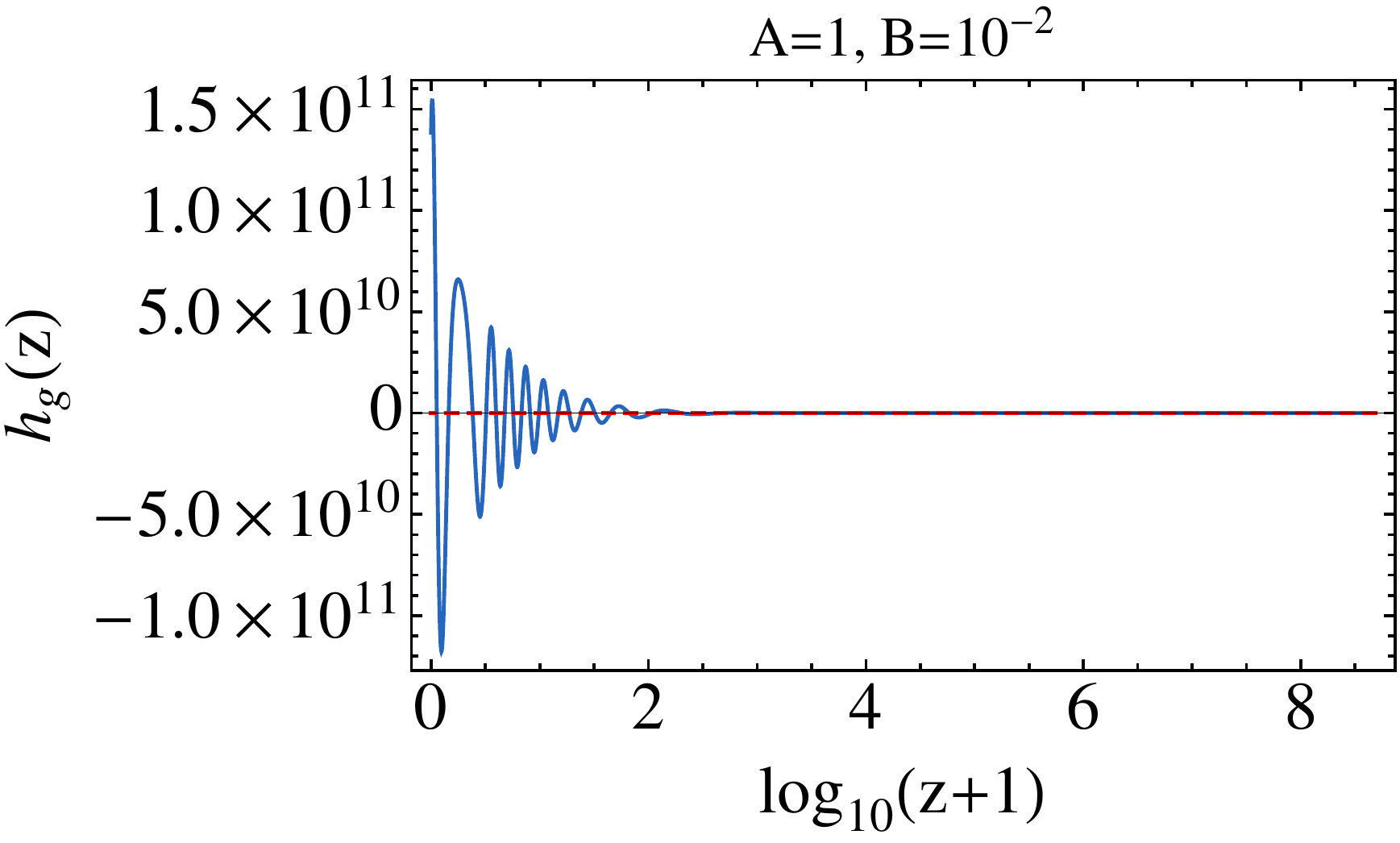}}
 \subfigure[\label{logA1-B10^-2-k50-hg}]
      {\includegraphics[scale=0.40]{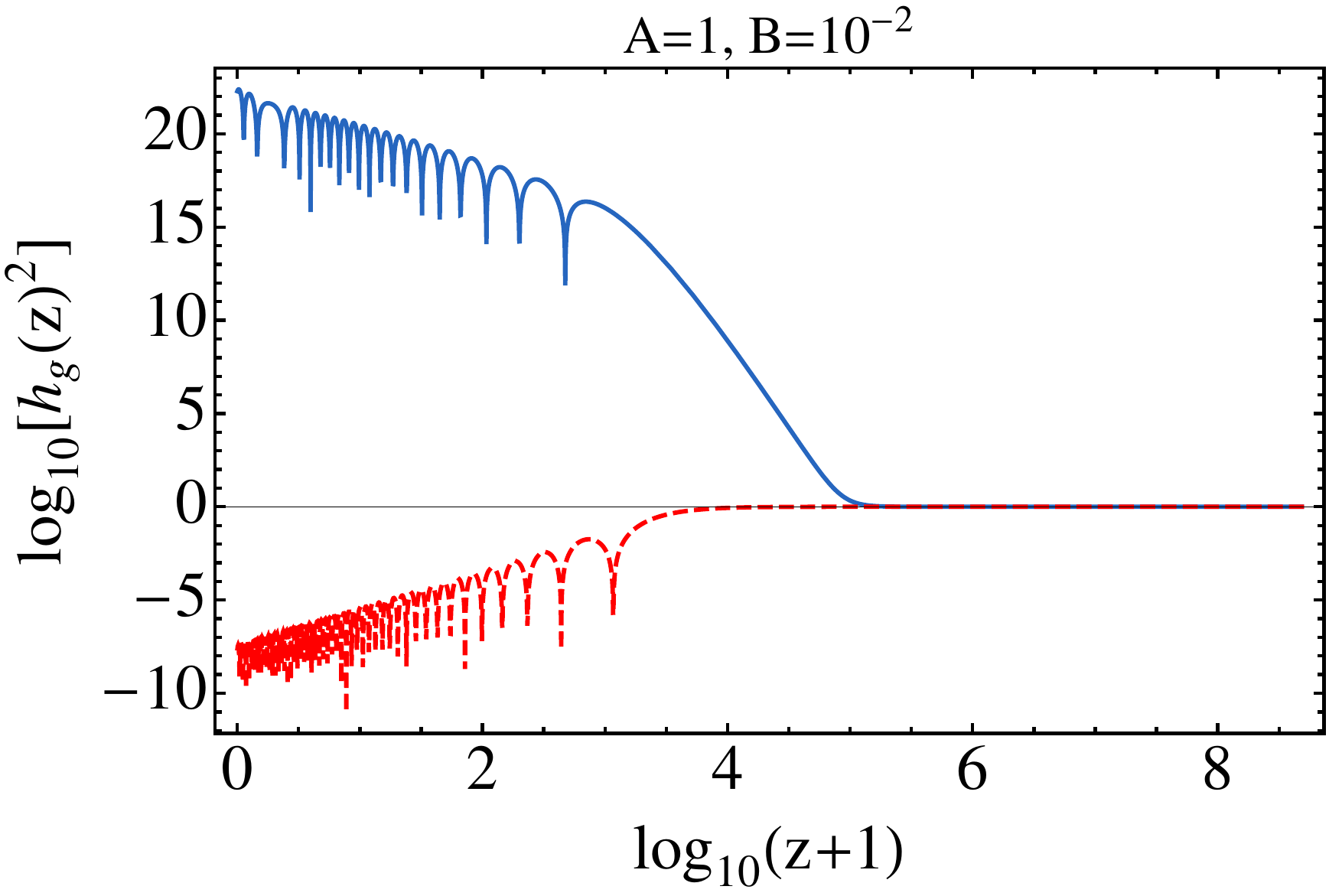}}\qquad
 \subfigure[\label{A1-B10^-10-k50-hg}]
      {\includegraphics[scale=0.40]{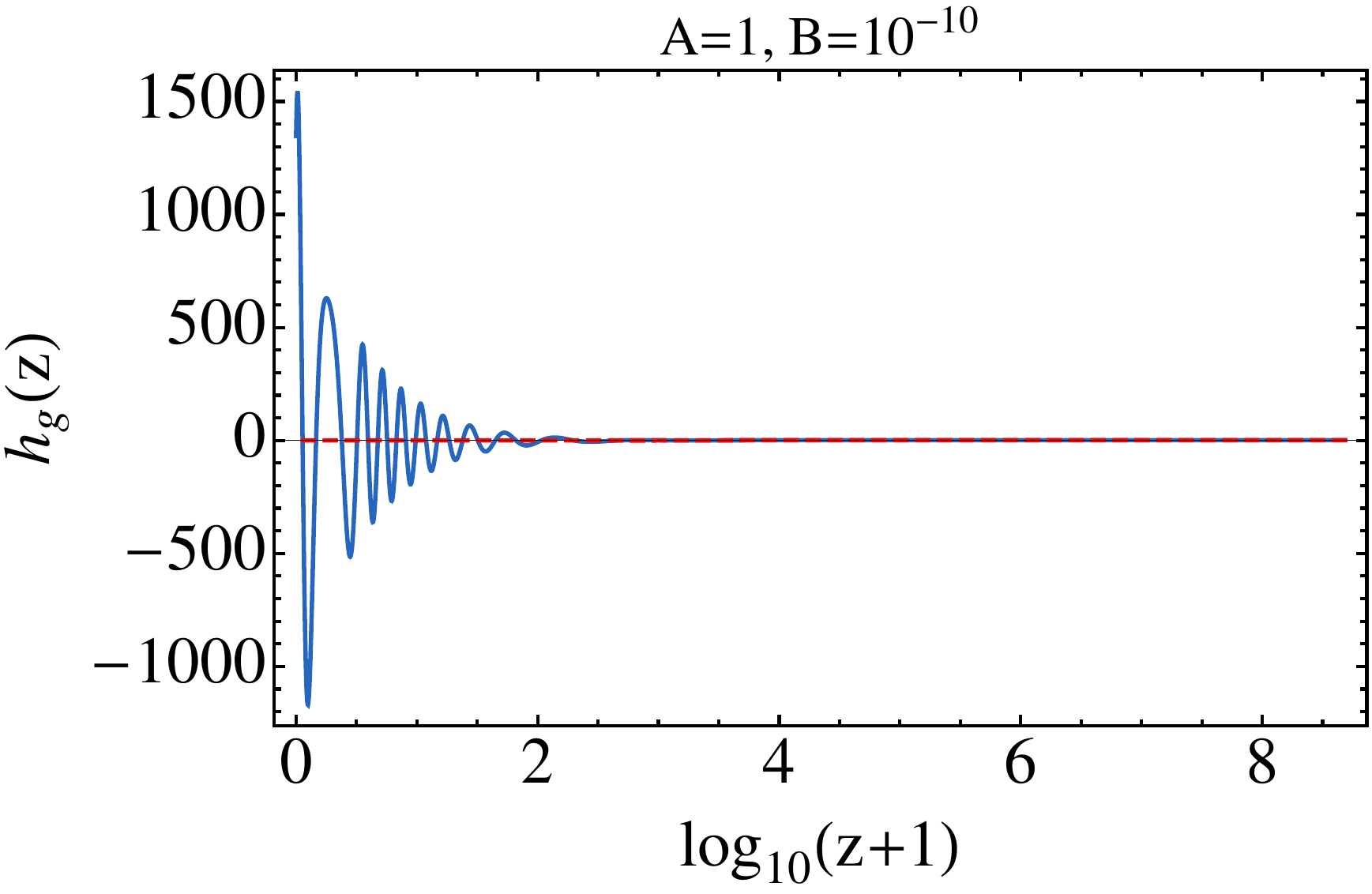}}
 \subfigure[\label{logA1-B10^-10-k50-hg}]
      {\includegraphics[scale=0.40]{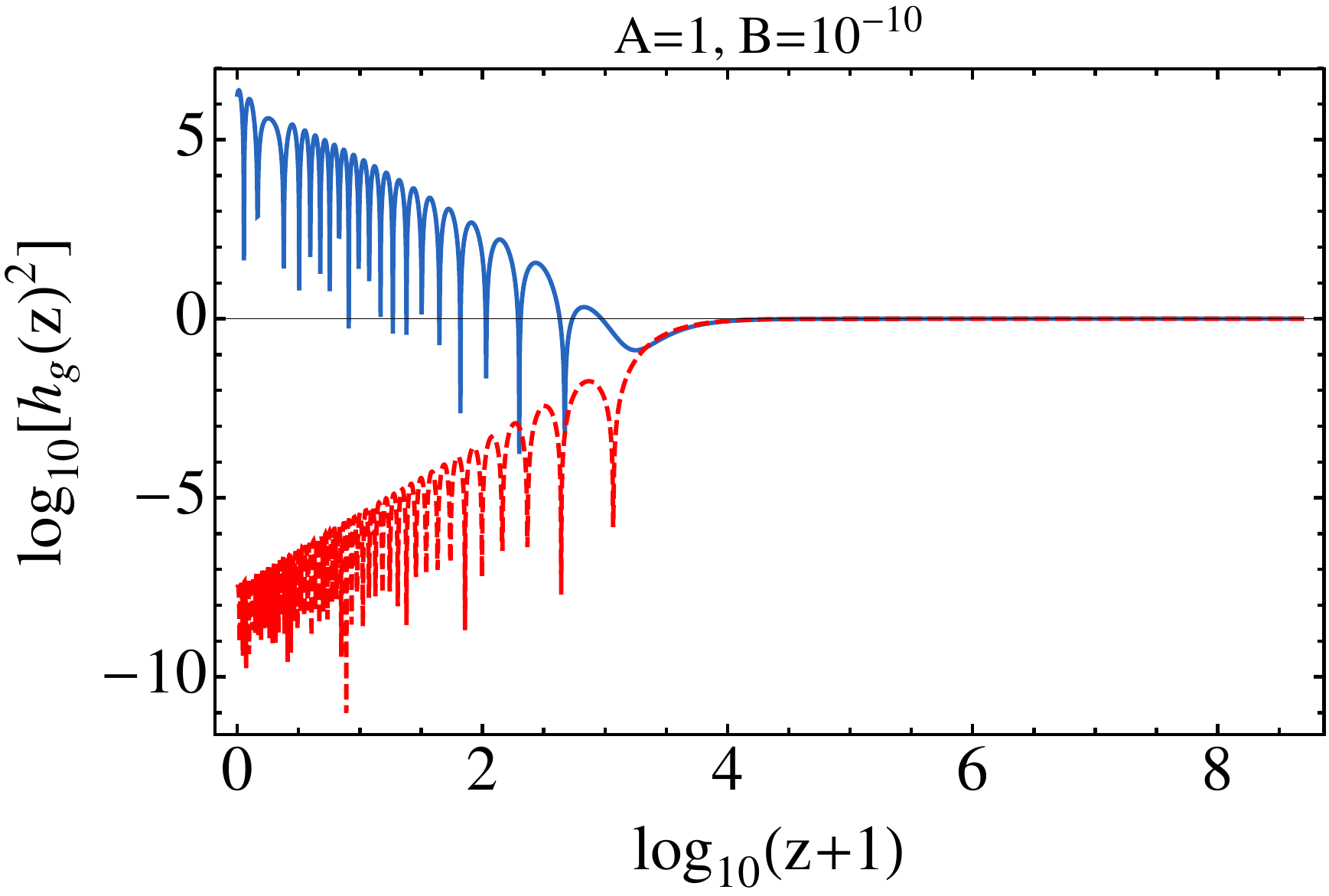}}\qquad
 \subfigure[\label{A1-B10^-13-k50-hg}]
      {\includegraphics[scale=0.40]{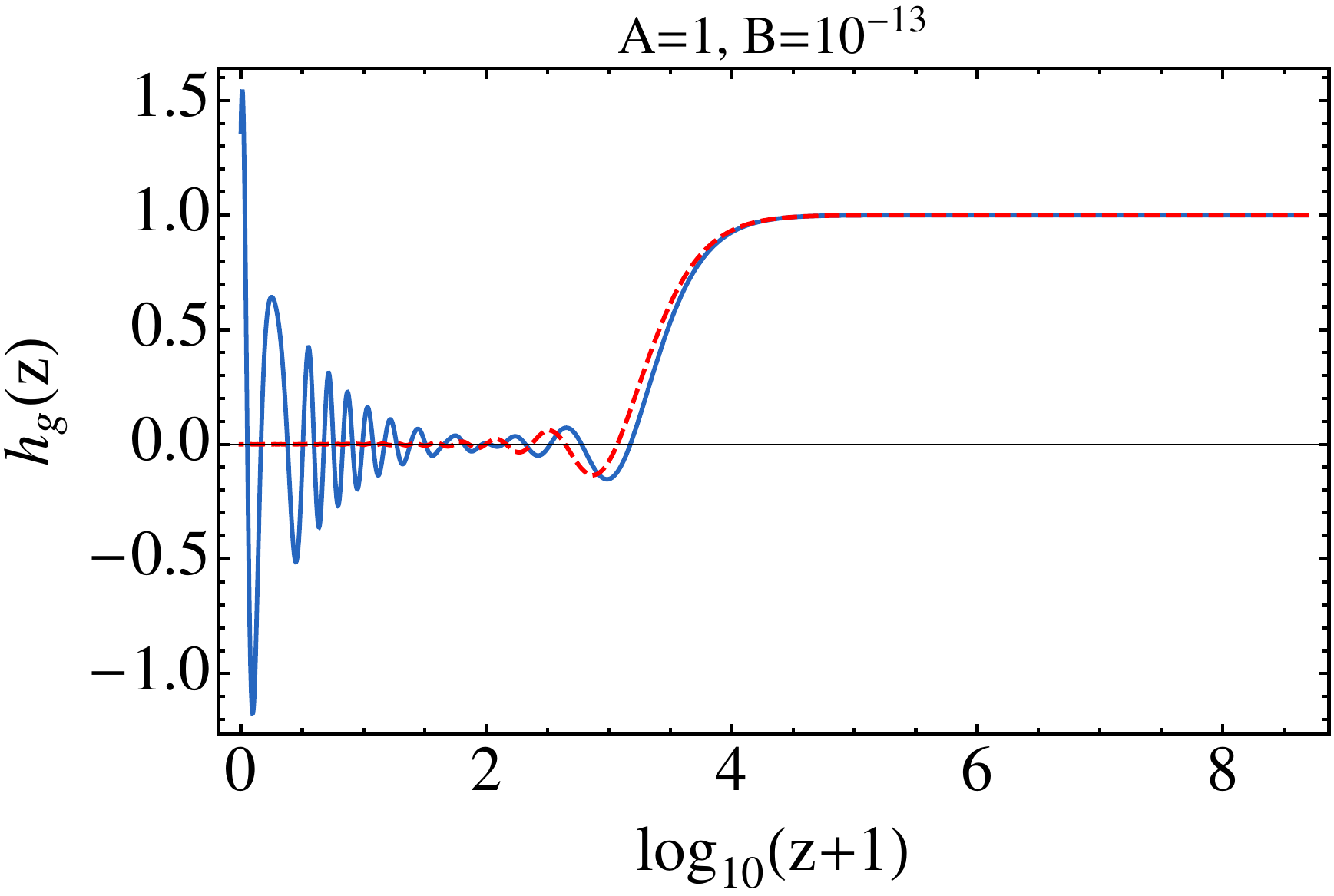}}
 \subfigure[\label{logA1-B10^-13-k50-hg}]
      {\includegraphics[scale=0.40]{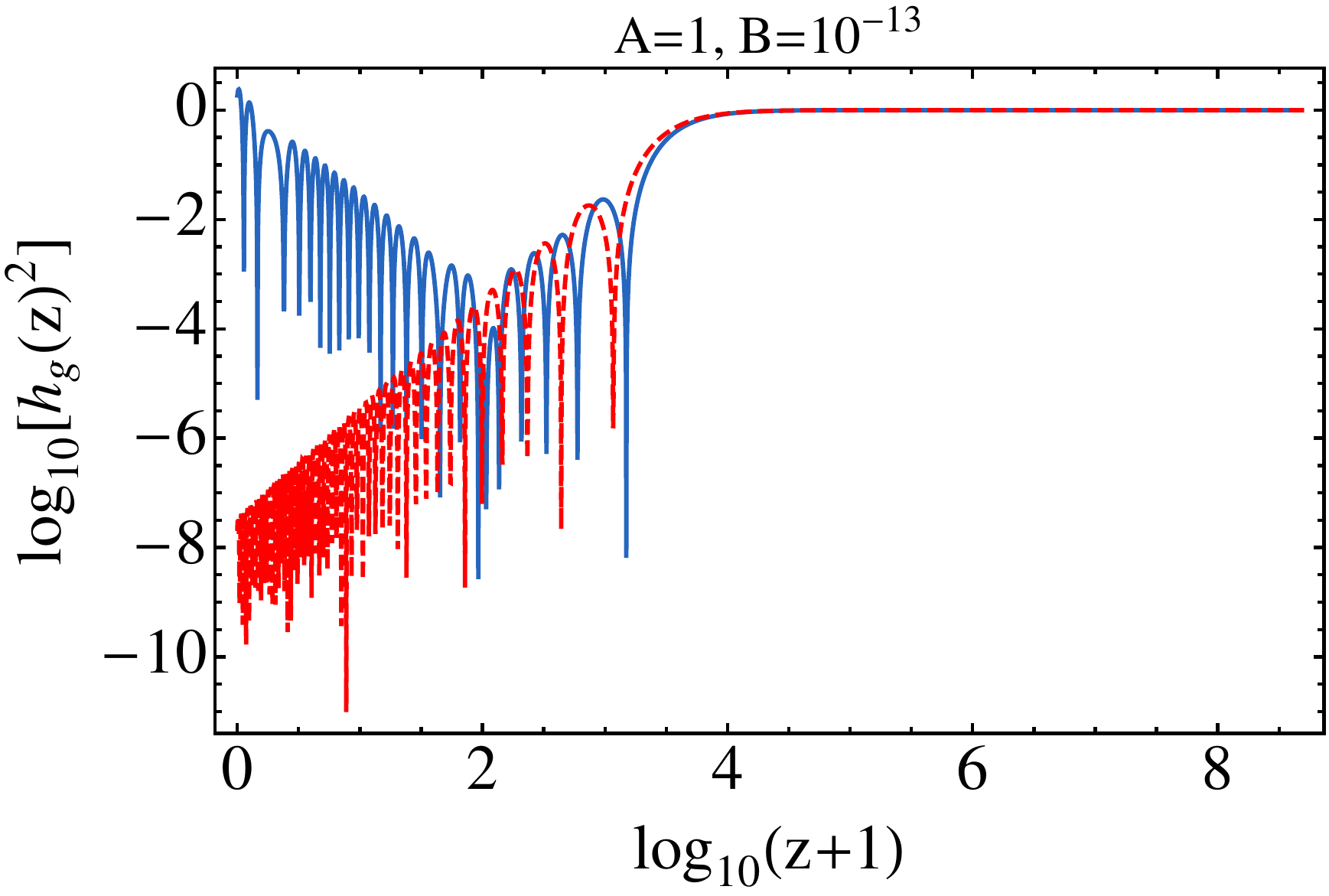}}\qquad
 \subfigure[\label{A1-B0-k50-hg}]
      {\includegraphics[scale=0.40]{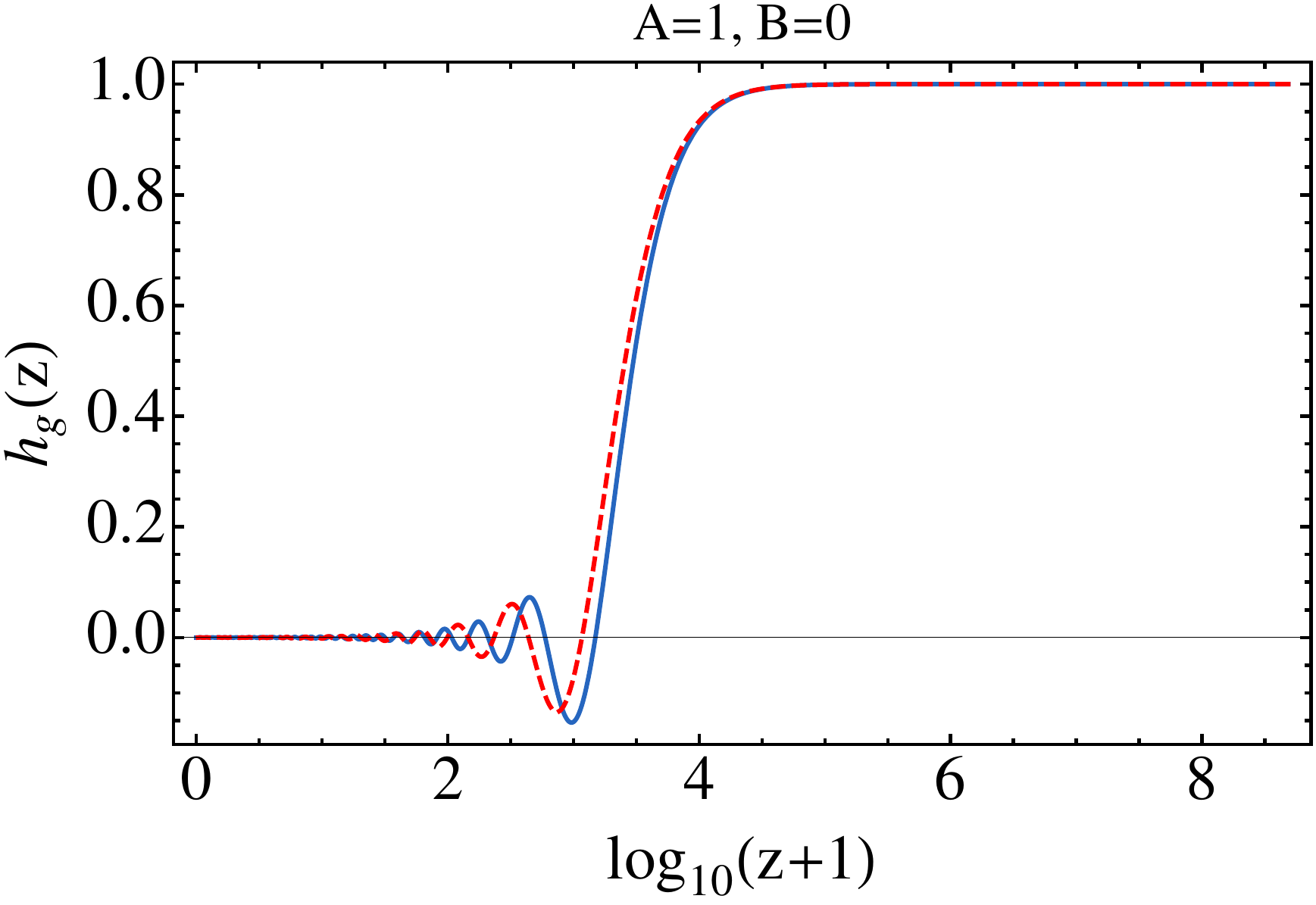}}
 \subfigure[\label{logA1-B0-k50-hg}]
      {\includegraphics[scale=0.40]{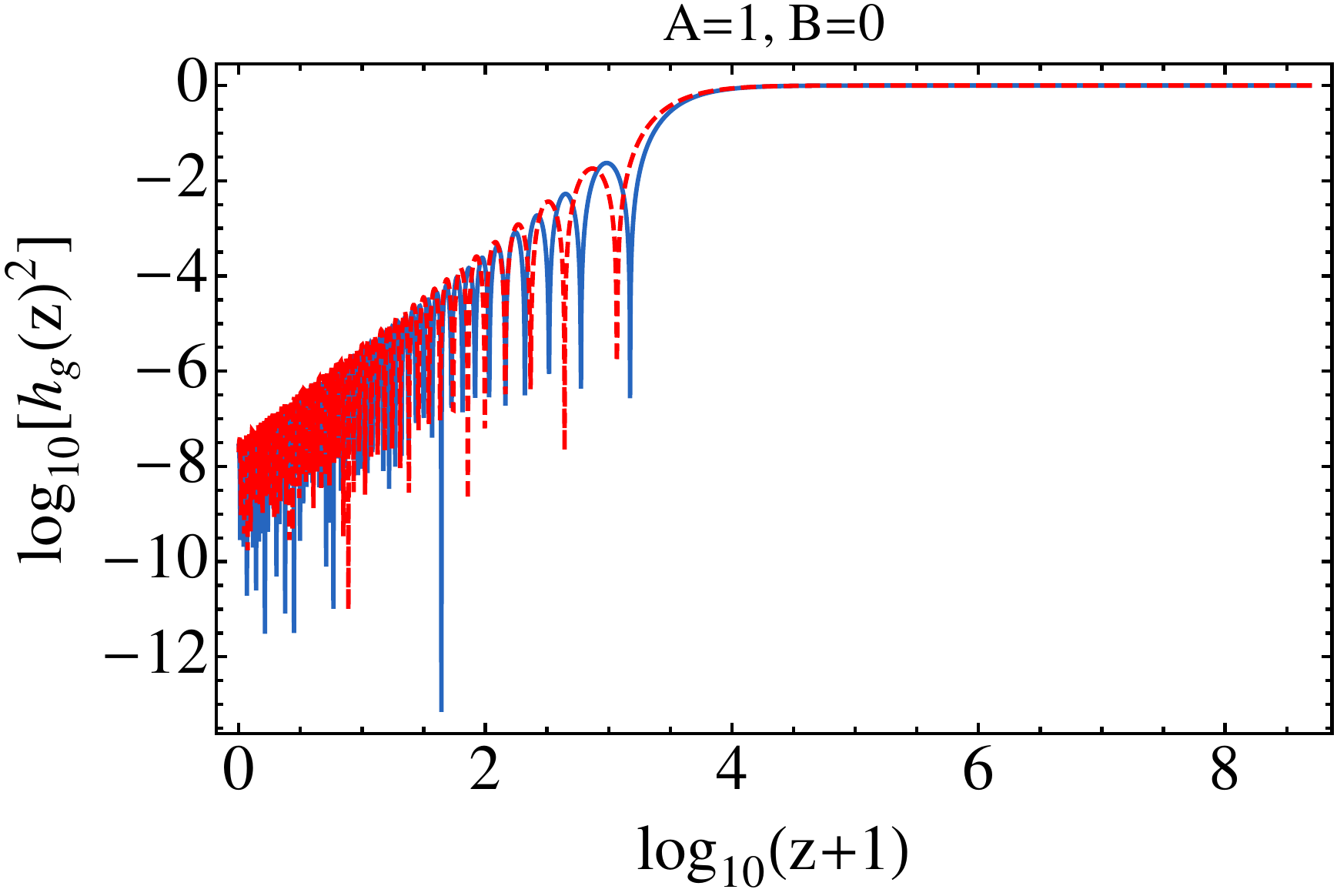}}\qquad
						
\caption{\label{gw3}Evolution of tensor perturbations with wave vector $k=50H_0$ for the metric $g$ for  $A=1$ and $B=10^{-2},
B=10^{-10}, B=10^{-13}$ and $B=0$. In Figs. \ref{A1-B10^-2-k50-hg} and Figs. \ref{A1-B10^-10-k50-hg}, \ref{A1-B10^-13-k50-hg} and \ref{A1-B0-k50-hg} respectively . The red dashed line represents the evolution of tensor perturbation in $\Lambda CDM$, with initial condition after inflation $h_{GR}(\tau_{\text{in}})=1$, $h_{GR}'(\tau_{\text{in}})=0$.}
 \end{figure}

We want  to check whether  there exists a choice of the initial conditions such that we recover an evolution of tensor perturbations similar to the one of $\Lambda CDM$ and an amplitude of tensor perturbations today which is of order of the one of $\Lambda CDM$. We find that if we tune the initial conditions for  $h_f$  to be very small, i.e. $B\ll A$, the instability can be avoided and we can recover an evolution of tensor perturbations at late times (i.e. during the matter era and later) that is similar to the standard gravitational wave evolution of General Relativity.  In 
Fig.~\ref{gw3}  we show how the evolution of tensor perturbations is affected by decreasing  $B$. The evolution of tensor perturbation $h_g$ in the bigravity model for different initial values $B$ at fixed $A=1$ is superimposed to the $\Lambda CDM$ result with initial condition $h_{GR}(\tau_{\text{in}})=1$, $h'_{GR}(\tau_{\text{in}})=0$. 

The amplitude of $h_g$ at late times is proportional to $B$ for values of $B/A\gsim 10^{-16}$. For smaller values of $B$ it converges to the GR result and becomes independent of $B$. In other words, if $h_f$ is not about 16 orders of magnitude smaller than $h_g$ initially, the value of the latter at late times is entirely determined by $h_f$.

\begin{figure}[ht!]
    \centering
    \subfigure[\label{imbr}]
      {\includegraphics[scale=0.40]{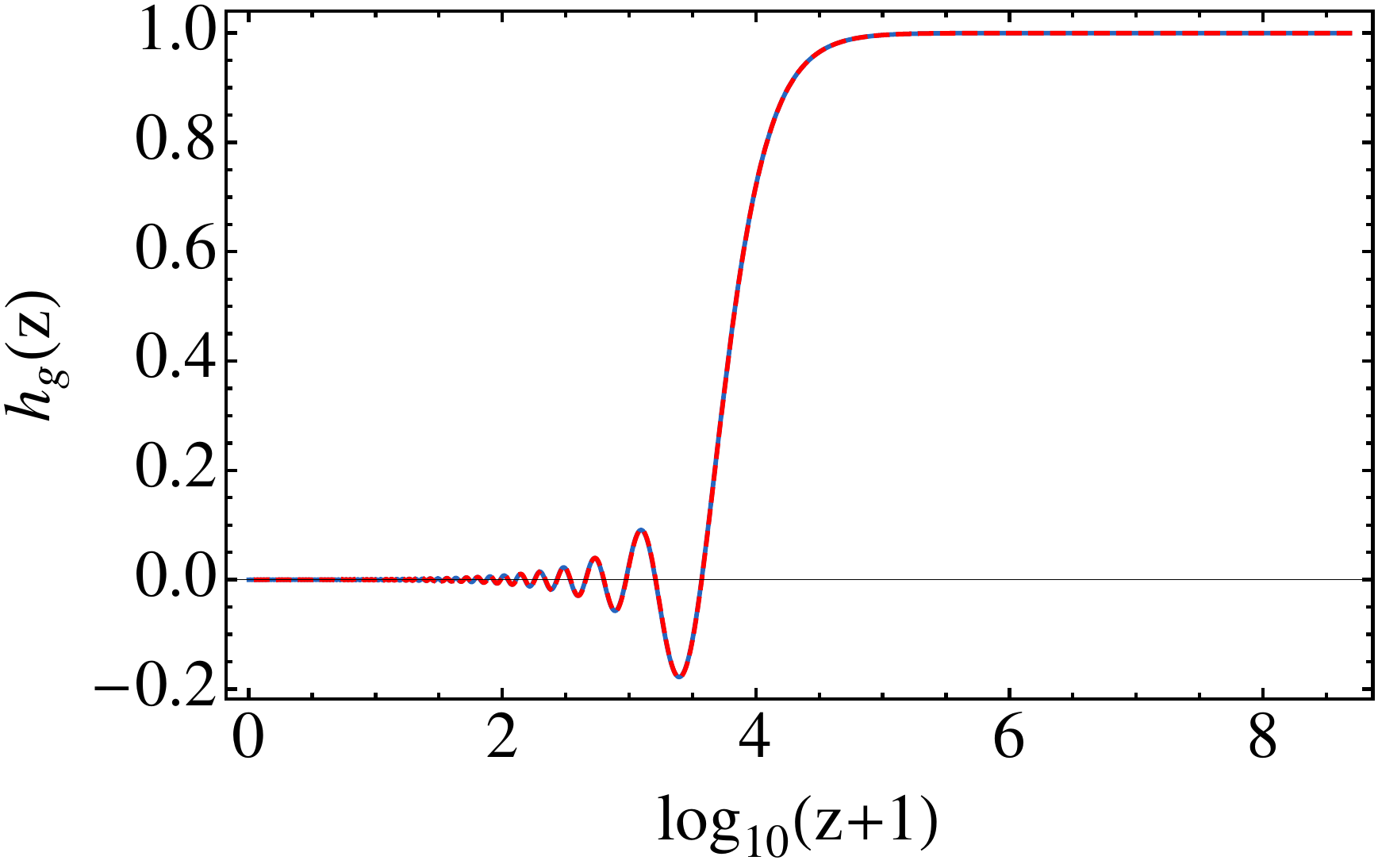}}\qquad
 \subfigure[\label{imbrlate}]
      {\includegraphics[scale=0.40]{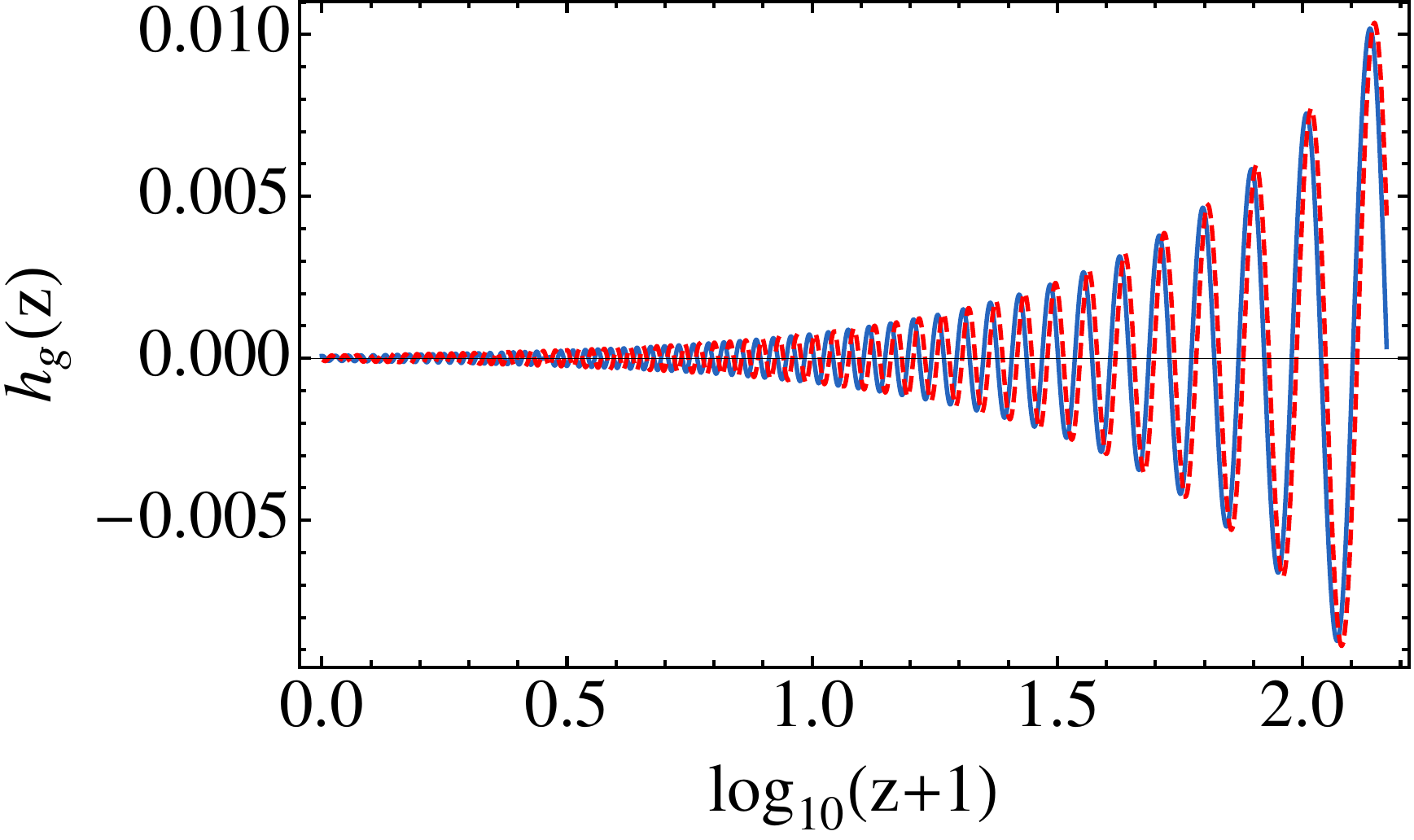}}
\caption{\label{imbrtot}Evolution of tensor perturbations for the metric $g$ in the case  $A=1$ and  $B=0$. The red dashed line represents the evolution of pure GR tensor perturbation, with initial condition after inflation $h_{GR}(\tau_{\text{in}})=1$, $h_{GR}'(\tau_{\text{in}})=0$ calculated on a bigravity background (i.e. we choose the evolution of the scale factor to be the one of the $\beta_1$-$\beta_4$ model).}
 \end{figure}

The small shift in redshift of the bimetric $h_g$ spectrum for $B=0$ with respect to the one of $\La \rm CDM$  is due to the presence of a slight difference between the evolution of the scale factor in the $\beta_1$-$\beta_4$  bigravity model compared to  $\Lambda CDM $ (see Fig. \ref{Hplot}), while the coupling of the tensor mode $h_g$ with $h_f$ in the perturbation equation (\ref{e:hg}) is effectively negligible.  This can be checked easily  comparing the spectrum of tensor perturbations of the $B=0$ bigravity model with the one of $\Lambda CDM$, calculated on a bigravity background: the two spectra overlap with a very good precision\footnote{In other words, if we choose fine-tuned initial conditions for tensor perturbations, $B<10^{-16}A$, the coupling between the two tensor modes in  (\ref{e:hg}) is effectively negligible and the fact that the evolution of tensor perturbations of the physical metric differs form the one in $\Lambda CDM$ can be simply ascribed to a slightly different background evolution. }, as shown in Fig. \ref{imbrtot}.

   \begin{figure}[ht!!]
    \centering
 \subfigure[\label{A0-B1-k50-hg}]
      {\includegraphics[scale=0.40]{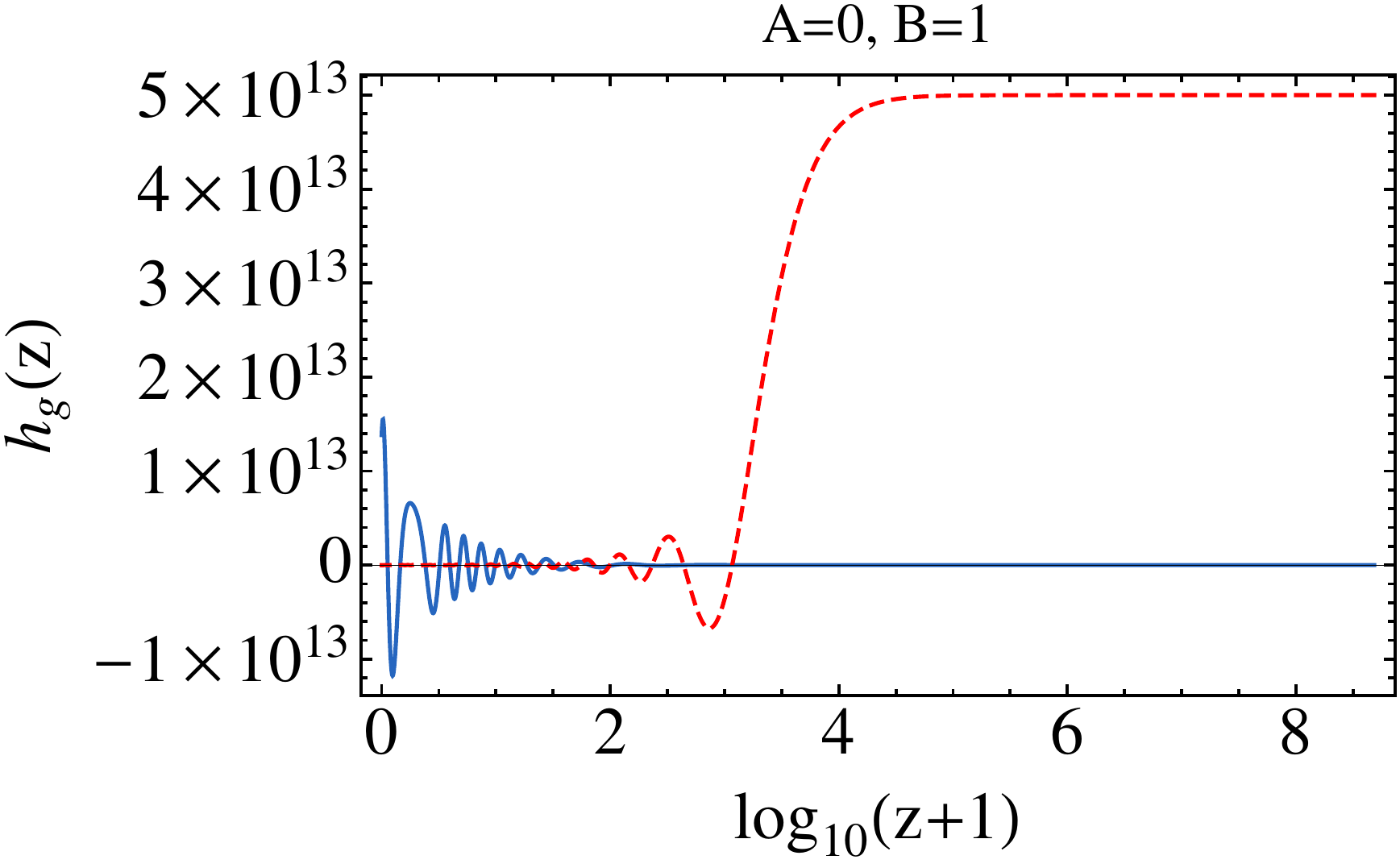}}\qquad
 \subfigure[\label{logA0-B1-k50-hg}]
      {\includegraphics[scale=0.40]{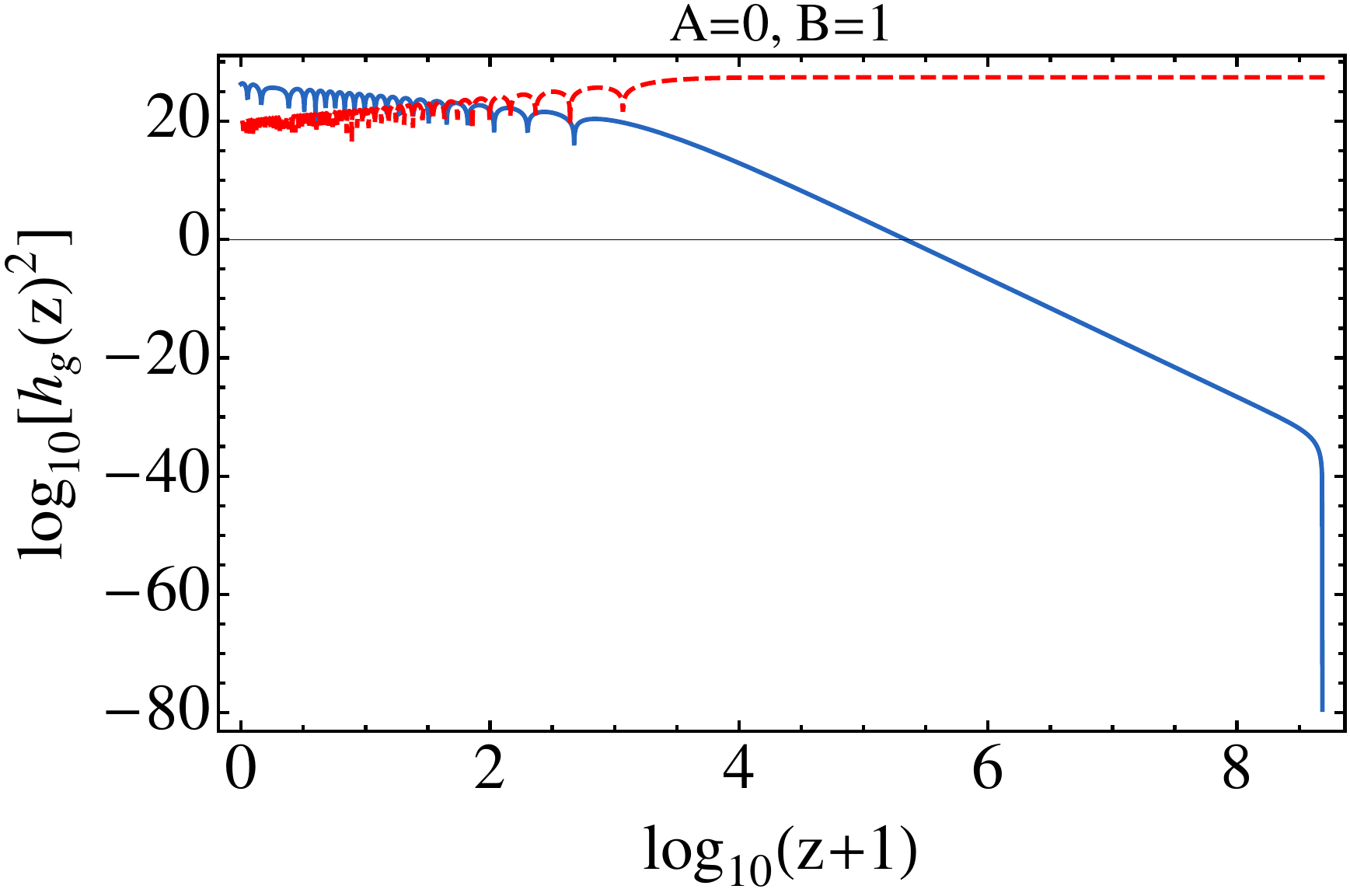}}
\subfigure[\label{A0-B1-k200-hg}]
      {\includegraphics[scale=0.40]{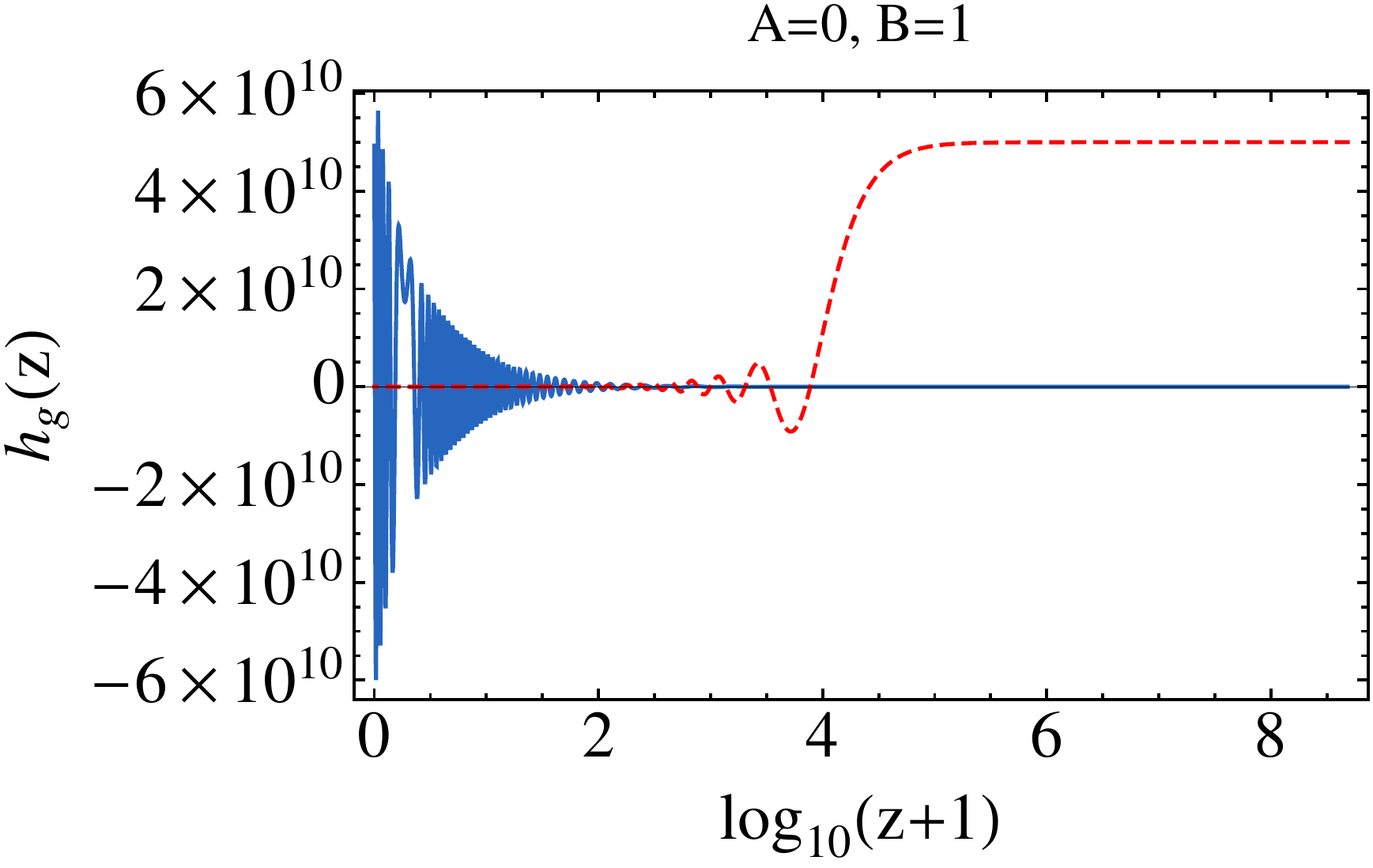}}\qquad
 \subfigure[\label{logA0-B1-k200-hg}]
      {\includegraphics[scale=0.40]{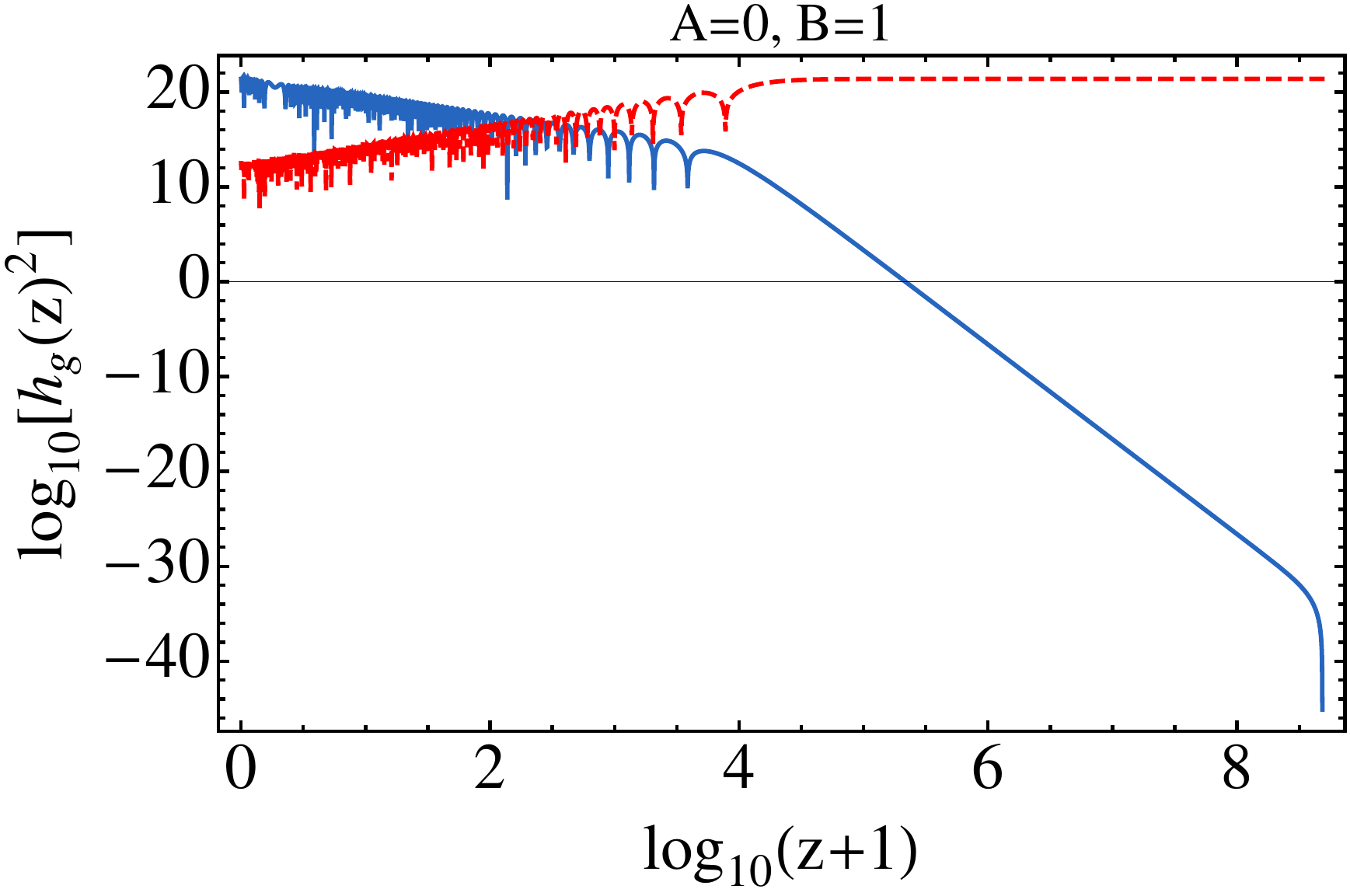}}		
\caption{\label{gw4}Evolution of tensor perturbations for the metric $g$ for the case  $A=0$\,,  and $B=1$ for $k=50H_0$ (Figs. \ref{A0-B1-k50-hg} and \ref{logA0-B1-k50-hg}) and $k=200H_0$  (Figs. \ref{A0-B1-k200-hg} and \ref{logA0-B1-k200-hg}). The red dashed line represents the rescaled  evolution of tensor perturbation in $\Lambda CDM$, with initial condition after inflation $h_{GR}(\tau_{\text{in}})=1$, $h_{GR}'(\tau_{\text{in}})=0$. The rescaling is $5\cdot 10^{13}$  and $5\cdot 10^{10}$ for $k=50H_0$ and $k=200H_0$, respectively.}
 \end{figure}
 
\FloatBarrier
 
\begin{figure}[ht!]
     \centering
    \subfigure[\label{A=1B=0decoupling}]
      {\includegraphics[scale=0.38]{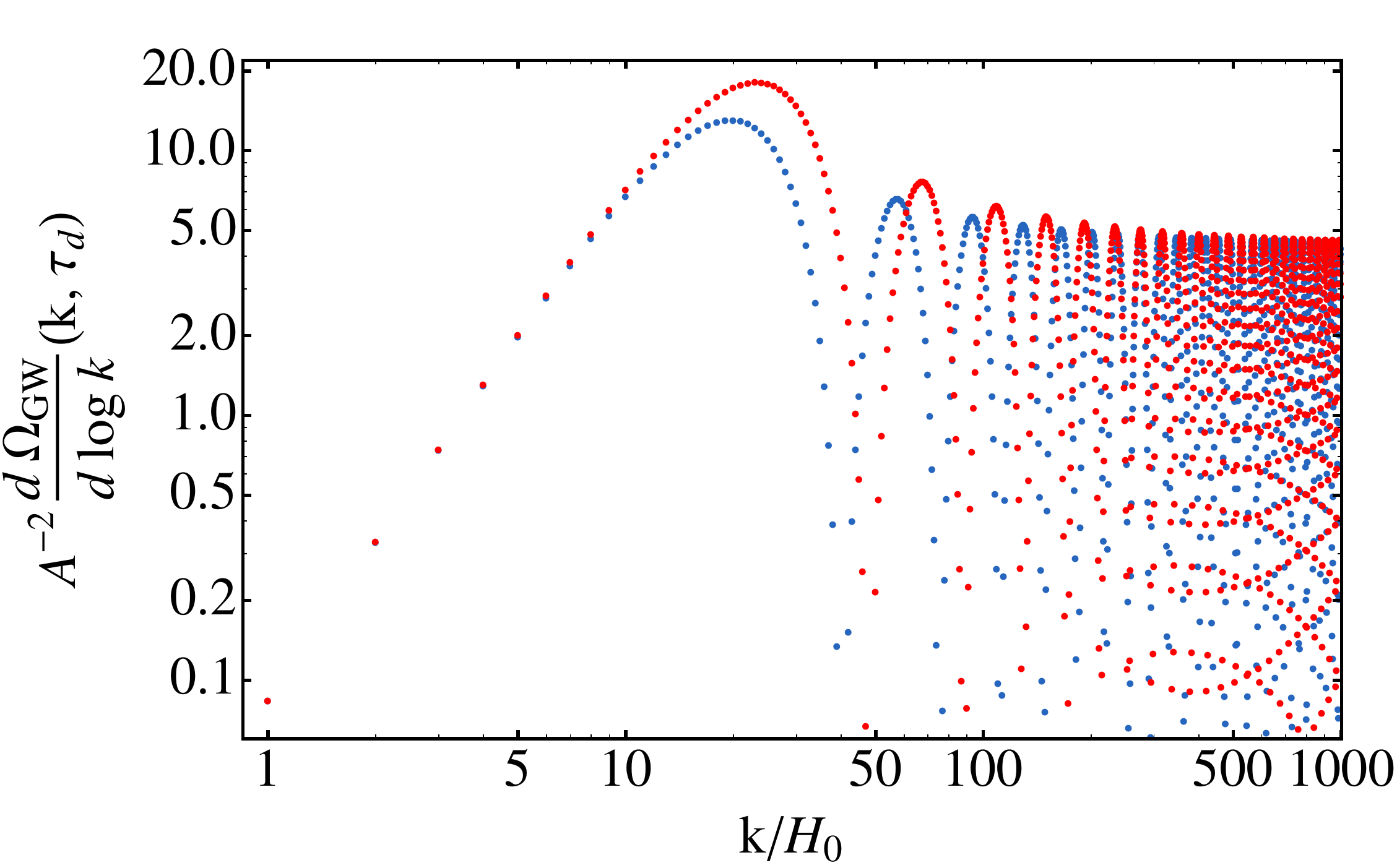}}
 \subfigure[\label{A=1B=0today}]
      {\includegraphics[scale=0.38]{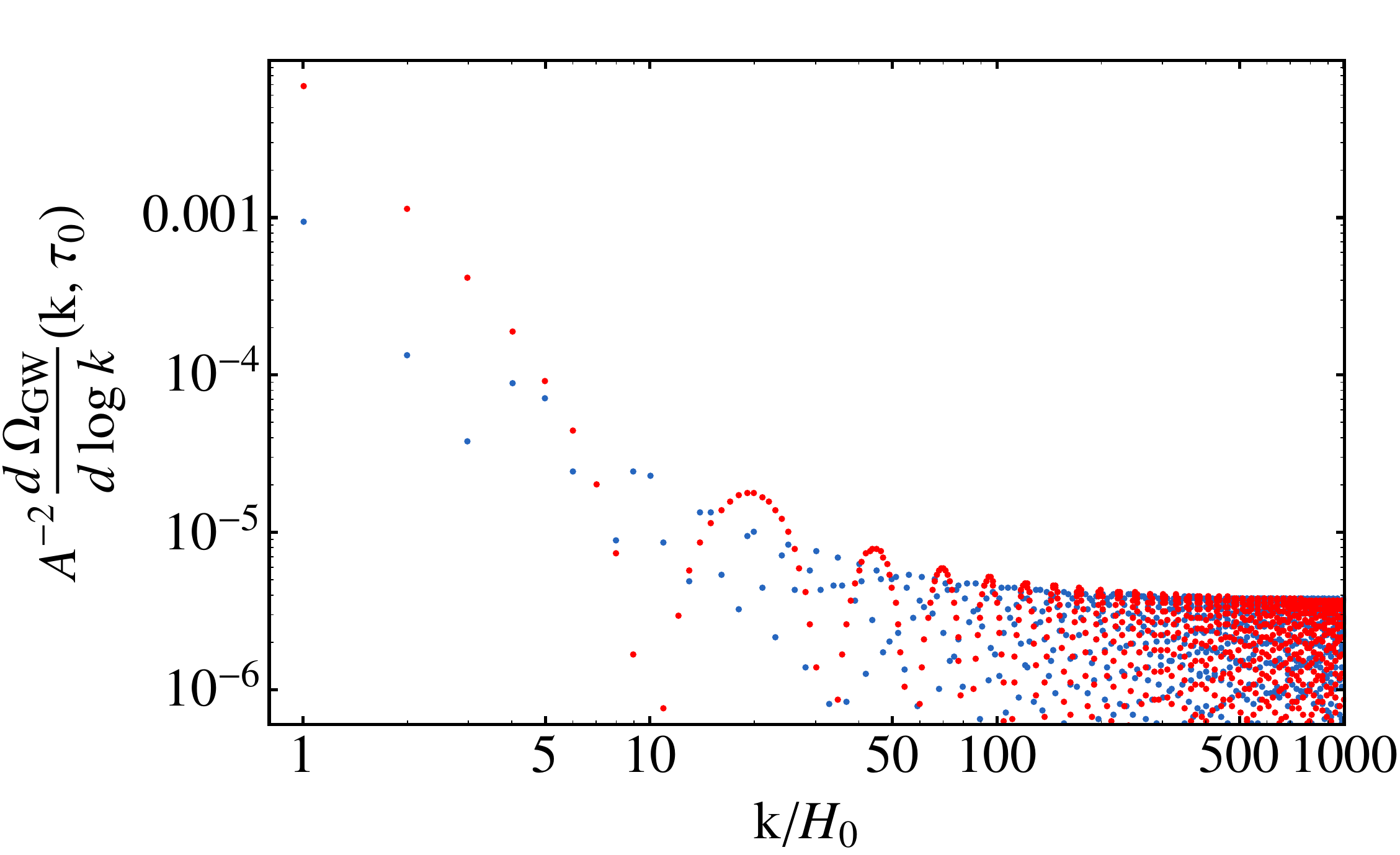}}
          \subfigure[\label{A=1B=-15decoupling}]
      {\includegraphics[scale=0.38]{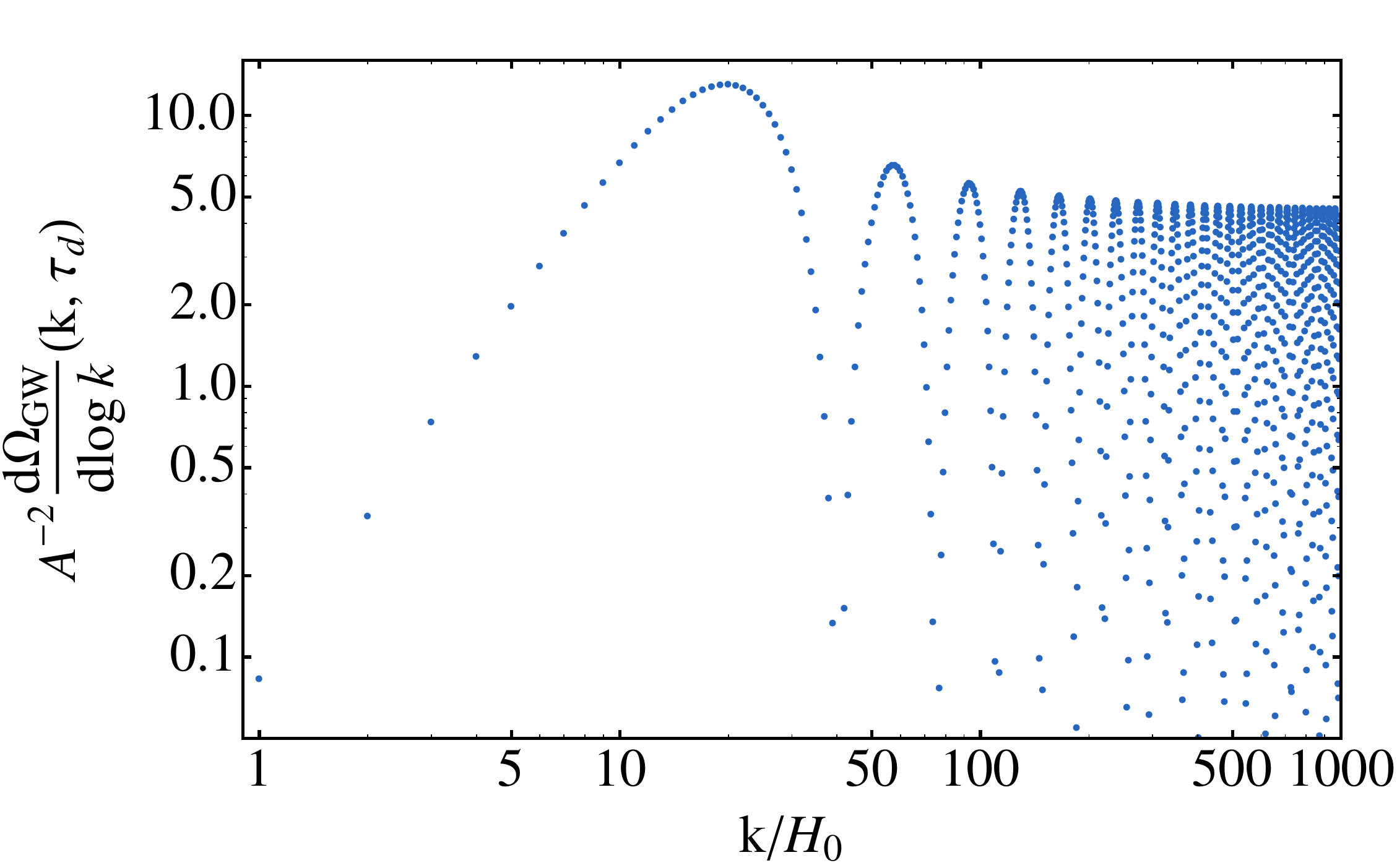}}
 \subfigure[\label{A=1B=-15today}]
      {\includegraphics[scale=0.38]{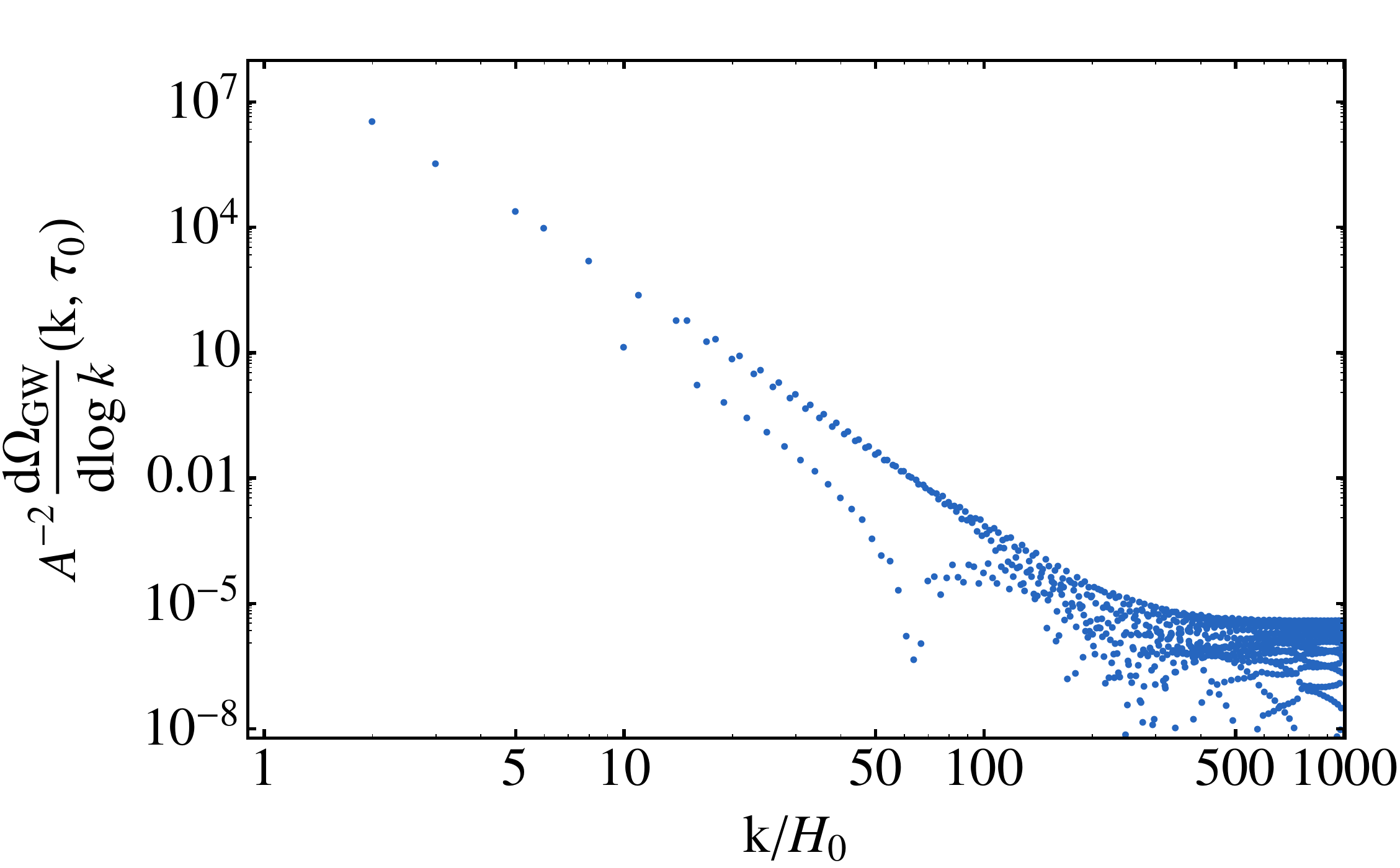}}
          \subfigure[\label{A=0B=1decoupling}]
      {\includegraphics[scale=0.38]{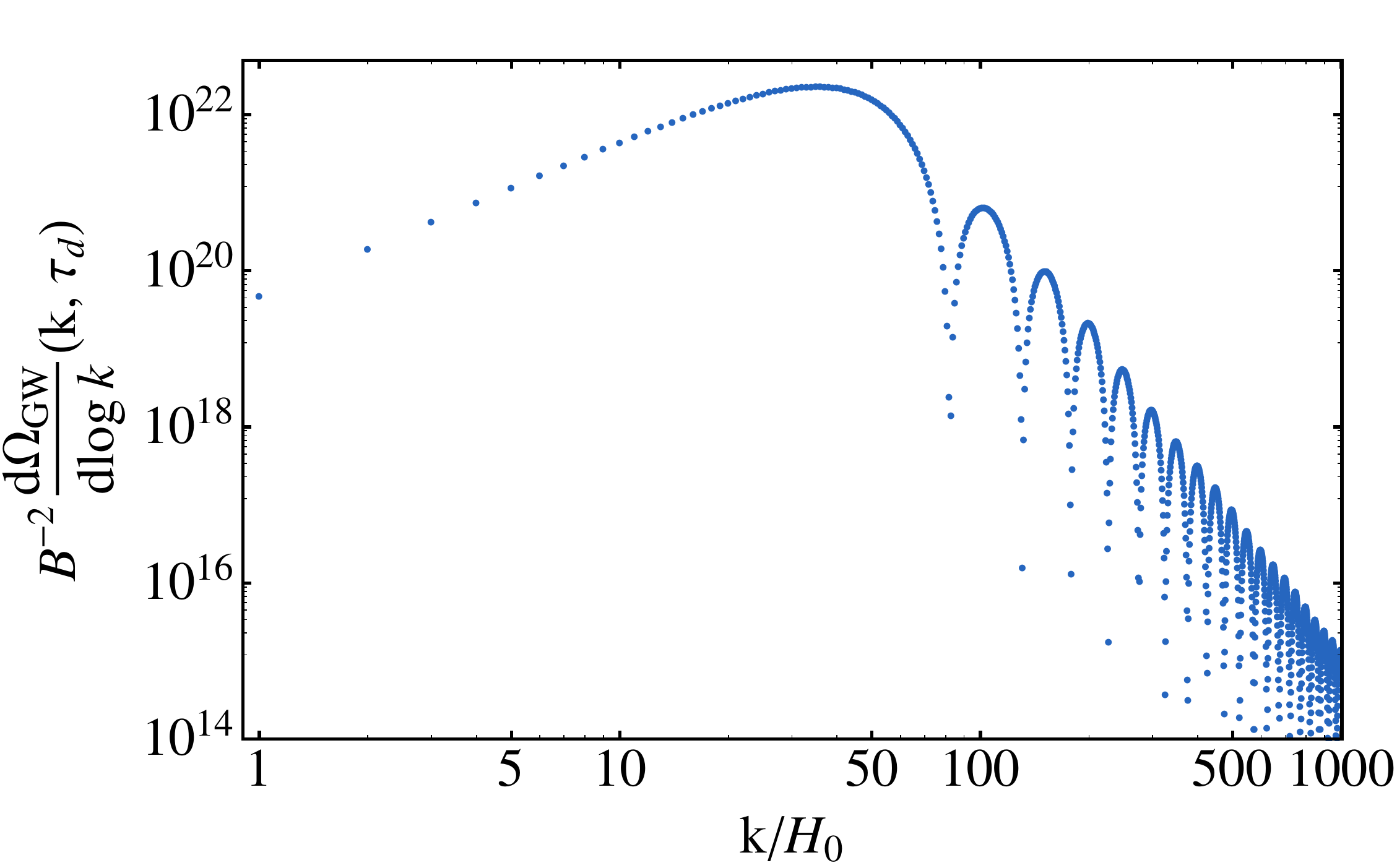}}
 \subfigure[\label{A=0B=1today}]
      {\includegraphics[scale=0.38]{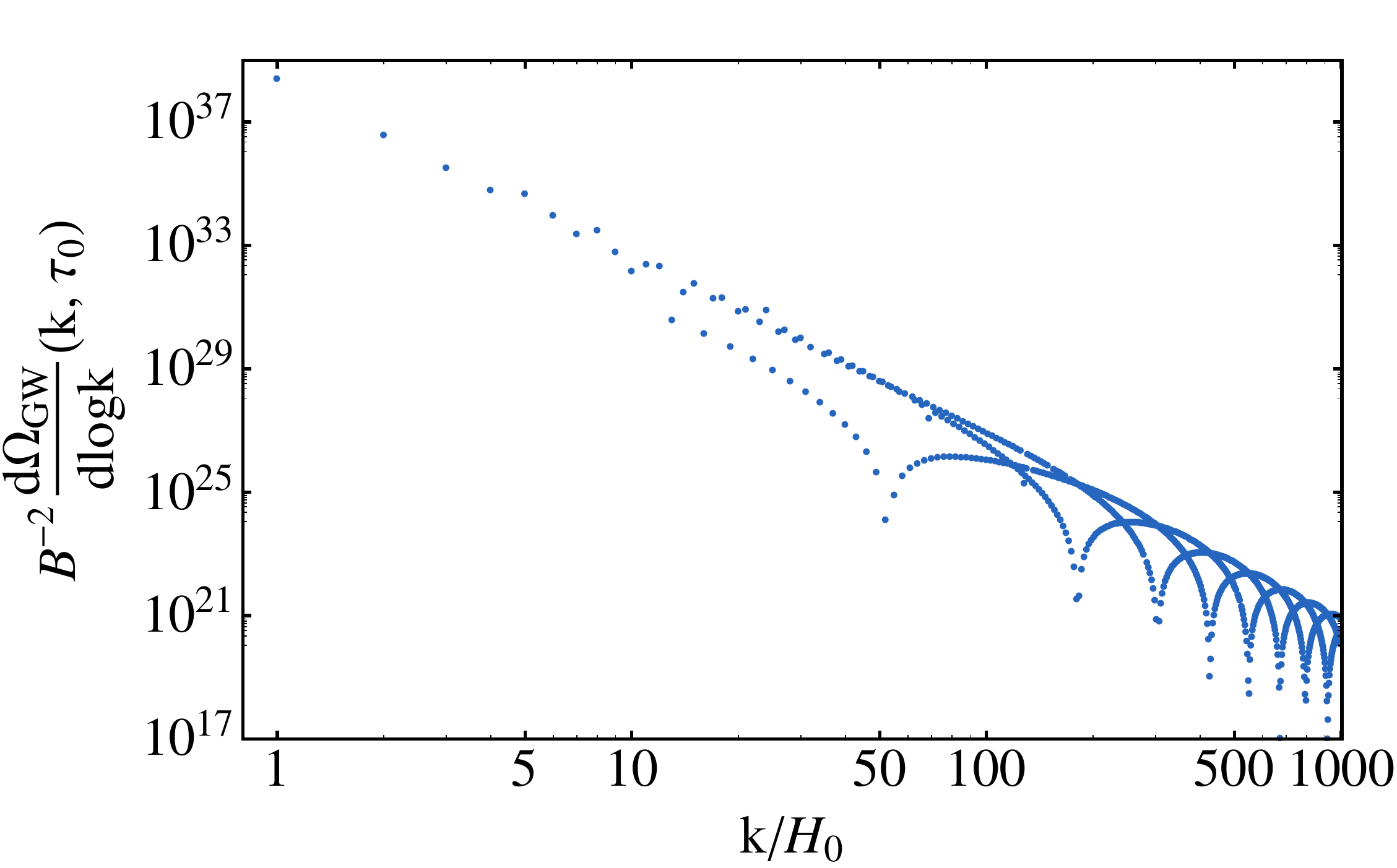}}
 \caption{\label{power}Spectra of the energy of gravitational waves $h_g$ at the redshift of decoupling (left) and today (right) for the cases $B=0$ and  $B=10^{-15}A$ and $A=0$,   in
 Figs. \ref{A=1B=0decoupling}, \ref{A=1B=0today},  Figs.  \ref{A=1B=-15decoupling}, \ref{A=1B=-15today} and Figs.  \ref{A=0B=1decoupling}, \ref{A=0B=1today}\,, respectively. In the first case, the spectrum for the bigravity model at a given redshift is superimposed to the spectrum of $\Lambda CDM$ at the same redshift (red points).}
 \end{figure}

In Fig.~\ref{power} we show the energy spectrum of the gravitational waves $h_g$ in units of the critical density
\be\label{omega}
\frac{d \Omega_{GW}}{d \log k}(\tau, k)\equiv\frac{k^5 |h_g|^2(\tau, k)}{12\, H_0^2}\,,
\ee
for the cases $B=0$ and $B=10^{-15}A$ and  $A=0$ at the redshift of decoupling and today. 
\,

\begin{figure}[ht!]
    \centering
    \subfigure[\label{czero}]
      {\includegraphics[scale=0.38]{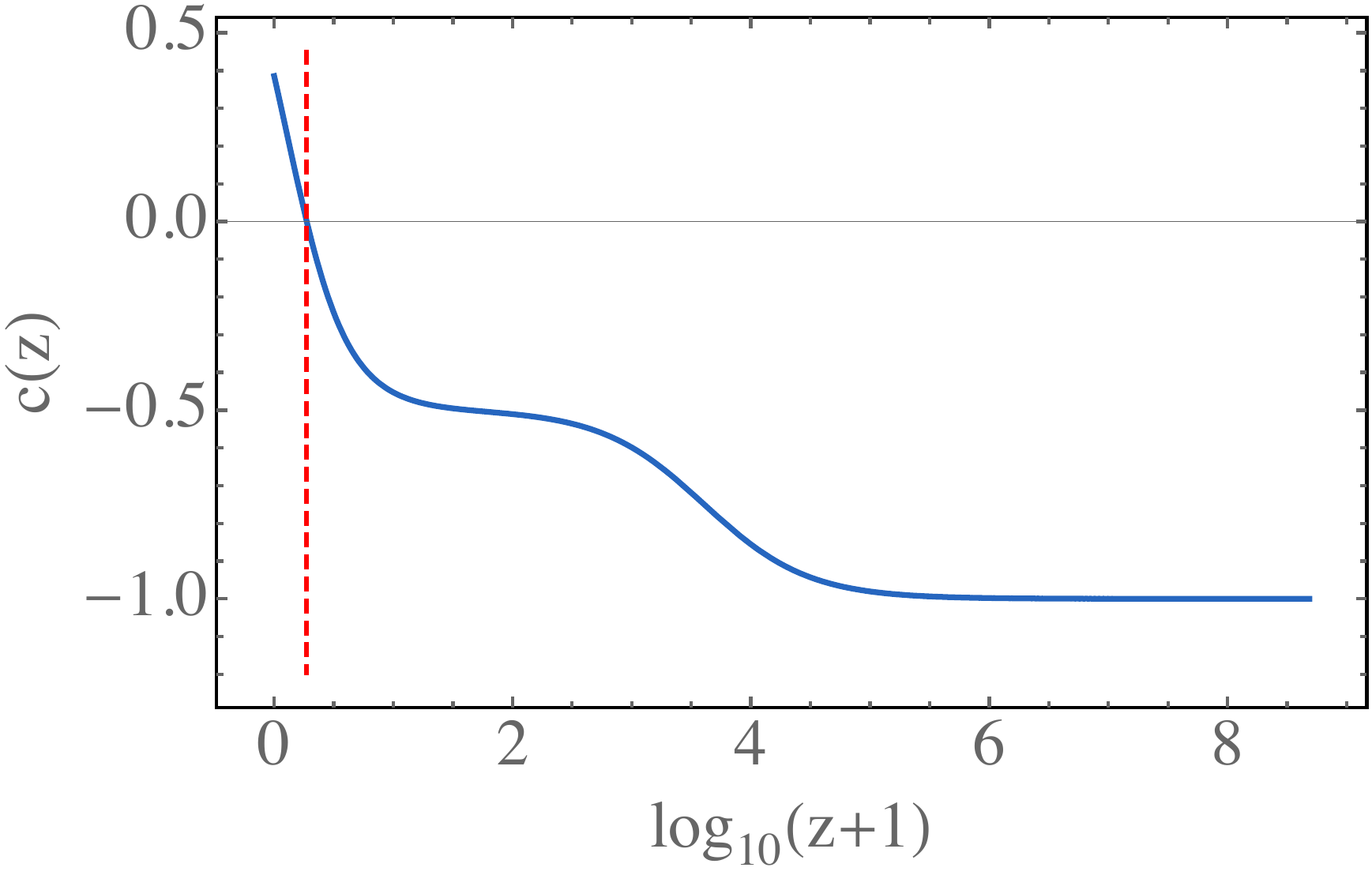}}
 \subfigure[\label{hfzero}]
      {\includegraphics[scale=0.38]{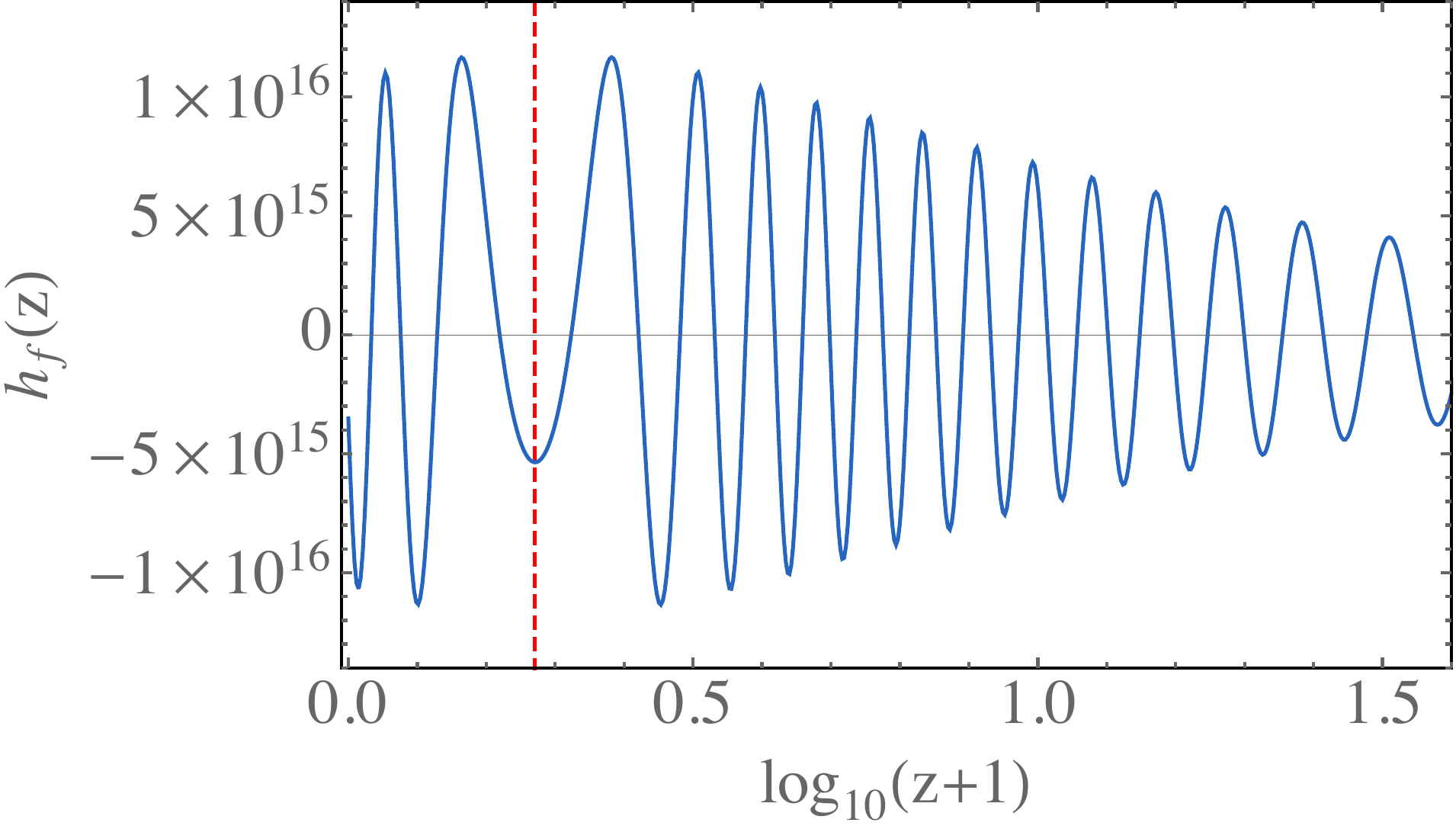}}
\caption{\label{zero}The plots of $c(z)$ and $h_f(z)$ in Fig.~\ref{czero} and~\ref{hfzero} respectively, for a mode $k\simeq100\HH_0$, are shown together with the value of $z$ for which $c$ becomes zero (indicated by the red, dashed line in both the plots), $z_c\sim 0.9$, which corresponds to the value for which the derivative of $h_f$ changes sign.}
 \end{figure}

Let us also make the following remark: one might worry about the singularity of the term   $c'/c$ in eq.~(\ref{e:Bian}) when the lapse function of the $f$ metric, $c$,  passes through zero (see Fig.~\ref{f:lapse}). It can actually be shown by a simple analytic argument that this singularity is just an apparent one. First, by using eq.~(\ref{e:c}), we find that eq.~(\ref{e:hf}) can also be written as
\be\label{e:hf_bis}
h_f''+\left[2\,c\,\HH-\frac{c'}{c}\right]\,h_f'+c^2 k^2\,h_f-m^2\beta_1\frac{c\, a^2}{r}\, \left(h_g-h_f\right)=0\,.
\ee
When $c\sim 0$, eq.~(\ref{e:hf_bis}) can be approximated by
\be\label{e:hf_bis_simple}
h_f''-\frac{c'}{c}\,h_f'=0\,,
\ee
which is solved by $h_f'\propto c$. Therefore the singularity in the term $c'/c$ in eq.~(\ref{e:hf_bis_simple}) when $c= 0$ is cancelled by the factor $h_f' \propto c$ and the differential equation for $h_f$ is regular for all values of $\tau$. The fact that $h_f'$ passes through zero when $c=0$  at $z=z_c \simeq 0.9$ is well visible in Fig.~\ref{zero}.

\section{Discussion and conclusions}\label{s:con}
\subsection{Higuchi bounds}
In cosmology and in particular in theories of modified gravity, it is important to check whether the theory may contain 'ghosts'. In this context a ghost is a degree of freedom with a kinetic term of the wrong sign.  The energy of such a degree of freedom is not bounded from below and via its coupling to other degrees of freedom it can pass to them unlimited amounts of energy, rendering the theory unstable and therefore unphysical. In a static spacetime this instability is exponential.  In an expanding spacetime it is typically a milder power law instability.

As pointed out for the first time by Higuchi~\cite{Higuchi:1986py}, even though the sixth degree of freedom of generic massive gravity, which is always a ghost, is absent, in the dRGT theory of massive gravity the helicity-0 mode of the massive graviton can behave like a ghost for particular values of the theory on a de Sitter background, leading to instabilities of the theory beyond the classical linear regime. The condition for having the kinetic term positive definite is known as Higuchi bound. The study of the stability of massive gravity linearized around a de Sitter background (with flat reference metric) was continued in \cite{Woodard} where it has been shown that the helicities ($\pm 1$ and $\pm 2$) of the massive graviton are stable and unitary since they are immune to the helicity-0 constraint. 

The requirement that the helicity-0 mode on a FRW background has a positive-definite kinetic term is referred to as the \emph{generalized} Higuchi bound. This has been studied for the first time in the bigravity theory in \cite{Fasiello:2012rw} and in \cite{Fasiello:2013woa} (see also \cite{DeFelice} for an alternative analysis of the scalar sector). 

In the background  branch with $\mathcal{H}_f=\mathcal{H}$, the generalised Higuchi bound for the helicity-0 mode can be written as
\be\label{Higuchi}
\tilde{m}^2\left(1+\frac{1}{r^2}\right)-2H^2\ge 0\,,
\ee
where 
\be
\tilde{m}^2\equiv m^2 r \left( \beta_1+2 \beta_2 r +\beta_3 r^2\right)\,.
\ee
For the vector modes we find instead the condition
\be
\tilde{m}^2\ge 0\,,
\ee
which is always satisfied in the $\beta_1$-$\beta_4$ model. This is not the case for the Higuchi bound for the helicity-0 mode as has been noted also in Ref.~\cite{Lagos:2014lca}. Indeed, using the background  constraints, eqs. (\ref{11}-\ref{33}), the bound (\ref{Higuchi}) can be written as 
\be
\beta_1 r-\frac{2}{3} \beta_4 r^2+\frac{1}{3} \frac{\beta_1}{r}\ge 0\,.
\ee
For the best-fit values with $2\beta_1\simeq\beta_4$, this constraint is satisfied only in the asymptotically de Sitter phase of the cosmological expansion, where $r=1$ so that  the bound is saturated. Hence, the scalar sector is affected by a ghost instability.
In an expanding Universe with time dependent Hubble parameter, this instability is not exponential like in the de Sitter case but it
 manifests itself by the presence of a power law growing mode 
 in the scalar sector of perturbations, as found in Sec. \ref{s:scal}. 

In the context of bigravity, the Higuchi bound in the tensor sector has not been properly addressed in the literature. If we write the quadratic kinetic part of the action for the tensor modes from eq. (\ref{startingaction}), we find 
\be\label{tensor-Higuchi}
S_{\rm kin}^{(\pm 2)}\propto M_g^2\int d^4 x\,a^2\left((h'_g)^2+r^2\,\frac{\sqrt{c^2}}{c^2}(h_f')^2\right)\,,
\ee
where $\sqrt{c^2}$ comes from the square root of the determinant of the $f$-metric. Here we can choose either $c$ or $-c$ for $\sqrt{c^2}$, but we are not allowed to choose $|c|$ in order to have a differentiable action \footnote{For a detailed discussion of this point, see also Refs.~\cite{Gratia:2013gka,Gratia:2013uza}.}. To reproduce the phenomenology discussed in this paper we have to choose the positive square root\footnote{We could also choose $-c$, but then we would have to change the sign of $c$ to reproduce the $\La$CDM phenomenology,  so that in the end it does not change our finding that there is a ghost in the tensor sector.}.
Only with this we obtain the correct equations of motion, e.g. eq.~(\ref{e:c}).
Therefore the correct action is
\be
S_{\rm kin}^{(\pm 2)}\propto M_g^2\int d^4 x\,a^2\left((h'_g)^2+r^2\,\frac{1}{c}(h_f')^2\right)\,,
\ee
and the kinetic term for the tensor mode of the $f$ metric is positive definite only if
$ c\ge 0$.

In the background branch that we consider, $c$ is negative and crosses zero at recent time, $z_c\sim 0.9$. This means that along the 
entire cosmological evolution, the helicity-2 sector is affected by a ghost instability. This instability is connected with the one we have observed in the study of perturbations. Actually, writing the $h_f$-equation in the form~(\ref{e:hf_bis}) shows that in the epochs of $c\simeq$ constant, the sign of $c$ indicates whether we have a damped ($c>0$) or anti-damped ($c<0$) evolution. At late time, $z<z_c\simeq 0.9$ the lapse function $c$ changes sign and the tensor sector becomes healthy. This is clearly visible in the numerical solutions shown in Fig.~\ref{f:gw1} where one sees a decay of the amplitude of $h_f$ at very late times.

The physical interpretation to the negative sign of the lapse $c$ is that the time for the $f$-metric sector goes in the opposite direction with respect to the time for the physical sector. 
The scale factor $b$ is decreasing when $a$ is increasing since $\HH_f = b'/bc = \HH =a'/a$. As a consequence, instead of decreasing, the amplitude of tensor perturbations for the $f$-metric are growing in time. 
 
We have chosen the lapse $c$ negative at early times and crossing zero going to positive value only at very recent times. We observe that we could have done the opposite choice, taking $c>0$ at early times. This choice however does not give rise to a viable cosmological evolution. 

Finally, we stress that a  violation of the generalized Higuchi bound 
in a Friedmann universe is not as devastating as it is in a de Sitter universe since the instability it gives rise to is power law and not exponential. Nevertheless, in order to agree with observations which are 
well reproduced with the GR behaviour, we need to fine tune these unstable modes so that their initial conditions are significantly suppressed
compared to the usual GR modes. 

\subsection{Non-linearities}
There is an additional subtlety which 
becomes relevant as soon as there are unstable modes in a theory which is intrinsically non-linear. It is a simple choice of initial conditions to set the unstable modes to zero initially and within linear perturbation theory we have found that their coupling to the other modes is sufficiently suppressed so that they are not generated significantly.

However, once we go beyond linear perturbation theory it is to be expected that the unstabler modes should acquire  amplitudes of the order of $\Phi^2$ where $\Phi \sim 10^{-5}$ is a typical linear mode which we expect to couple to all other modes at the next order.
Therefore, even if a given inflationary model does not generate any tensor perturbations we expect tensor perturbations induced from scalar perturbations on the level of  $10^{-10}$. For the case of general relativity these induced perturbations have been calculated in 2nd order perturbation theory and numerically~\cite{Ananda:2006af,Adamek:2014xba}.

However, the coupling of the $g$-metric to the $f$-metric is suppressed by a factor $m^2\sim \HH_0^2$ which makes it very small.  As we have seen, at least at linear order the coupling of the $g$-metric to $f$-perturbations is nearly always negligible. Therefore, an inflationary model with nearly vanishing initial conditions for the $f$-metric may actually remain viable.

\subsection{Conclusions}
We have found that in bimetric cosmology the tensor perturbations of the second metric, the one that does not couple to matter, exhibits a power law instability, $h_f \propto \tau^3$ on super Hubble scales and $h_f\propto \tau$ on sub Hubble scales. For `natural'  initial conditions with $h_g \sim h_f$, the time evolution of $h_g$ is very different from the behavior in $\La$CDM cosmology. Due to its coupling to $h_f$ it grows rapidly and the final gravitational wave spectrum is determined entirely by the initial amplitude of $h_f$. Only if the initial amplitude of $h_f$ is suppressed by a factor of about $\tau_{\rm in}^3\tau_{0}/k^4$ w.r.t $h_g$ we can recover the standard behavior of gravitational waves.
This opens up new possibilities to test bimetric cosmology via the gravitational wave sector. Not only the final gravitational wave spectrum shown in Figs.~\ref{A=1B=0decoupling} to~\ref{A=0B=1today} can be very different from the standard GR result, but also its time evolution differs leading to a different signature in the CMB. 

To determine the initial conditions $A$ and $B$ we would have to specify an inflationary phase which generates them. Assuming an agnostic point of view as we have done in this work, no firm predictions can be made. Nevertheless, if inflation reheats to about $T_{\rm in}=10^{10}$GeV, the gravitational wave amplitude on very large scales $k\sim \HH_0$ at late times is of the order of $B(T_{\rm in}/T_{\rm eq})^3(T_{\rm eq}/T_0)^{3/2}\simeq 10^{32}B$ unless $B<10^{-32}A$.  
In other words, unless there is a very significant suppression of gravitational waves of the $f$-metric, their amplitude and time evolution will completely dominate the gravitational wave signal and show up in the CMB.

This finding has yet another consequence: we may obtain a significant gravitational wave signal even from low energy inflation. For an inflationary Hubble parameter $H_{\rm in}$, the gravitational wave amplitude is typically $A\simeq H_{\rm in}/M_{p}$, leading to a tensor to scalar ratio $r=16\ep$. Assuming a bimetric theory with $A\sim B$ we
now obtain a scalar to tensor ratio from inflation given by
\be
r  = 16\ep\left(\frac{T_{\rm in}}{T_{\rm eq}}\right)^6\left(\frac{T_{\rm eq}}{T_{\rm 0}}\right)^3 \,.
\ee
Since the scalar perturbation amplitude is
$$ A^2_s \simeq \frac{H^2_{\rm in}}{\ep M^2_p} \simeq 10^{-9}$$ 
this requires $$\ep \simeq 10^9 \frac{H^2_{\rm in}}{ M^2_p}\,.$$
For standard inflation $r=16\ep$ requires $H_{ \rm in}\simeq 10^{-3}M_p$ for a tensor to scalar ratio of $r\sim 0.1$.  

Setting $T^2_{\rm in}\simeq H_{\rm in}M_p$
we obtain for our bimetric cosmology
\be
r  \simeq 2\times10^{10} \frac{T^4_{\rm in}}{M^4_p}\left(\frac{T_{\rm in}}{T_{\rm eq}}\right)^6\left(\frac{T_{\rm eq}}{T_{\rm 0}}\right)^3 \simeq 0.3 \left(\frac{T_{\rm in}}{1{\rm GeV}}\right)^{10} \,.
\ee
For arbitrary values of $B$ we obtain correspondingly
\be
r \simeq 0.3 \left(\frac{T_{\rm in}}{1{\rm GeV}}\right)^{10}\left[\frac{B}{H_{\rm in}/M_p}\right]^2 \,.
\ee
This rules out all simple well motivated inflationary models which cannot provide a mechanism to suppress the generation of $f$-perturbations during inflation.

To conclude, we have found that both, the scalar and the tensor sectors of $\beta_1$-$\beta_4$ bimetric theories, exhibit a power law instability which is related to the Higuchi ghost. Depending on the inflationary model, this instability can render the theory in serious conflict with observation. On the other hand, it may also open a new possibility to obtain significant tensor perturbations from low scale inflation.

\subsection*{Acknowledgments}

We thank Julian Adamek, Jens Chluba, Yves Dirian, Stefano Foffa, Michele Maggiore and Ignacy Sawicki for interesting discussions and suggestions. This work is supported by the Swiss National Science Foundation.

\newpage

\bibliographystyle{utphys}
\bibliography{bi-grefs}
\end{document}